\documentclass{article}

\usepackage{PRIMEarxiv}

\usepackage[utf8]{inputenc} 
\usepackage[T1]{fontenc}    
\usepackage{hyperref}       
\usepackage{url}            
\usepackage{booktabs}       
\usepackage{amsfonts}       
\usepackage{nicefrac}       
\usepackage{microtype}      
\usepackage{lipsum}
\usepackage{fancyhdr}       
\usepackage{graphicx}       
\graphicspath{{media/}}     

\usepackage{graphicx}%
\usepackage{multirow}%
\usepackage{amsmath,amssymb,amsfonts}%
\usepackage{amsthm}%
\usepackage{mathrsfs}%
\usepackage[title]{appendix}%
\usepackage{xcolor}%
\usepackage{textcomp}%
\usepackage{manyfoot}%
\usepackage{booktabs}%
\usepackage{algorithm}%
\usepackage{algorithmicx}%
\usepackage{algpseudocode}%
\usepackage{listings}%
\usepackage{subfigure}
\usepackage[authoryear]{natbib}

\usepackage{tikz}
\usetikzlibrary{shapes.geometric, arrows.meta, positioning}

\tikzstyle{block} = [rectangle, rounded corners, minimum width=1.0cm, minimum height=1.5cm, text centered, draw=black, fill=blue!15, font=\normalsize]
\tikzstyle{arrow} = [thick,->,>=stealth]

\newcommand{\edt}[1]{{\color{black}#1}} 
\newcommand{\rev}[1]{{\color{black}#1}} 
\newcommand{\ed}[1]{{\color{black}#1}} 

\title{Optimized Fish Locomotion using Design-by-Morphing and Bayesian Optimization}

\author{
  Hamayun Farooq \\
  Department of Mathematics and Statistics \\
  Emerson University \\
  Multan, 60700, Punjab, Pakistan\\
   \And
  Imran Akhtar \\
  School of Interdisciplinary Engineering \& Sciences \\
  National University of Sciences \& Technology \\
  Islamabad, 44000, Pakistan\\
  \And
  Muhammad Saif Ullah Khalid \\
  Nature-Inspired Engineering Research Lab \\
  Department of Mechanical and Mechatronics Engineering \\
  Lakehead University \\
  ON P7B 5E1, Thunder Bay, Canada
  \And
  Haris Moazam Sheikh\thanks{Corresponding Author: h.m.sheikh@soton.ac.uk}\\
  Department of Aeronautical and Astronautical Engineering \\
  University of Southampton \\
  Southampton, SO16 7QF, United Kingdom \\
}

\begin{document}
\maketitle

\begin{abstract}
Nature has always inspired scientists and engineers to understand the underlying mechanism leading to optimal design in bio-inspired dynamics. This study presents a computational framework for optimizing undulatory swimming profiles using a combination of Design-by-Morphing and Bayesian optimization strategies. The swimming profile are expressed by \textit{morphing} five baseline bio-inspired profiles using Design-by-Morphing to create an exploratory design space. The optimization objective is to find the optimal swimming profile, wavelength and undulation frequency to maximize propulsive efficiency. {Arbitrary Lagrangian--Eulerian formulation is employed to simulate the unsteady flow around two-dimensional undulating swimmers.} \ed{The optimized swimming profiles demonstrate a marked improvement in propulsive efficiency relative to the reference anguilliform and carangiform modes. The best-performing optimized cases achieve peak efficiencies in the range of 49\%--57\% over a broad range of kinematic conditions, representing an overall enhancement of 16\%--35\% compared to reference anguilliform and carangiform modes. The improved performance is attributed to favorable surface stress distributions and enhanced energy recovery mechanisms.} A detailed force decomposition reveals that the optimal swimmer minimizes resistive drag and maximizes constructive work contributions, particularly in the anterior and posterior body regions. Spatial and temporal work decomposition indicates a strategic redistribution of input and recovered energy, enhancing performance while reducing energetic cost relative to propulsive force. 
These findings demonstrate that morphing-based parametric design, when guided by surrogate-assisted optimization, offers a powerful framework for discovering energetically efficient swimming gaits, with significant implications for the design of autonomous underwater propulsion systems and the broader field of bio-inspired locomotion.
\end{abstract}


\section{Introduction}
\label{sec:introduction}

Optimization of \edt{undulatory kinematic} profiles for self-propelling systems, such as fish-like robots, has become a pivotal area of research in bio-inspired robotics and computational fluid dynamics (CFD) \citep{abouhussein2023computational,li2025optimization,fernandez2022numerical, lin2019performance, cui2018complex}. These systems aim to mimic the natural motion of aquatic organisms, generating thrust for propulsion by effectively converting undulatory motion into mechanical energy. Such systems show significant promise in various applications, including autonomous underwater vehicles (AUVs) \citep{rattanasiri2015numerical,cui2018complex}, where energy efficiency and maneuverability are critical performance metrics \citep{sun2024investigation, yu2017numerical}. \rev{To enhance the efficiency of these systems, optimizing the swimming profile is essential, ensuring maximum propulsion while minimizing energy consumption \ed{or maximizing the propulsive forces to propel faster. In bio-inspired systems, the propulsion efficiency is often enhanced through fin-fin interaction or fish schooling ~\citep{khalid2018hydrodynamics,akhtar2007hydrodynamics,akhtar2005biologically,mittal2003towards,akhtar2003thrust}}. Such optimization efforts are a key source of inspiration for engineering applications, including energy-efficient AUVs, flow-energy harvesting devices \citep{farooq2023comparative, salazar2018optimal}, and bio-inspired swimming microrobots for targeted drug delivery \citep{tabak2018hydrodynamic, bunea2020recent}, where achieving optimal propulsion translates directly into improved performance and extended operational capability.}

The fish-like swimming profiles traditionally used in CFD are categorized primarily based on the mode of body deformation, which is often quantified by the non-dimensional body wavelength $\lambda^*$ (the ratio of body wavelength to body length) \citep{sfakiotakis2002review, triantafyllou2000hydrodynamics}. These include anguilliform ($\lambda^* \ll 1$) motion, where the entire body undergoes large-amplitude undulations (e.g., eel); subcarangiform and carangiform ($\lambda^* \approx 1$) modes, where deformation is progressively confined to the posterior half or third of the body (e.g., trout and mackerel, respectively); and thunniform ($\lambda^* \gg 1.0$) motion, characterized by minimal body deformation with oscillations largely restricted to the caudal fin (e.g., tuna) \citep{khalid2021larger,khalid2021anguilliform}. These deformation-based classifications are widely adopted in CFD to assess the hydrodynamic performance, thrust production, and propulsive efficiency of bio-inspired systems. In this study, we aim to systematically optimize the swimming profile by leveraging these existing bioinspired modes via a strongly coupled \emph{fluid–structure–optimization} framework to achieve enhanced propulsive performance. Over the years, numerous studies have focused on understanding the fluid-structure interaction (FSI) in bio-inspired robotic systems \citep{ying2022parameter}, particularly in the context of fish-like robots \citep{koiri2025comprehensive,khalid2016hydrodynamics, Khalid2020a, Khalid2020b}. These studies have primarily concentrated on the optimization of the kinematic and geometric properties of the robotic swimmer to improve its hydrodynamic performance. 



In particular, the propulsive performance of such systems is governed by key parameters including oscillation frequency, amplitude, wavelength, tail geometry, and swimming profile. \citet{wang2017optimization} investigated the effects of tail structure and oscillation parameters on the hydrodynamic performance of a robotic fish using CFD. Their study confirmed that reverse Kármán vortex shedding at optimal Strouhal numbers (0.42 -- 0.55) \edt{generated} high thrust with minimal drag, highlighting the importance of tuning kinematic parameters to achieve efficient propulsion. In a complementary effort, \citet{ma2021oscillatory} conducted a parametric optimization study using COMSOL-based simulations to analyze how frequency, amplitude, and body wavelength \edt{affected the} swimming efficiency. Their results \edt{reinforced} the idea that systematic variation and optimization of motion profiles \edt{could} yield significantly improved performance metrics. \citet{maertens2016optimal} conducted a detailed investigation of optimal undulatory propulsion through both two-dimensional (2D) and three-dimensional (3D) simulations. Their study identified key parameters\edt{,} such as Strouhal number, the phase lag between heave and pitch at the trailing edge, and the angle of attack, to maximizing efficiency. In 2D simulations of self-propelled swimming, efficiency improved from approximately 40\% for a measured carangiform profile to 57\% with optimized body bending. In 3D simulations of a danio-like swimmer, efficiency increased from 22\% to 35\%, with angular recoil and upstream bending amplitude playing a significant role. Extending the analysis to a pair of interacting swimmers, they demonstrated that the downstream fish could exploit the wake of the upstream fish for energetic benefits. When synchronized with the upstream vortices, the downstream fish achieved up to 66\% efficiency in an in-line configuration and up to 81\% in an offset configuration. \ed{More recently, \citet{abouhussein2023optimized} investigated the hydrodynamics of accelerating phalanx fish schools coupled with a gradient-free surrogate-based optimization algorithm. By examining midline kinematics, inter-swimmer separation distance, and phase synchronization, they showed that the optimal kinematics of accelerating schools closely resemble those of solitary accelerating swimmers. For thunniform bio-inspired swimmers, an optimal separation distance of approximately twice the body length was identified. Furthermore, separation distance was found to exert a stronger influence on propulsion efficiency than phase synchronization when other parameters were held constant.}

\ed{Despite extensive research efforts, most optimization studies on bio-inspired swimmers are conducted around fixed baseline geometries or through localized parametric variations, which inherently limit the exploration of broader design spaces and constrain geometric flexibility.} In contrast, the \textit{Design-by-Morphing} (DbM) methodology enables continuous interpolation across multiple baseline geometries, allowing exploration of a broader and more expressive design space. For example, \citet{sheikh2023airfoil} introduced DbM for \edt{an} airfoil design, generating radical new shapes by combining a small set of baseline airfoils and optimizing them for maximal lift-to-drag ratio and stall margin using genetic algorithms. They demonstrated that DbM substantially outperforms traditional shape parameterizations in terms of aerodynamic efficiency and diversity of optimal designs. Similarly, \citet{raj2023flow} applied camber line morphing in a NACA airfoil, showing up to 22\% increase in aerodynamic efficiency through skin-morphed trailing-edge deformation paired with gradient-based optimization. These authentic studies illustrate DbM’s potential to discover high-performance morphing configurations, principles that can be translated into hydrodynamic profile optimization for robotic swimmers.

Despite recent advances, a gap remains in the literature regarding the simultaneous optimization of both geometric (morphing weights) and kinematic parameters (such as oscillation frequency and wavelength) in self-propelling bio-inspired systems, particularly under realistic hydrodynamic loading conditions with a focus on \ed{maximizing propulsive efficiency}. This study addresses this gap by introducing a DbM framework coupled with Bayesian optimization (BO) to jointly optimize the swimming profile and kinematics of a fish-like robotic swimmer. \rev{A recent study by \citet{lee2024bayesian} demonstrated the growing use of combined DbM and BO (DbM–BO) frameworks for fluid dynamic design problems. In their work, DbM was employed to generate a diverse family of riblet surface geometries, which were evaluated using high-fidelity large-eddy simulations to quantify drag-reducing performance. The design space was explored with a BO algorithm, which efficiently handled the mixed continuous and discrete variables inherent in riblet shape definition. Remarkably, within only 125 optimization epochs, the study identified three novel riblet configurations. This effort highlights how the integration of DbM with BO provides a powerful and computationally feasible route for discovering unconventional, high-performance designs in complex turbulent flow applications.}

\rev{We employ 2D CFD simulations to evaluate the propulsive performance of candidate swimming profiles. Although real swimming is inherently three-dimensional, two-dimensional simulations, which are essentially slices of a biological swimmer, are widely adopted in the literature due to their ability to capture the essential hydrodynamic mechanisms with significantly reduced computational cost.} Swimming profiles are constructed as linear combinations of five baseline geometries, enabling exploration of a high-dimensional continuous design space. To assess the propulsive performance of candidate configurations, full-order CFD simulations are employed, capturing the underlying FSI dynamics and evaluating metrics\edt{,} such as propulsive efficiency and power consumption.

To efficiently navigate this computationally intensive design space, we integrate BO as a surrogate-assisted global optimization strategy. BO leverages probabilistic models, such as Gaussian Processes, to predict the performance landscape and iteratively sample the most promising candidates with minimal CFD evaluations. Through this integrated DbM–BO framework, our aim is to identify optimal combinations of morphing parameters and swimming kinematics that maximize propulsive efficiency, thereby advancing the design of energy-efficient, self-propelling fish-like robotic systems.

The remainder of this paper is organized as follows: Section \ref{sec:nummethd} describes the methodology used for optimizing the swimming profile, including the design-by-morphing technique, CFD simulations, and performance metrics. The validation of the proposed numerical strategy is presented in Section~\ref{sec:valid}. Section \ref{sec12} presents the results of the optimization process, followed by a discussion of the findings. Finally, Section \ref{sec13} summarizes the key conclusions.

\section{Methodology}
\label{sec:nummethd}
This section outlines the integrated methodology used to identify optimal swimming profiles that maximize propulsive efficiency. The process involves three key steps: (i) design-by-morphing, where swimmer shapes are generated using a set of parametric design variables; (ii) CFD-based performance evaluation, where each design is tested through 2D CFD simulations to compute the corresponding efficiency; and (iii) BO, where a surrogate-guided search strategy iteratively explores the design space to improve performance.

\begin{figure}[htbp]
    \centering
    \begin{tikzpicture}[>={Stealth[length=3mm,width=4mm]},node distance=2.8cm]

        \node (start) [block, fill=gray!20, text width=1.8cm, align=center] 
            {Initial Design Sampling};
        \node (design) [block, right of=start, text width=1.4cm, align=center] 
            {Design by Morphing};
        \node (cfd) [block, right of=design, fill=green!20, text width=2.cm, align=center] 
            {CFD Simulation \& Efficiency Evaluation};
        \node (bo) [block, right of=cfd, fill=orange!20, text width=2.cm, align=center] 
            {Bayesian Optimization (MixMOBO)};
        \node (opt) [block, right of=bo, fill=yellow!20, text width=1.5cm, align=center] 
            {Optimum Profile};

        \draw [->] (start) -- (design);
        \draw [->] (design) -- (cfd);
        \draw [->] (cfd) -- (bo);
        \draw [->] (bo) -- (opt);
        \draw [->] ([xshift=7mm,yshift=0mm] bo.south west) -- ++(0,-0.8) -- ++(-5.18,0) node[midway, above] {Not Converged} -- (design.south) ;
        \draw [->] ([xshift=-7mm,yshift=0mm]bo.south east) -- ++(0,-0.8) -- ++ (2.38,0) node[midway, above] {Converged} -- (opt.south);

    \end{tikzpicture}
    \caption{Optimization framework combining DbM, CFD simulation, and BO to maximize propulsive efficiency.}
    \label{fig:optimization_flowchart}
\end{figure}
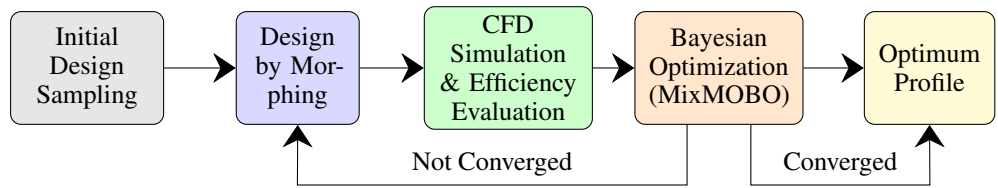

These steps form an iterative loop where each cycle refines the design based on previously gathered performance data. This approach balances exploration and exploitation, allowing us to efficiently navigate the complex design space with a limited number of expensive CFD evaluations. The overall workflow of this iterative process is illustrated in Fig.~\ref{fig:optimization_flowchart}. In the subsequent subsections, we provide a comprehensive explanation of each step involved in the optimization framework.

\subsection{Design-by-Morphing}
\begin{figure}[h] 
    \centering
    \begin{minipage}{0.45\textwidth} 
        \centering
        \subfigure[]{\includegraphics[width=\textwidth]{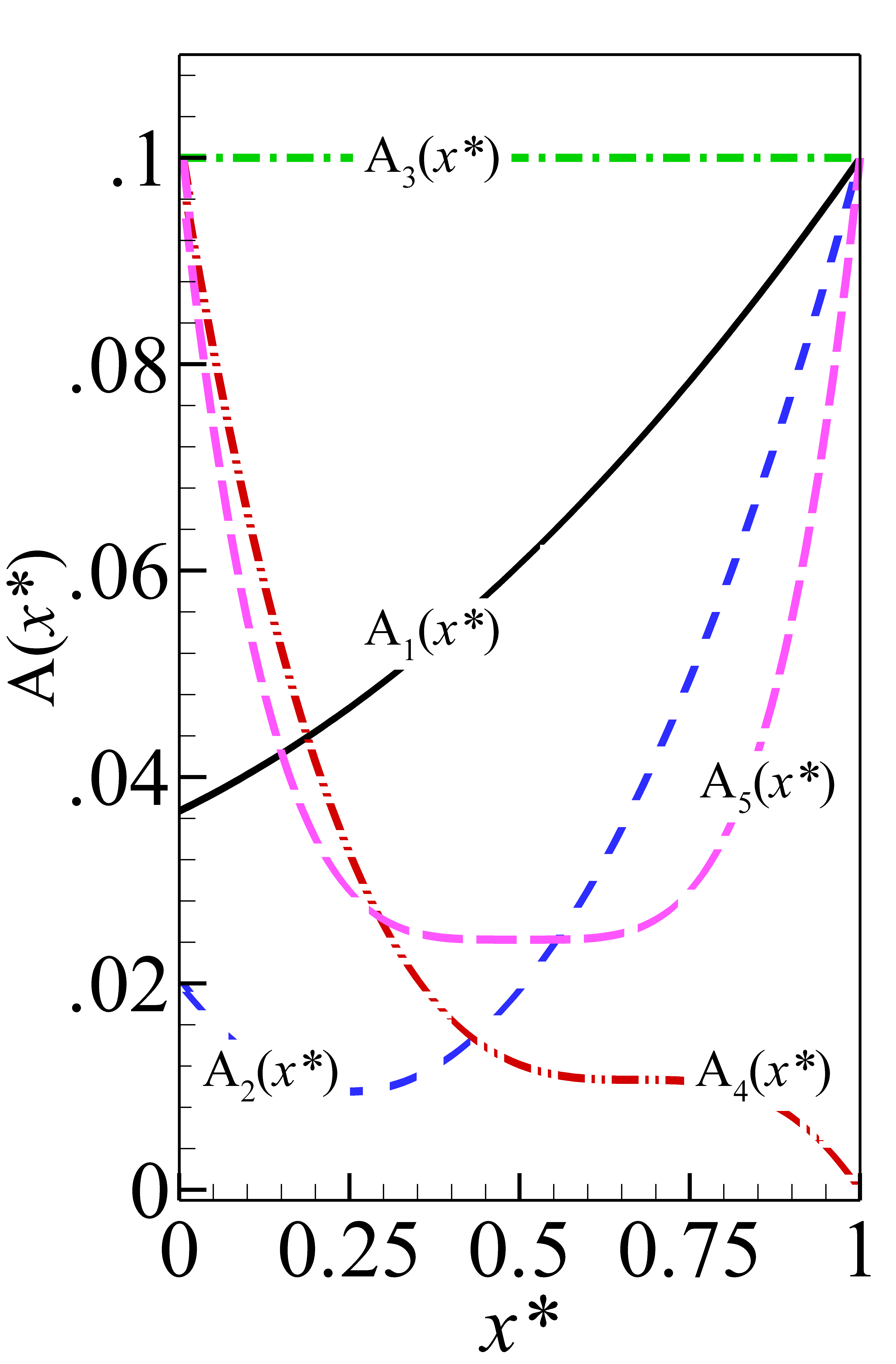}\label{fig:kina}}
        
    \end{minipage}
    \hfill 
    \begin{minipage}{0.48\textwidth}
        \centering
        \includegraphics[width=\textwidth]{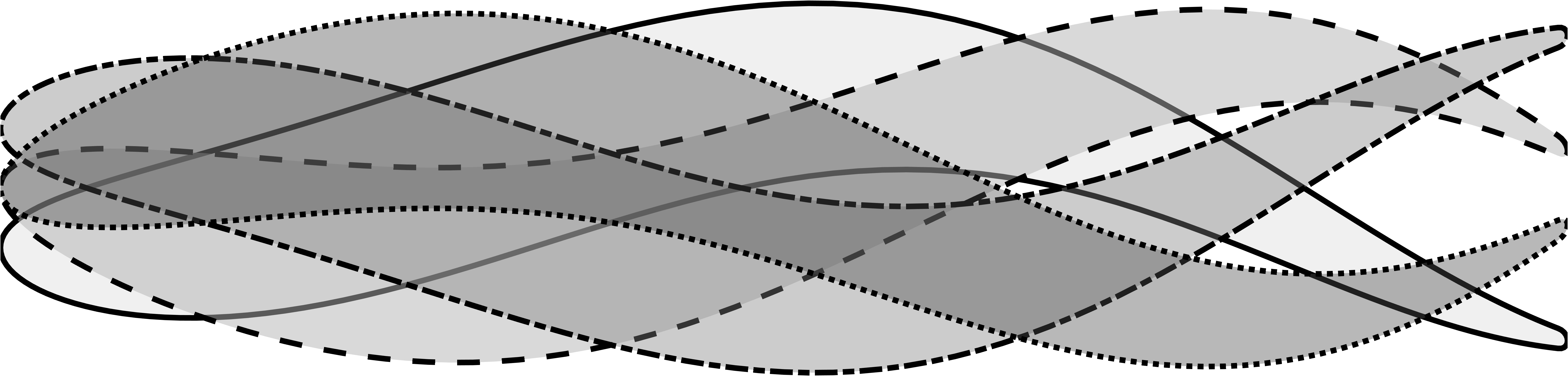}
        \includegraphics[width=\textwidth]{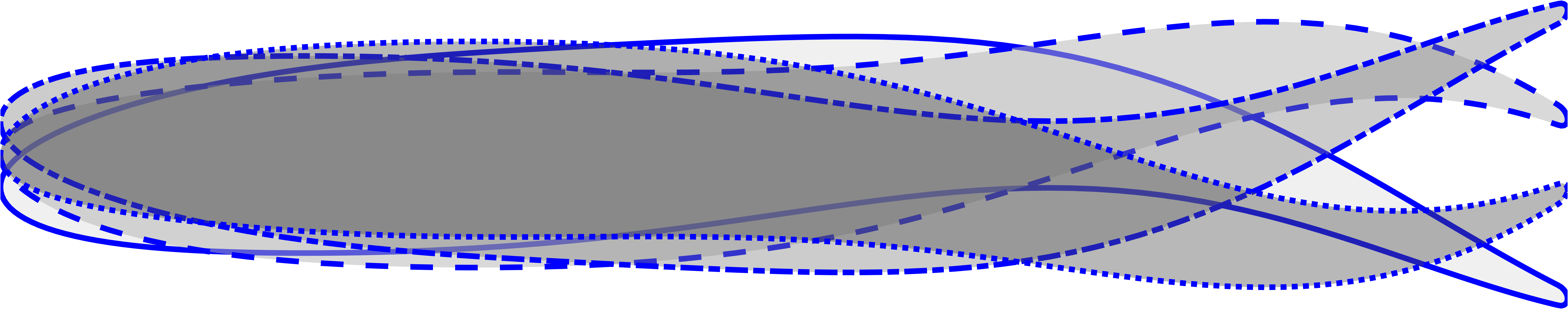}
        \includegraphics[width=\textwidth]{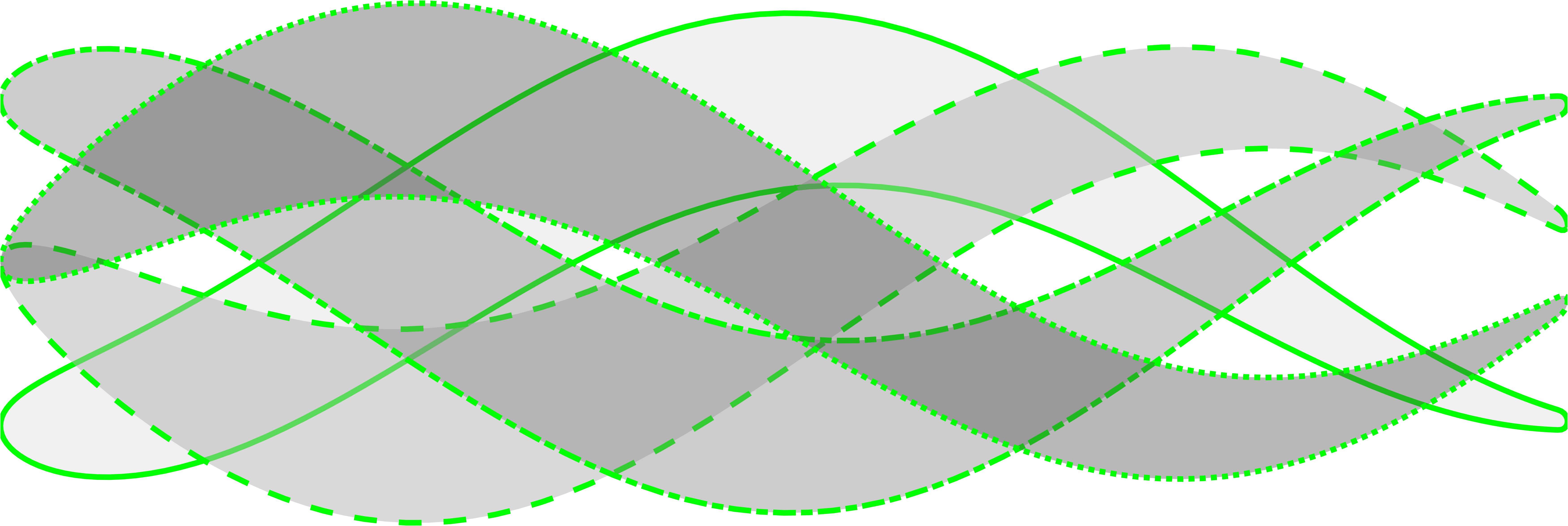}
        \includegraphics[width=\textwidth]{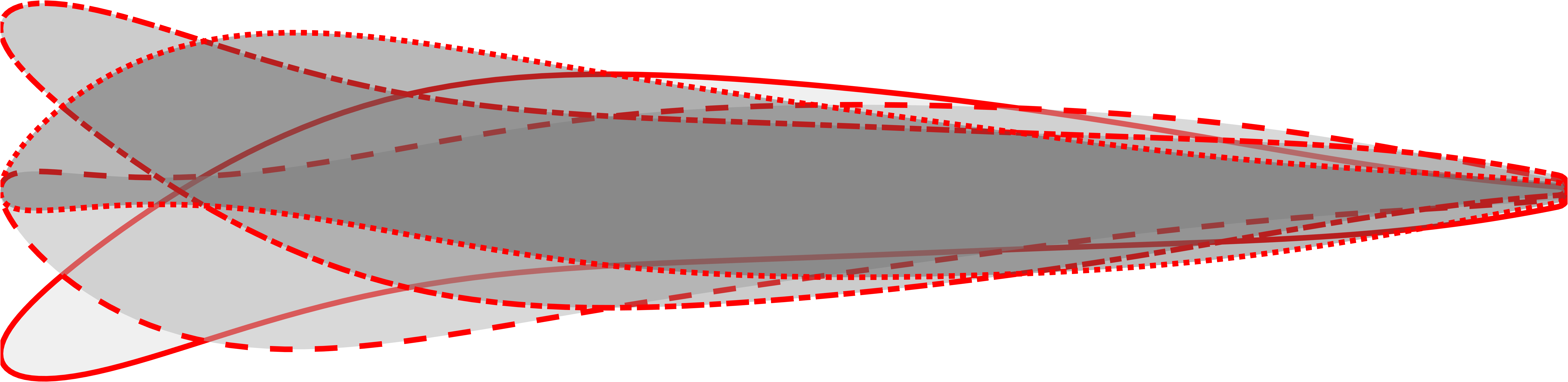}
        \subfigure[]{\includegraphics[width=\textwidth]{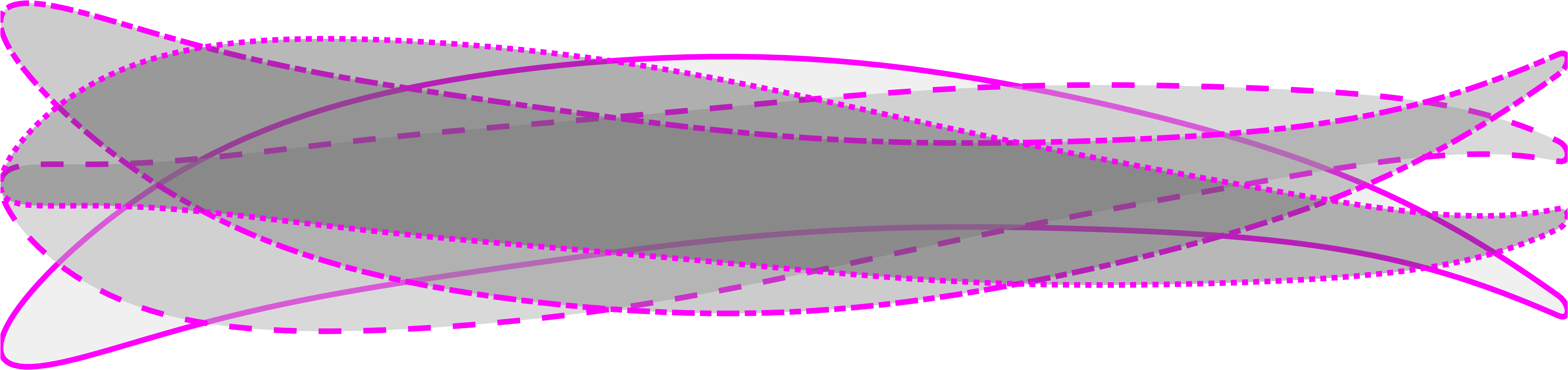} \label{fig:kinb}}
    \end{minipage}
    
    \caption{(\textbf{a}) Swimming amplitudes and (\textbf{b}) kinematic description of the NACA-0012 airfoil for five baseline shapes. \ed{The kinematic motion for all profiles is illustrated at $\lambda^* = 1$ to ensure consistent comparison.}}
    \label{fig:kin}
\end{figure}

\begin{figure}[h]
\centering
   \includegraphics[angle=0,width=0.7\linewidth]{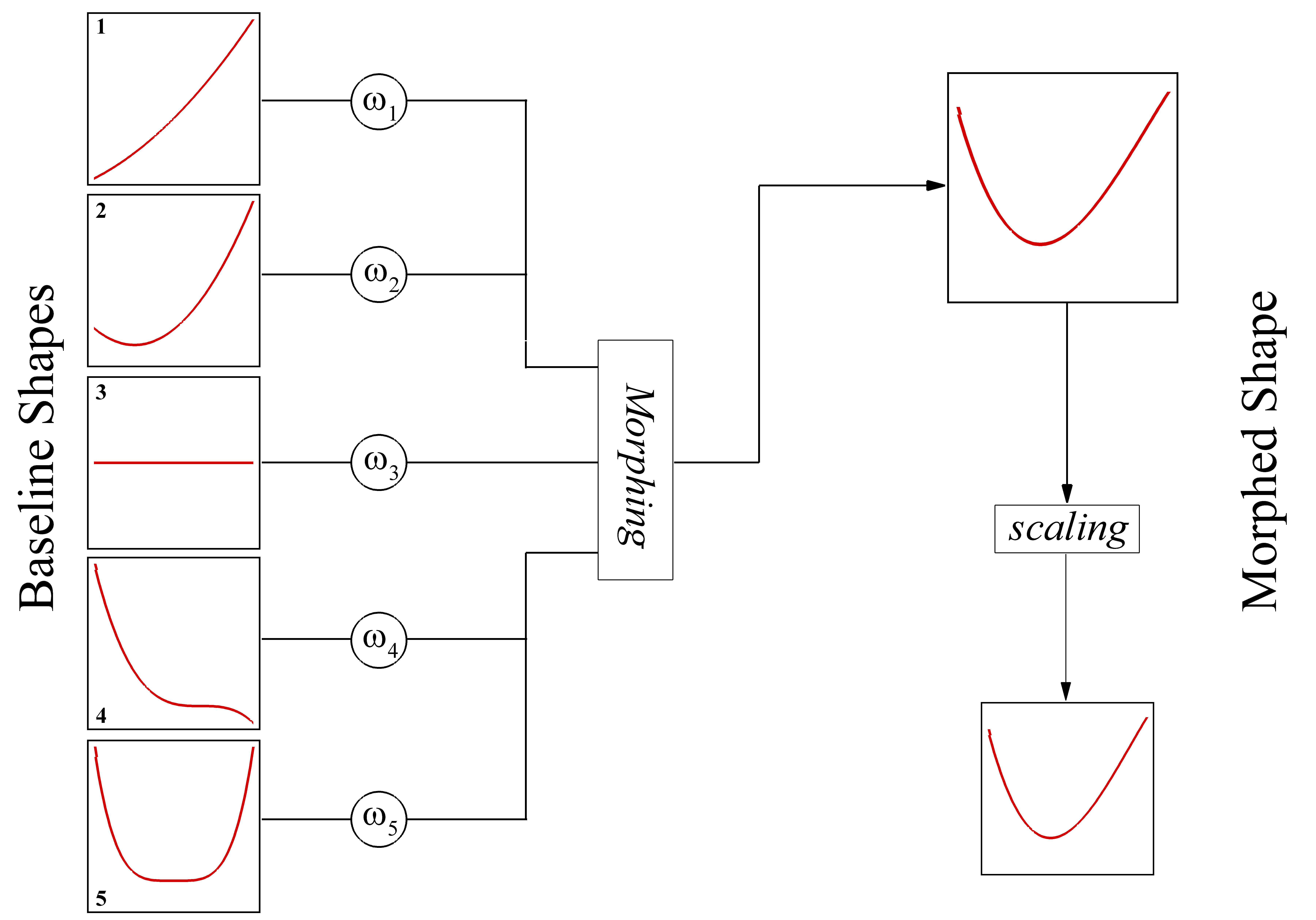} 
\caption{A schematic representation of the DbM strategy}
\label{fig:DbM}
\end{figure}

The swimming profile $A(x^*)$, representing the lateral displacement along the fish body, is optimized using a DbM technique. This approach models the swimming profile as a linear combination of multiple baseline shapes to explore a wide range of design spaces, \ed{as illustrated in Fig.~\ref{fig:DbM} for five baseline shapes}. By varying the coefficients associated with these baseline profiles, we generate new swimming kinematics that aim to maximize propulsive efficiency. We considered five baseline shapes, including commonly used swimming profiles\edt{,} such as anguilliform and carangiform \citep{khalid2021anguilliform,maertens2016optimal}, along with the unconventional (or weird) shapes introduced to enhance diversity in the design space. These additional shapes, which differ significantly from natural swimming patterns, expand the design space and allow for a broader exploration of potential swimming profiles. \ed{Furthermore, to ensure the independence of the swimming profiles, polynomials of varying degrees \edt{are} employed, except that both the anguilliform and carangiform profiles \edt{are} represented using second-degree polynomials.} Table~\ref{tab:baseprofiles} presents the mathematical expressions for the baseline shapes, whereas the swimming amplitudes and kinematic description of the NACA-0012 airfoil in swimming motion are depicted in Fig.~\ref{fig:kin}. The swimming profile based on the design-by-morphing technique is mathematically defined as:

\begin{align}
A(x^*) = \frac{a_n}{\gamma}\sum_{j=1}^N \omega_jA_j(x^*)
\end{align}

\noindent where $N$ represents the number of baseline shapes, which in our case is five. The term $\gamma$ is the normalization factor, defined as $\gamma = \max_{i}{\left| \sum_{j=1}^N \omega_j A_j(x^*_i) \right|}$, and $a_n$ denotes the maximum lateral amplitude of the swimming body. In the present case, we set it as $0.1/L$\ed{, consistent to maximum tail amplitude of frequenctly used carangiform and anguilliform profiles}. Furthermore, the weighting coefficients $\omega_j$, referred to as ``weights'' or ``design variables'' satisfy the condition $\sum_j \omega_j^2 = 1$. These weights serve as independent parameters that define the search space for optimizing the swimming profile. Enforcing the summation condition on the weights reduces the degrees of freedom by one. By utilizing a unit hypersphere centered at the origin, the five design space variables (i.e. weights) can be mapped to four hyperspherical coordinates. The transformation is expressed as:
\begin{align}
\omega_i =
\left\{
\begin{array}{ll}
\cos(\varphi_1), & i = 1 \\[2pt]
\displaystyle \prod_{k=1}^{i-1} \sin(\varphi_k) \cos(\varphi_i), & i = 2, 3, 4 \\[2pt]
\displaystyle \prod_{k=1}^{i-1} \sin(\varphi_k), & i = 5
\end{array}
\right.
\end{align}

\noindent Here the $\varphi_i$ represents the angular coordinates. Since the morphed swimming shape based on the negative weights (i.e. $-\omega_j$) is simply the mirror image of the corresponding shape with positive weights (i.e., $\omega_j$), it is reasonable to consider only one-half of the hypersphere. Consequently, angular coordinates are restricted to the range $\varphi_i \in \left[0, \pi\right]$.

\begin{table}

    \centering
    \def~{\hphantom{0}}
    \caption{Mathematical expressions for the baseline shapes.}
    \begin{tabular}{l l}
    \hline
    Profile Type &  Mathematical Expression  \\ 
    \hline
    \hline
        Anguilliform & $A_1(x^*) = 0.0367+0.0323x^*+0.0310{x^*}^2$ \\
        Carangiform & $A_2(x^*) = 0.02-0.0825x^*+0.1625{x^*}^2$  \\
        Horizontal & $A_3(x^*) = 0.1$  \\
        Third-degree (weird) & $A_4(x^*) = 0.2970(0.67-x^*)^3+0.01067$  \\
        Fourth-degree (weird) & $A_5(x^*) = 1.21212(x^*-0.5)^4+0.024242$  \\
        
    \end{tabular}

    \label{tab:baseprofiles}
\end{table}

\subsection{Governing Model for Fluid Flows}

\rev{The unsteady flow over the undulating foils is governed by the incompressible Navier–Stokes equations under laminar flow conditions, subject to the continuity constraint.} The non-dimensional forms of the continuity and momentum equations are given by:  

\begin{align}
\frac{\partial u_j}{\partial x_j} = 0, \quad \text{for} \quad j = \{1,2\}  
\label{eqn:CartContinuity}
\end{align}

\begin{align}
\frac{\partial u_i}{\partial t} + u_j\frac{\partial u_i}{\partial x_j} = -\frac{1}{\rho}\frac{\partial p}{\partial x_i} + \frac{1}{\mbox{Re}}\frac{\partial^2 u_i}{\partial x_j \partial x_j}, \quad \text{for} \quad i, j = \{1,2\}  
\label{eqn:CartMomentum}
\end{align}

\noindent Here the \(u_i\), \(p\), and \(\rho\) denote the \(i\)-th \edt{flow} velocity component, pressure, and density of the fluid, respectively. The Reynolds number is defined as $\mbox{Re} = {\rho L U_{\mbox{max}}}/{\mu}$, where $\mu$ is the dynamic viscosity. In this context, the spatial variables and flow velocity are non-dimensionalized using the characteristic length $L$, which is defined as the backbone of the undulating fish, and the maximum undulatory velocity $U_{\mbox{max}}$, respectively.

To model the effects of an undulating foil interacting with the fluid, modifications to the original fluid model are necessary for fluid-structure coupling. Two computational techniques are widely employed for FSI problems: the immersed boundary (IB) method and the arbitrary Lagrangian-Eulerian (ALE) method. 


In this study, we employed the ALE method to simulate flows over self-propelling foils. The ALE formulation of the momentum equation is expressed as:

\begin{align}
\frac{\partial u_i}{\partial t} + \widetilde{u}_j\frac{\partial u_i}{\partial x_j} = -\frac{1}{\rho}\frac{\partial p}{\partial x_i} + \frac{1}{\mbox{Re}}\frac{\partial^2 u_i}{\partial x_j \partial x_j}, \quad \text{for} \quad i, j = \{1,2\}  
\label{eqn:CartALEMomentum}
\end{align}

\noindent Here, \(\widetilde{\mathbf{u}} = \mathbf{u} - \mathbf{u}_\mathsf{g}\) represents the relative flow velocity with respect to the grid node velocity, \(\mathbf{u}_\mathsf{g}\). The computation of \(\mathbf{u}_\mathsf{g}\) is detailed in \citep{farooq2022numerical, farooq2022nonlinear}. It is important to note that Eq.~\eqref{eqn:CartALEMomentum} reduces to Eq.~\eqref{eqn:CartMomentum} in the Eulerian description (\(\widetilde{\mathbf{u}} = \mathbf{u}\)), where the grid remains stationary. In the Lagrangian description (\(\widetilde{\mathbf{u}} = 0\)), the equation tracks the variation of \(\mathbf{u}\) for a fixed fluid particle.

The optimization objective in this study is defined in terms of the propulsive efficiency of the undulating foil. This measure is particularly suitable, as it reflects the balance between the useful forward propulsion and the energetic cost of the imposed motion. It \edt{is} demonstrated that the definition of propulsive efficiency is controversial for propelling bodies, as conventional thrust-based efficiency leads to zero net thrust in the steady-state phase. Therefore, to evaluate propulsive efficiency, \cite{kern2006simulations} proposed an alternative definition of efficiency, $\eta$, which is expressed as:

\begin{align}
\eta = \frac{E}{W},
\end{align}

\noindent where

\begin{subequations}
\begin{gather}
E = \frac{m\bar{U}_o^2}{2}, \quad
W = -\int_t^{t+T}\oint_s {\sigma} \cdot \hat{\textbf{n}}\cdot \textbf{V} ds, \tag{\theequation a-b}
\end{gather}
\label{eqn:FluidForces}
\end{subequations}

\noindent with $E$ denoting the time-averaged kinetic energy of the forward translational motion and $W$ the work performed by the undulatory actuation over the interval $[t,\;t+T]$. Here, $\sigma$ represents the stress tensor, $\hat{\textbf{n}}$ is the unit outward normal vector, and $\textbf{V}$ denotes the local velocity of the surface of the swimming foil. In addition, the hydrodynamic forces acting on the propelling body can be formulated in terms of the surface stresses. The axial and lateral forces are obtained as

\begin{subequations}
\begin{gather}
    F_A = \oint_S \left( \sigma \cdot \hat{\mathbf{n}} \right) \cdot \hat{\mathbf{e}}_x \; ds,  \quad
    F_L = \oint_S \left( \sigma \cdot \hat{\mathbf{n}} \right) \cdot \hat{\mathbf{e}}_y \; ds, \tag{\theequation a-b}
\end{gather}
\end{subequations}

\noindent where $\hat{\mathbf{e}}_x$ and $\hat{\mathbf{e}}_y$ denote the unit vectors along the axial and transverse directions, respectively. \ed{Note that the drag force acting on the propelling body is defined as $F_D = -F_A$.}

\subsubsection{Modeling Undulatory Motion for Propulsion}

Undulatory motion is a fundamental mechanism utilized by aquatic animals and swimmers to generate propulsion through synchronized oscillatory kinematics. To systematically explore and optimize this mechanism, we adopt a simplified harmonically oscillating model to represent undulatory kinematics, expressed as:

\begin{equation}
y\left(x^*, t\right) = A\left(x^*\right)\cos\left[2\pi\left(\frac{x^*}{\lambda^*}-ft\right)\right]; \quad 0 \leq x^* \leq 1,
\end{equation}

\noindent where \( y(x^*,t) \) denotes the lateral displacement of the body during oscillation, \( 2\pi/\lambda^* \) is the wave number, \( f \) is the oscillation frequency, and \( A(x^*) \) represents the swimming profile non-dimensionalized by the total body arc-length \( L \). The swimming profile \( A(x^*) \) plays a central role in determining the propulsive performance and efficiency and is optimized in this study using the DbM technique.

\ed{The coordinate \( x^* \) is a normalized material coordinate that describes the position along the swimming body, independent of its physical length. The physical body length is determined by enforcing the preservation of the arc-length (and thus the conservation of volume). Accordingly, the physical $x$-coordinate is obtained by scaling the material coordinate as \( x = \alpha x^* \), where \( \alpha \) denotes the instantaneous physical length of the body. The head and tail positions are fixed at \( x(0)=0 \) and \( x(1)=\alpha \), respectively.

To enforce a prescribed arc-length \( L \), the scaling parameter \( \alpha \) is obtained at each time step by solving the following nonlinear constraint:
\begin{equation} 
\int_{0}^{1} \sqrt{ \alpha^2({dx^*})^2 + ({dy})^2 } \, = L 
\end{equation}
}




The forward velocity (\( U_o \)) and central position (\( x_o \)) of the body along the \( x \)-axis are governed by the following coupled dynamical equations:

\begin{subequations}
\begin{gather}
m\frac{dU_o}{dt} = F_x,\quad \frac{dx_o}{dt} = U_o,
\tag{\theequation a-b}
\end{gather}
\end{subequations}

\noindent where \( m=\rho V_f \) is the mass of the body,  the density of the foil $\rho$ is regarded as a constant value equal to the density of the fluid in this work, $V_f$ is the volume of the foil, and \( F_x \) represents the force exerted by the fluid in the \( x \)-direction. The initial conditions are set as \( U_o = 0 \) and \( x_o = 0 \). The volume of the NACA 0012 airfoil is considered as \(0.08621 \, \mbox{m}^3 \), consistent with prior studies \citep{zhang2018effects}.

The FSI problem is solved using the Hamming fourth-order predictor-corrector technique \citep{Viv_Mehmood2}, ensuring precise coupling between the fluid and structural dynamics. This computational framework enables the evaluation of the optimized profile \( A(x^*) \) by maximizing propulsive efficiency and minimizing energy dissipation.

\subsubsection{Discretization Scheme}
\label{sec:discretization}
A pressure-correction algorithm \edt{is} employed to solve the governing equations for fluid dynamics. Initially, the momentum equations are solved for velocity components in an uncoupled manner, neglecting the continuity constraint and omitting the pressure term. The resulting velocity field, referred to as the \textit{intermediate} velocity, is then corrected to ensure mass conservation. This correction involves solving a pressure Poisson equation derived by imposing the continuity constraint on the velocity field. The computed pressure distribution is subsequently used to obtain a divergence-free velocity field by adjusting the \textit{intermediate} velocity. This approach is consistent with the methodologies outlined by \citet{kim1985application} for Cartesian coordinates and \citet{ImranD, cfd_Zang} for curvilinear grids. In this study, an `O'-type body-fitted nearly orthogonal grid has been employed over the NACA-0012 airfoil (see Fig.~\ref{fig:grida}), generated by numerically solving elliptic differential equations following the method of \citet{ryskin1983orthogonal}. A non-staggered grid arrangement \edt{is} used, where the pressure and velocity components are computed at the cell center, while fluxes are evaluated at the midpoints of the cell faces.

The spatial discretization employs finite difference approximations. All spatial derivatives, except for convective terms, are approximated using a second-order accurate centered difference scheme. For the convective terms, a quadratic upwinding interpolation for convective kinematics (QUICK) scheme \citep{leonard1979stable} is employed. This scheme utilizes a generic stencil based on positive and negative fluxes to enhance stability and accuracy. Temporal discretization is achieved using a semi-implicit scheme. The diagonal viscous terms are advanced using the Crank-Nicolson method for second-order accuracy, while all other terms are approximated using the explicit Adams-Bashforth method. For further details on the numerical schemes and boundary conditions, readers are referred to the following Refs. \cite{ImranD, farooq2022nonlinear, farooq2022numerical}.

\subsubsection{Re-meshing Strategy}
\label{sec:remeshing}
The effectiveness of the ALE method largely depends on the re-meshing algorithm employed \citep{donea2017arbitrary}. Various methods \edt{were} proposed for re-meshing both structured and unstructured meshes \citep{batina1990unsteady, degand2002three, farhat1998torsional}. In fluid dynamics, the most commonly used re-meshing techniques involve solving partial differential equations \citep{helenbrook2003mesh, lohner1996improved, dwight2009robust}. However, these techniques often struggle to maintain mesh quality, especially when the body undergoes large rotational deformations. 

Alternatively, a re-meshing technique developed by \cite{de2007mesh} utilizes an interpolation approach based on the \textit{radial basis function} (RBF) interpolation. This method \edt{is} proven to be highly effective in maintaining mesh quality even under large translational and rotational displacements \citep{bos2013radial}. Compared to other re-meshing strategies, the RBF method demonstrates superior efficiency and robustness. In our simulations, we employ this technique to determine the displacement of both boundary and interior nodes.

The interpolation function $f(\mathbf{x})$, based on the boundary nodes, is expressed as:

\begin{align}
f(\mathbf{x}) &= \sum_{i=1}^{N_c}{\alpha_{i}\phi(\|\mathbf{x}-\mathbf{xc}_{i}\|)} + \mathbf{p}(\mathbf{x}) \label{eqn:RBF}
\end{align}

\noindent where $N_c$ represents the number of control points, $\phi(\|\cdot\|)$ is the RBF as a function of the Euclidean norm. In this case, we use the global support thin-plate spline (TPS) radial basis function, $\phi(x) = x^{2}\log(x)$. The term $\mathbf{p}(\mathbf{x})$ is a minimal polynomial of degree one with coefficients vector $\boldsymbol{\beta}$, which must satisfy the following condition \citep{bos2013radial}:

\begin{align}
\sum_{i=1}^{N_c}{\alpha_{i}\mathbf{p}(\mathbf{xc}_{i})} = 0. \label{eqn:condition}
\end{align}

\noindent Here the $\alpha_{i}$'s are the coefficients of the basis function that are determined under the condition that at each control point, $f(\mathbf{x})$ interpolates the given displacements, i.e., $f(\mathbf{xc}) = \Delta\mathbf{xc}$, where $\Delta \mathbf{xc}$ is the prescribed displacement at the node $\mathbf{xc}$. The control points are selected from a subset of nodes on the airfoil boundary using a greedy algorithm \citep{rendall2009efficient}.

The system of linear equations for determining the coefficients $\boldsymbol{\alpha}$ and $\boldsymbol{\beta}$ is obtained from Equation~(\ref{eqn:RBF}) along with the condition on $\mathbf{p}$ and $f$, leading to the following system:

\begin{align}
\left[\begin{array}{cc} \mathbf{\Phi}  & \mathbf{q} \\ 
                              \mathbf{q}^{T} & \mathbf{0} \end{array} \right]
\left[ \begin{array}{c} \boldsymbol{\alpha} \\ \boldsymbol{\beta} \end{array} \right] &= \left[ \begin{array}{c} \Delta\mathbf{xc} \\ \mathbf{0} \end{array} \right] \label{eqn:RBFSystem}
\end{align}

\noindent where $\mathbf{\Phi}$ is the matrix of values of the RBF $\phi(\|\mathbf{x} - \mathbf{xc}_i\|)$ evaluated at the control points, and $\mathbf{q}$ is a vector of ones corresponding to the polynomial term. This system of equations is solved to determine the coefficients that enable interpolation of the displacement field across the entire mesh.

\begin{figure}[h]
\centering
   \subfigure[]{\includegraphics[angle=0,width=0.48\linewidth]{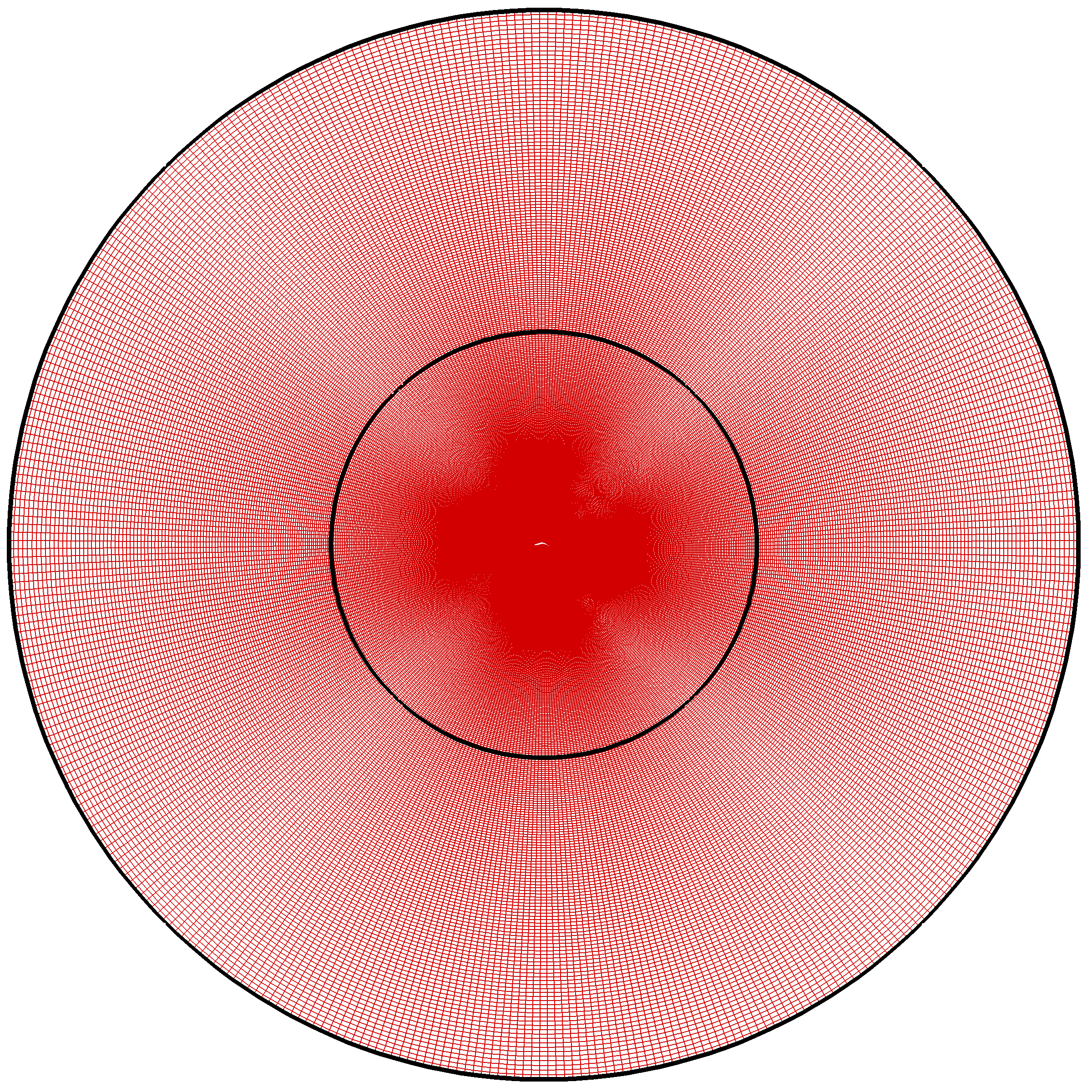} \label{fig:grida} } 
   \subfigure[]{\includegraphics[angle=0,width=0.48\linewidth]{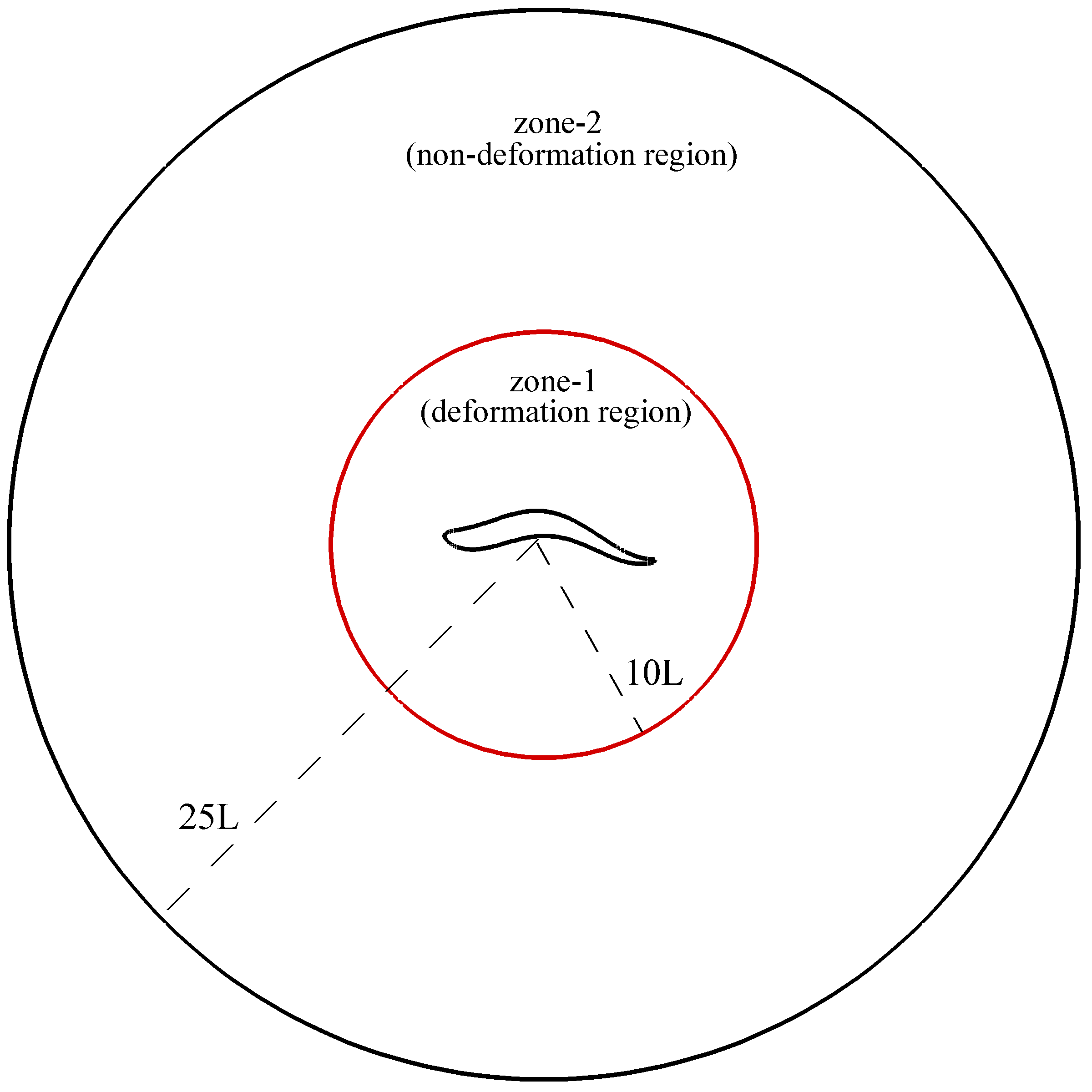} \label{fig:gridb}} 
\caption{(\textbf{a}) An `O'-type body fitted orthogonal grid between the airfoil and circular outer domain of radius $25L$ and (\textbf{b}) A schematic of the two-dimensional layout of the domain for the self-propulsion body.}
\label{fig:grid}
\end{figure}

\ed{To enable the self-propulsion of the undulating body, a dynamic mesh framework is employed, as illustrated in Fig.~\ref{fig:gridb}. The computational domain is divided into two zones, separated by an inner circular interface located at a radius of approximately \(10L\). The two zones are constructed independently, not in a manner analogous to a sliding-mesh strategy. The orientations of the two zones shown in the figure are presented solely for illustrative purposes, to highlight the mesh deformation strategy. Only the inner region (Zone~1) is allowed to deform during the simulation, with grid motion prescribed at each time step based on the displacement of the structural boundary nodes using the proposed RBF-based re-meshing algorithm. In contrast, the grid topology and cell shapes in the outer region (Zone~2) remain fixed. To realize self-propulsion, all grid nodes are translated with a uniform streamwise velocity \(U_o\), which is determined dynamically from the force balance on the body.}


\subsection{Bayesian Optimization}

Optimizing expensive black-box functions, particularly when involving mixed-variable or high-dimensional design spaces, \edt{is} an active area of research. BO \edt{is} widely recognized as a highly \rev{sample-}efficient strategy, as it achieves competitive performance with significantly fewer function evaluations compared to conventional optimization methods \citep{brochu2010tutorial}. In the present study, the evaluation of each design case involves high-fidelity simulations of self-propelling undulatory foils, which are computationally expensive. Therefore, BO provides an appropriate choice of optimizer to ensure that the optimization framework remains computationally feasible.

Several engineering design problems utilized BO in mixed-variable or high-dimensional design spaces, including architected material design \citep{sheikh2022systematic, vangelatos2021strength}, hyperparameter tuning in machine learning \citep{oh2018bock}, drug discovery \citep{korovina2020chembo}, and fluid-structure optimization \citep{sheikh2023airfoil}. In our case, the design space is parameterized by four morphing weights that define the swimming profile $A(x^*)$ through design-by-morphing, along with two undulation parameters: frequency and wavelength. Thus, the total design space constitutes six optimization input variables. For such expensive, noisy, and potentially multi-modal problems, local optima can trap conventional algorithms, while BO effectively balances exploration and exploitation.

To address this optimization problem, we employ the Mixed-Variable Multi-Objective Bayesian Optimization (MixMOBO) algorithm \citep{sheikh2022bayesian}. This framework is capable of handling both continuous and categorical variables, and has demonstrated efficiency in optimizing noisy black-box functions with limited evaluation budgets. In this study, MixMOBO is adopted in a single-objective discrete-parameter setting with the in-built HedgeMO strategy to ensure robustness, with the goal of maximizing propulsive efficiency. The design space is discretized to take advantage of MixMOBO's efficiency when dealing with mixed-variables and to ensure that the algorithm does not get stuck during exploration in a high-frequency response region \citep{sheikh2022optimization}. Following best practices \citep{sheikh2022bayesian, sheikh2022optimization}, an initial sampling size of approximately 10–20\% of the total budget is employed. 

\begin{figure}[!ht]
\begin{center}
\includegraphics[width=0.8\linewidth]{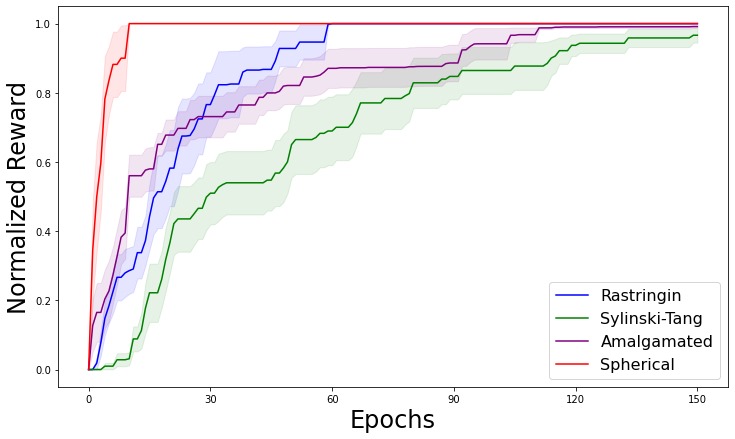}
\caption{MixMOBO performance benchmarking on four test functions.} 
\label{fig:benchmark}
\end{center}
\end{figure}

To validate the efficacy of MixMOBO for the present optimization problem, we also \edt{test} it on a set of standard 6D benchmark functions commonly used in the literature, namely the Spherical, Rastrigin, Styblinski–Tang, and anisotropic Amalgamated test functions, with design space similar to our swimming profile optimization design space. The test functions definitions and settings are same as \citep{sheikh2022optimization}. Each test function was run for 10 independent runs and performance was quantified using the normalized reward metric, defined as:

\[
R = \frac{f_{\text{opt}} - f_{\text{init}}}{f_{\text{global}} - f_{\text{init}}},
\]

\noindent where $f_{\text{opt}}$ is the optimum obtained by BO, $f_{\text{init}}$ is the optimum found by random search, and $f_{\text{global}}$ denotes the known global optimum. The results are demonstrated in Fig.~\ref{fig:benchmark} as the mean normalized reward with standard deviation over 10 runs. These results show MixMOBO can achieve consistent convergence within our prescribed computational budget of 150 epochs with 50 initial sample points for a design space of similar dimensionality.

\section{Numerical \ed{Verification \&} Validation}\label{sec:valid}
\ed{This section details the verification of the numerical solver through systematic grid-independence and time-step refinement studies. In addition, the validation of the proposed methodology is performed by conducting simulations for both self-propelled and tethered swimming configurations and comparing the results with available reference data.

A detailed grid-independence study is performed using four grid resolutions: $\mbox{G1}$ with $250\times200$ grid points, $\mbox{G2}$ with $374\times300$ grid points, $\mbox{G3}$ with $512\times400$ grid points, and $\mbox{G4}$ with $614\times480$ grid points. Flow simulations are performed to compute the time-averaged drag coefficient (defined as $\bar{C}_D= \bar{F}_D/0.5\rho L U_\infty^2$) and the power ($\bar{P}=W/T$) for a tethered NACA 0012 airfoil undergoing carangiform undulation at an undulation frequency of 4~Hz and a non-dimensional wavelength of 0.5. The computed parameters are reported in Table~\ref{tab:Indep}.}
No significant changes in the computed parameters \edt{are} observed when the resolution \edt{is} increased beyond that of grid $\mbox{G3}$, indicating that the medium mesh size provides satisfactory results. 

Using grid $\mbox{G3}$, a time-step independence study \edt{is also} performed by considering three levels of time-steps: $\mbox{TS1}$, 3000; $\mbox{TS2}$, 4000; and $\mbox{TS3}$, 5000 time-steps per oscillating cycle. As shown in Table~\ref{tab:Indep}, $\mbox{TS2}$ \edt{is} chosen for time-marching in the simulations since the variation in computed parameters with respect to different time-steps is minimal.

\begin{table}
\centering
\caption{Validation of \edt{the} existing solver through \edt{mesh} \& time refinements}
\begin{tabular}{cccc}
    \hline
{\footnotesize Grid Resolution} &{\footnotesize Time Steps} &    \\
{\footnotesize $N_x \times N_y$} &{\footnotesize per Cycle} & $\bar{C}_D$ & $\bar{P}$ \\
    \hline
    \hline
 G1: 250$\times$200 &  4000  &  -0.3441 & 0.4163\\
 G2: 374$\times$300 &  --  &  -0.3230 & 0.3873\\
 G3: 512$\times$400 &  --  &  -0.3144 & 0.3779\\ 
 G4: 614$\times$480 &  --  &  -0.3118 & 0.3749\\
 G2: 512$\times$400 &  3000  &  -0.3178 & 0.3829\\
 G2: 512$\times$400 &  5000  &  -0.3140 & 0.3763

\end{tabular}

\label{tab:Indep}
\end{table}

\begin{figure}[!ht]
\begin{center}
\subfigure[]{\includegraphics[width=0.48\linewidth]{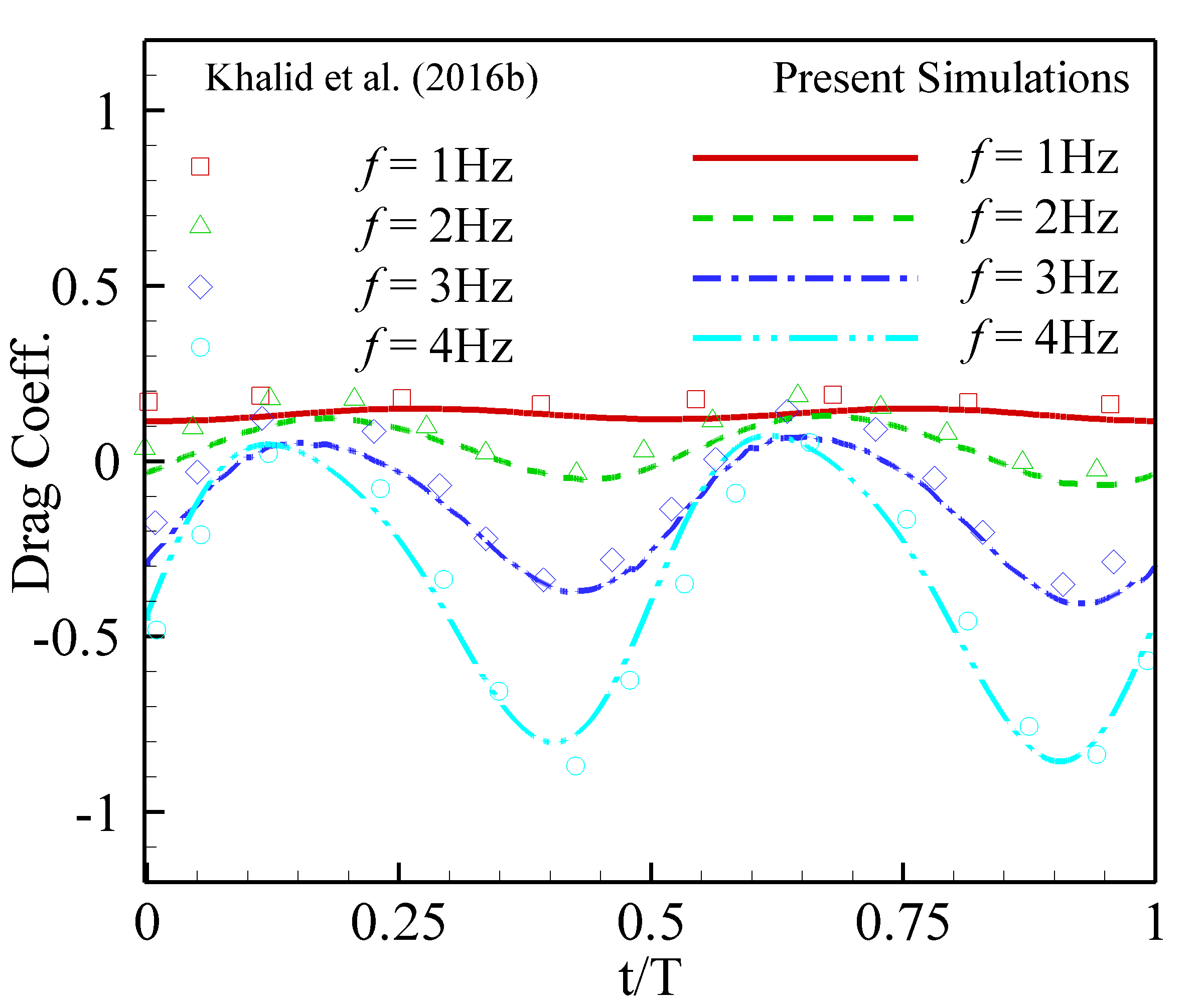} \label{fig:validtetha}}
\subfigure[]{\includegraphics[width=0.48\linewidth]{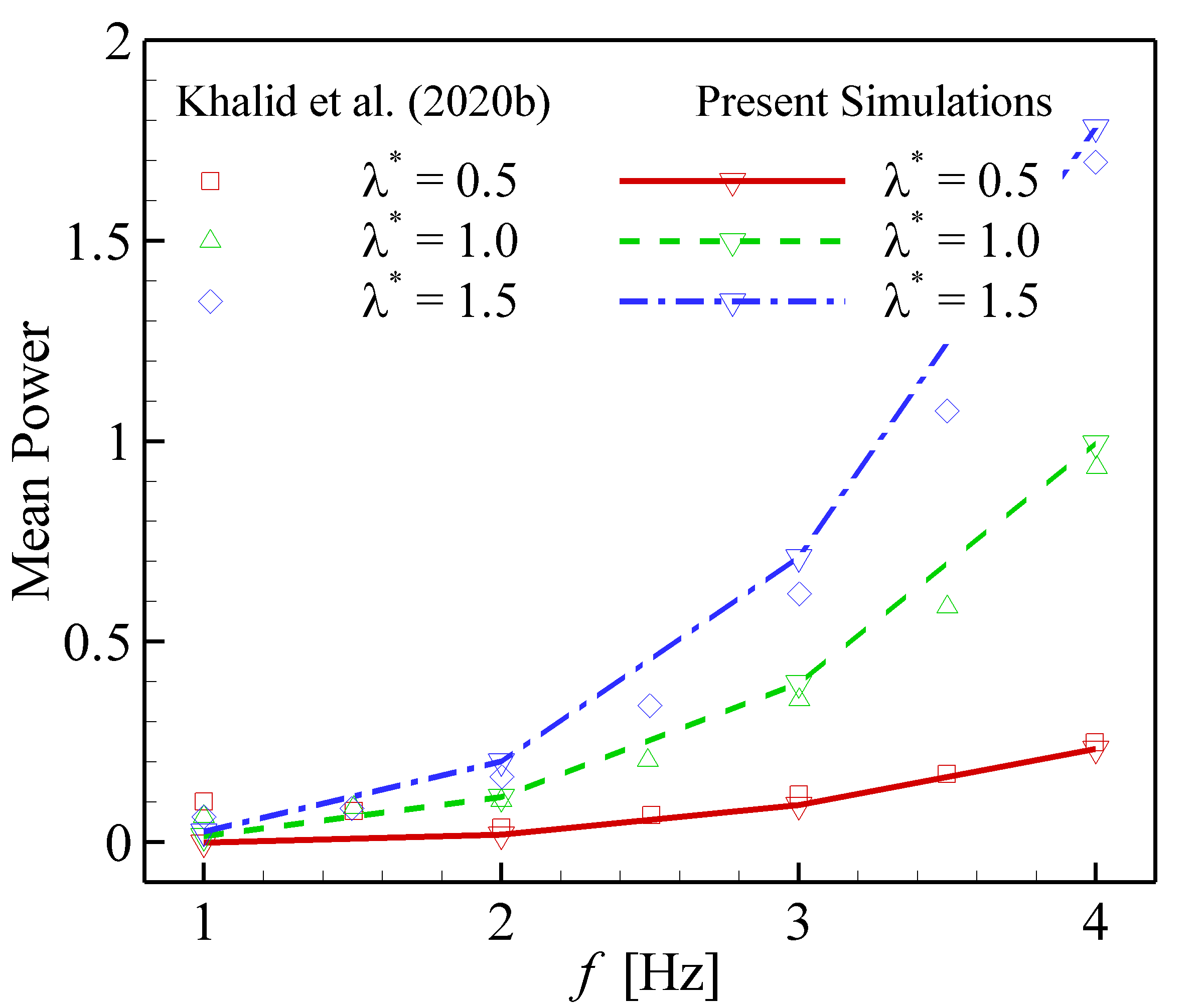} \label{fig:validtethb}}
\caption{Validation of the present results: (\textbf{a}) time history of drag coefficients compared with the data of \citet{khalid2016hydrodynamics}, and (\textbf{b}) time-averaged power compared with the results of \citet{khalid2020flow}.}
 
\label{fig:validteth}
\end{center}
\end{figure}

\begin{figure}[!ht]
\begin{center}
\subfigure[]{\includegraphics[width=0.32\linewidth]{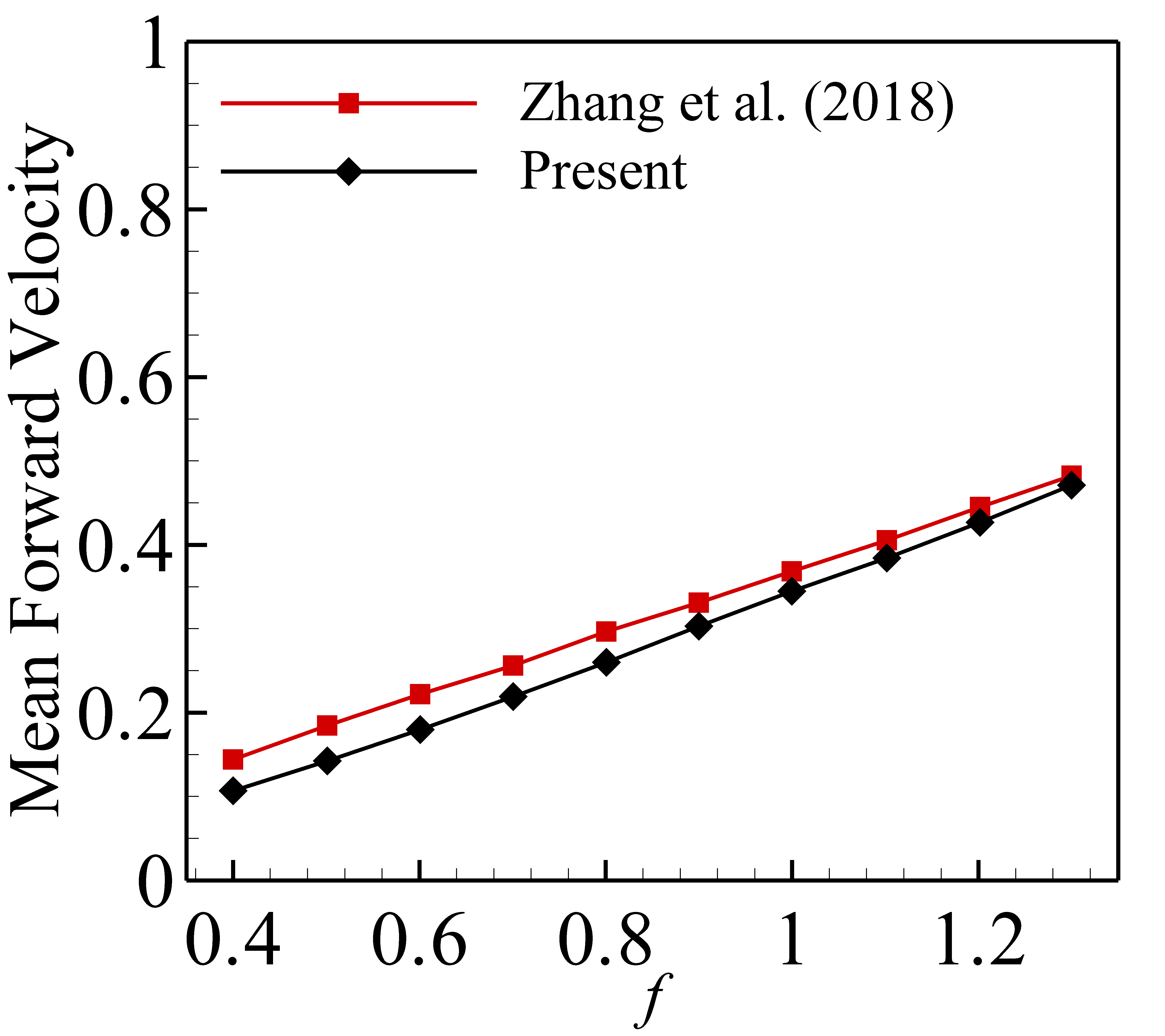}}
\subfigure[]{\includegraphics[width=0.32\linewidth]{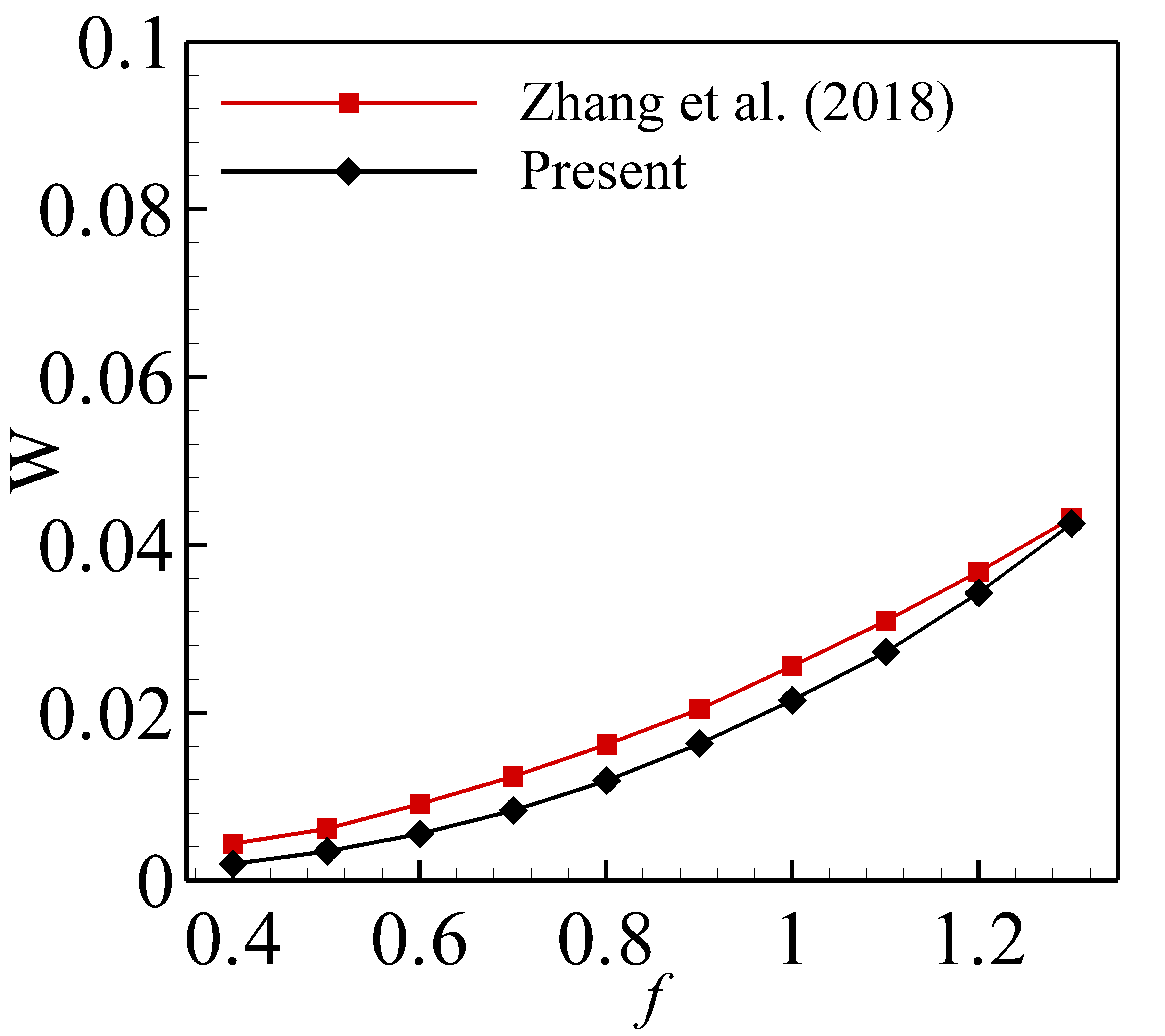}}
\subfigure[]{\includegraphics[width=0.32\linewidth]{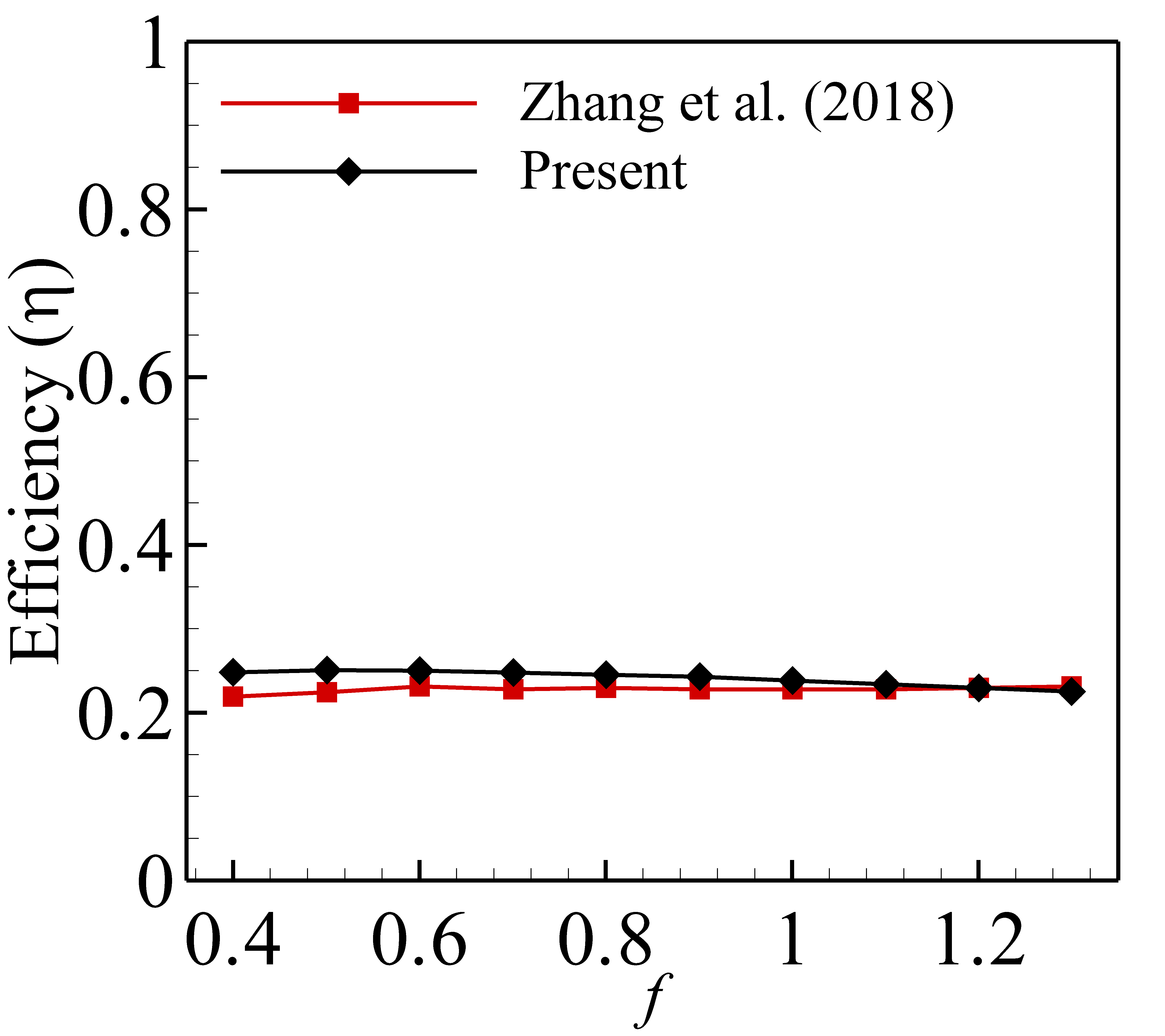}}
\caption{Comparison of (a) \edt{the} mean forward velocity $\bar{U}_{o}$, (b) work consumption of \edt{the} undulating foils, and (c) propulsive efficiency with the published data reported in \citep{zhang2018effects} with $\lambda^*=1$.} 
\label{fig:MFVel}
\end{center}
\end{figure}

\ed{For the validation of the numerical solver, we first compare our results for a tethered NACA 0012 airfoil undergoing carangiform undulation at $\mbox{Re}=500$ and $\lambda^*=1$ over a range of undulation frequencies from 1 to 4~Hz. Our results, in terms of the time history of drag coefficient, along with those reported by \citet{khalid2016hydrodynamics}, is plotted in Fig.~\ref{fig:validtetha}, showing good agreement across all considered frequencies. In addition, simulations of a tethered NACA 0012 airfoil undergoing anguilliform undulation at $\mbox{Re}=1000$ are performed over a range of undulation frequencies and non-dimensional wavelengths. The resulting time-averaged power, as shown in Fig.~\ref{fig:validtethb}, is compared with the results of \citet{khalid2020flow}, confirming the accuracy of the solver in terms of power computation.

Furthermore, we perform numerical simulations of flow around a self-propelling NACA 0012 airfoil under the same configuration as considered by \citet{zhang2018effects} and compare the obtained results with the data reported therein. These simulations are conducted at $\mbox{Re}=1000$ for undulatory frequencies ranging from 0.4 to 1.3~Hz. The mean forward velocity $U_o$, work consumption, and propulsive efficiency are computed and plotted as functions of the undulatory frequency in Fig.~\ref{fig:MFVel}, together with the corresponding results of \citet{zhang2018effects}. Excellent agreement is observed between the present results and those available in the literature, further validating the coupling of the dynamical system with the numerical framework.}

\section{Results \& Discussion}\label{sec12}
\edt{Our present work includes} 2D simulations of self-propulsive \edt{undulating bodies for} over a wide range of parameter space, involving four design space variables and two undulation parameters, namely frequency and wavelength. The specifications of the governing parameters, structural kinematics, design variables, and flow regimes \edt{are} presented in Table~\ref{tab:Param}. \ed{All the cases of self-propulsion are simulated for 24 oscillating cycles to mitigate the transient effects.} Furthermore, the subsequent sections present a detailed discussion of the findings from the optimization study.

\begin{table}
\centering
\caption{Detail of \edt{the} governing parameters and their relevant specifications}
\begin{tabular}{cc}
\hline
Parameters &  Specifications\\
 \hline
 \hline
 Geometry & \quad NACA 0012 \\

 Kinematics & \quad Five baseline shapes \\
 $\mbox{Re}$ & \quad $1\times 10^3$ \\
 $f$ & \quad $\{0.5, 1.0, \cdots, 4.5\}$ \\
 $\lambda^{*}$ & \quad $\{0.5, 0.6, \cdots, 1.5\}$ \\
 $\varphi_i$ & \quad $\{0, \pi/6, \cdots, \pi\}$ \\
 Objective function & \quad $\eta$ \\
\end{tabular}

\label{tab:Param}
\end{table}

Before delving into the detailed findings of the objective optimization, it is essential to assess the performance of the commonly used carangiform and anguilliform swimming profiles in the frequency-wavelength space. To this end, high-fidelity simulations are conducted over the specified wavelength–frequency domain, resulting in \ed{162} simulations, \ed{81} for each profile. We compute the propulsive efficiency for both swimming modes. The results from these simulations serve as benchmarks for evaluating the performance of optimized swimming profiles generated by the DbM-BO approach. By establishing a baseline using these well-studied natural modes, we can better quantify the improvements achieved by our optimization framework in terms of enhanced propulsive efficiency across the design space.

\begin{figure}[!ht]
\begin{center}
\subfigure[]{\includegraphics[width=0.48\linewidth]{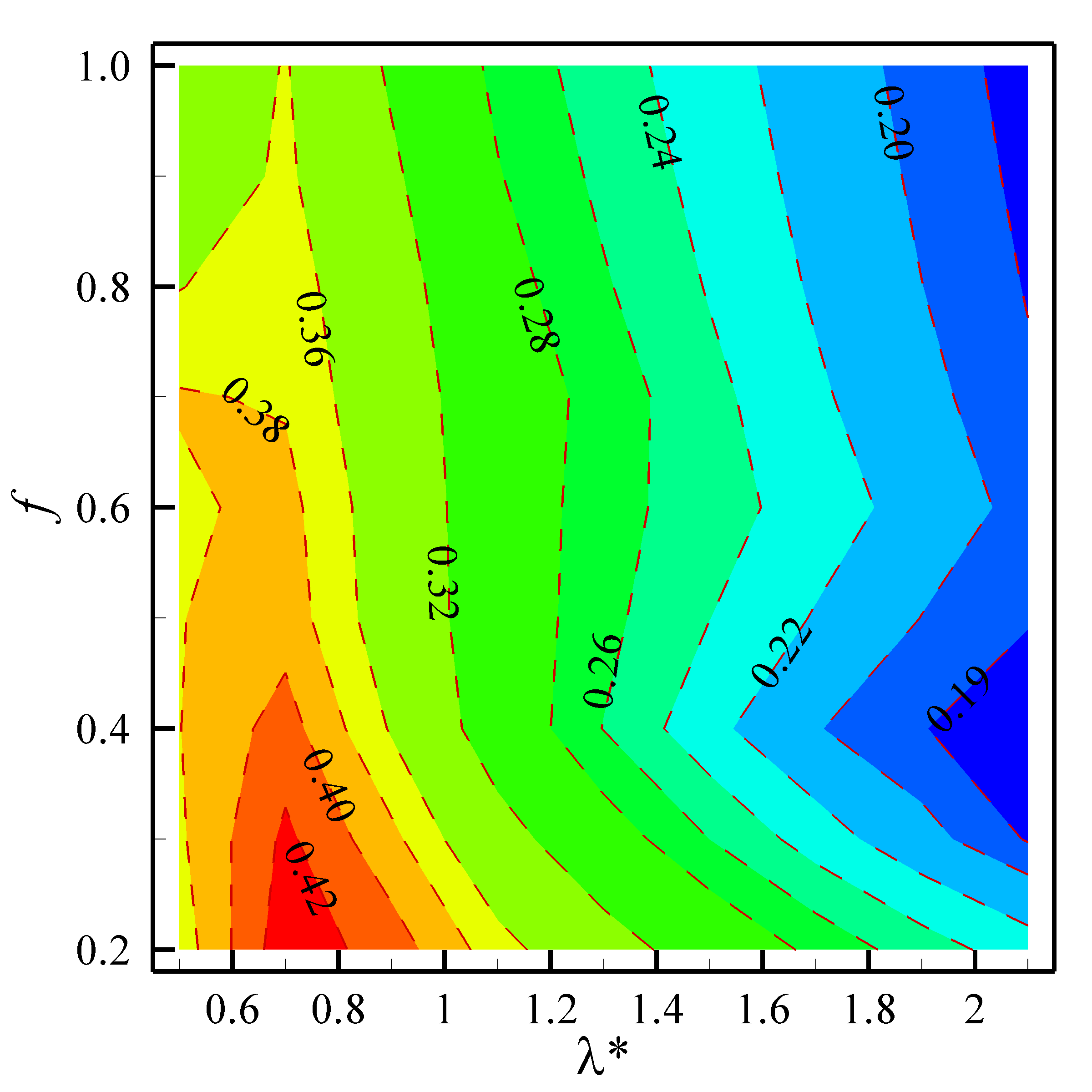}}
\subfigure[]{\includegraphics[width=0.48\linewidth]{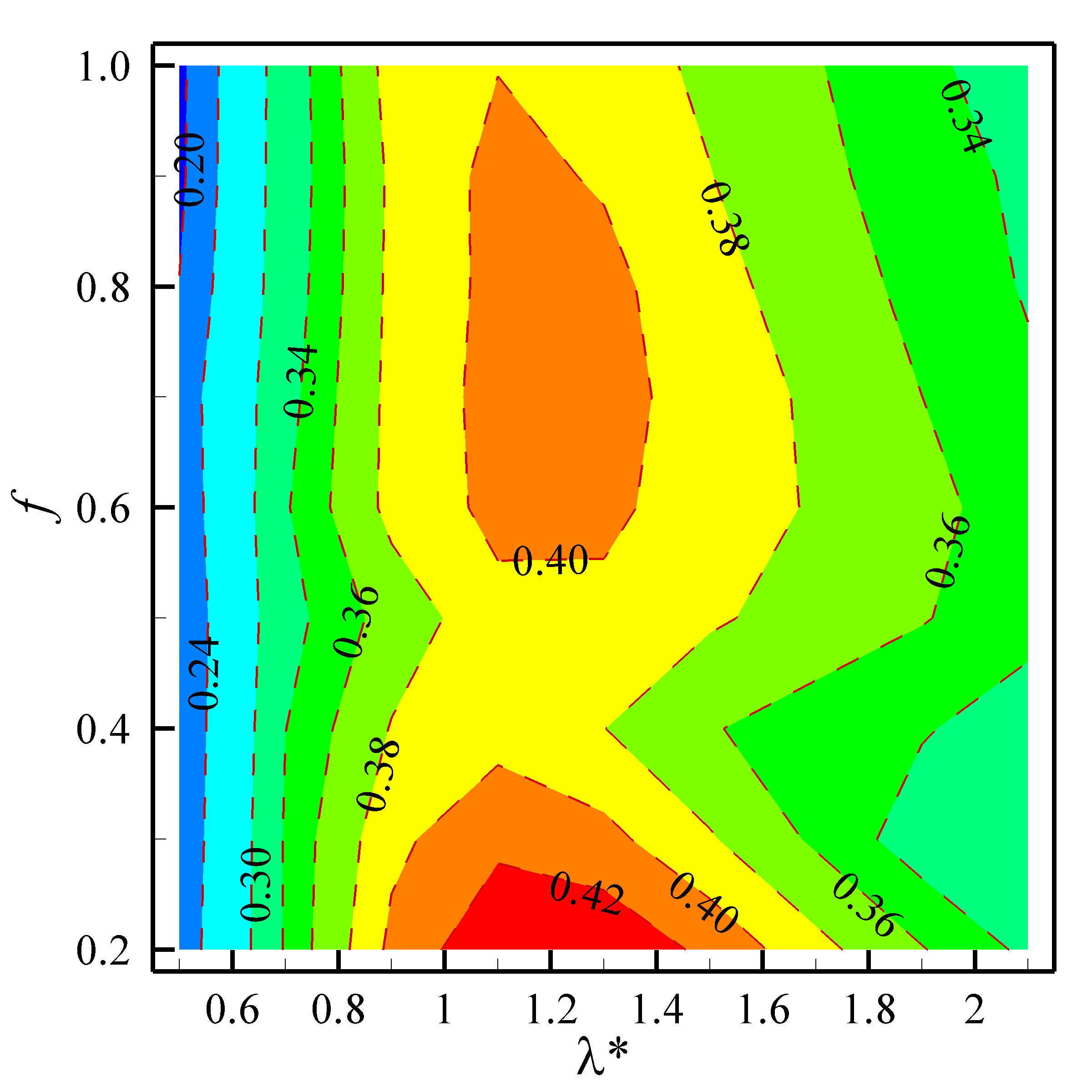}}
\caption{Contour map of the propulsive efficiency in the frequency–wavelength space for (\textbf{a}) carangiform and (\textbf{b}) anguilliform swimming modes at $\mbox{Re}=1000$.} 
\label{fig:eff}
\end{center}
\end{figure}

\ed{In Fig.~\ref{fig:eff}, the propulsive efficiency contours over the frequency–wavelength domain are shown for both carangiform and anguilliform swimming modes at \(\mbox{Re}=1000\). For both modes, the maximum efficiency lies in the range of approximately \(35\%\)--\(42\%\), representing the upper bound of performance under the considered flow conditions and kinematic parameters. The carangiform mode exhibits a clear trend, with higher efficiencies concentrated in the lower-left region of the parameter space, corresponding to relatively shorter wavelengths (\(\lambda^* \approx 0.5\)--\(1.1\)) and low to moderate undulation frequencies (\(f \approx 0.2\)--\(0.8\)). The peak efficiency reaches about \(42\%\), indicating that carangiform swimming achieves optimal energy conversion through compact body undulations at comparatively lower frequencies. In contrast, the anguilliform mode displays a broader region of elevated efficiency, primarily distributed along a vertical band spanning wavelengths \(\lambda^* \approx 0.9\)--\(1.5\) and frequencies \(f \approx 0.2\)--\(0.9\). This distribution is more symmetric than that observed for carangiform swimming, suggesting that anguilliform kinematics sustain high propulsive efficiency across a wider range of undulatory wavelengths and frequencies. These efficiency maps provide important baseline references for assessing the performance gains achieved by swimming profiles optimized using the DbM–BO framework within the same frequency–wavelength domain.}

\subsection{Performance Analysis of Optimal Profiles}

To identify the optimal swimming profile, we employ the MixMOBO algorithm within a six-dimensional design space. The objective of the optimization is to maximize the propulsive efficiency $\eta$.

\begin{figure}[!ht]
\begin{center}
\includegraphics[width=0.8\linewidth]{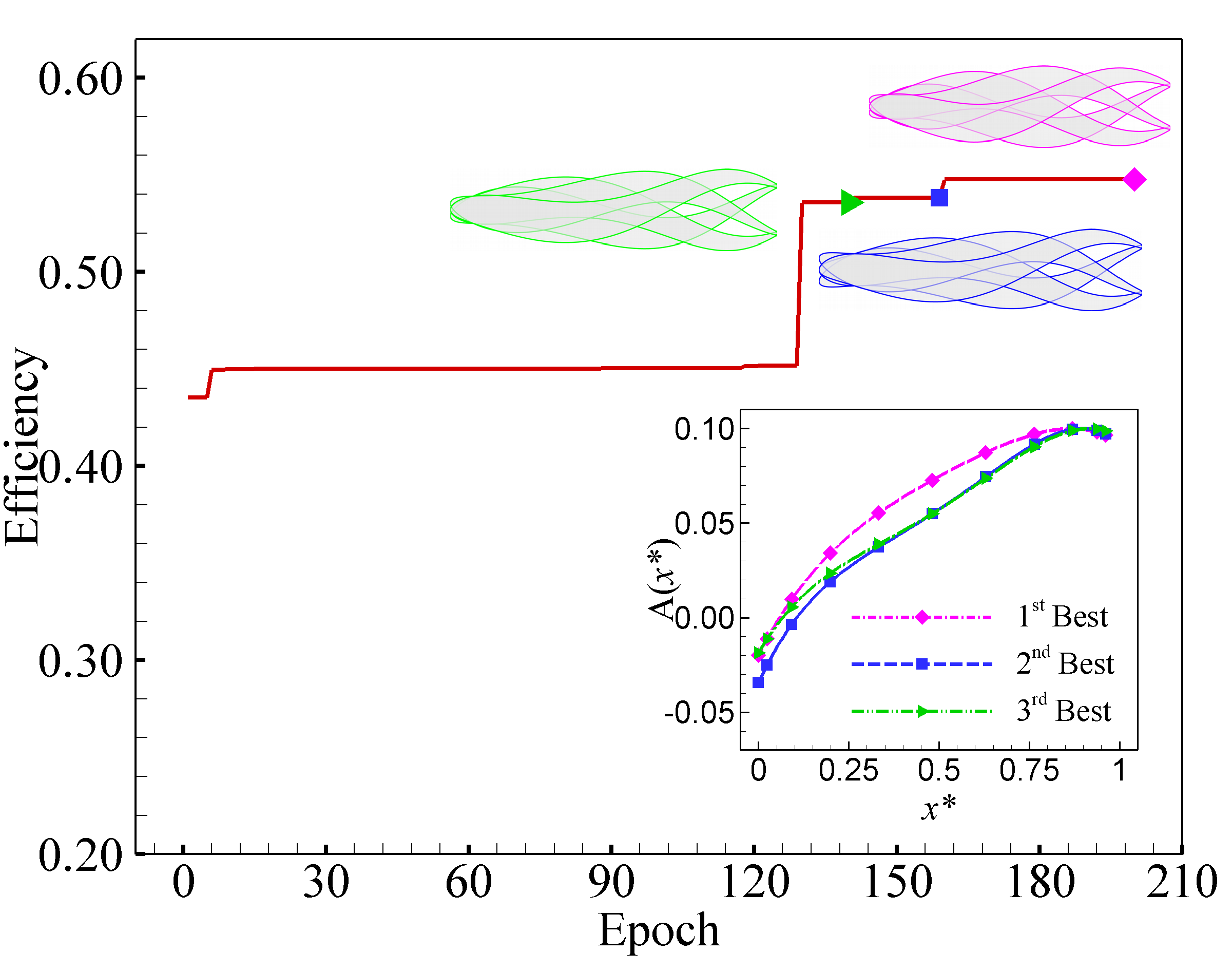}
\caption{Convergence history of the MixMOBO algorithm illustrating the evolution of the best-obtained propulsive efficiency over 200 search iterations. The inset shows the corresponding optimal swimming profile shapes along with the undulation kinematics of the NACA 0012 airfoil.} 
\label{fig:perf1}
\end{center}
\end{figure}

\ed{The optimization process begins with 50 randomly generated initial samples, followed by 150 optimization iterations (epochs) guided by the MixMOBO strategy. Figure~\ref{fig:perf1} illustrates the evolution of the best-obtained propulsive efficiency over the optimization cycle, together with the corresponding optimal swimming profiles and undulation kinematics. Key parameters of several high-performing cases are summarized in Table~\ref{tab:Parf}. The optimized profiles achieve efficiencies in the range of \(52\%\)--\(54\%\), indicating a noticeable enhancement of roughly \(24\%\)--\(28\%\) compared to the maximum efficiency (\(\sim 42\%\)) attained by the reference carangiform and anguilliform swimming modes within the same kinematic parameter space. Furthermore, the best-performing optimized profiles attain their maximum propulsive efficiency at low undulation frequencies, consistent with the trends observed for conventional carangiform and anguilliform swimming modes. Notably, the optimal wavelength values of these profiles fall within the high-efficiency wavelength range characteristic of anguilliform swimming, as shown in Fig.~\ref{fig:eff}. }

\begin{figure}[!ht]
\begin{center}
\subfigure[]{\includegraphics[width=0.48\linewidth]{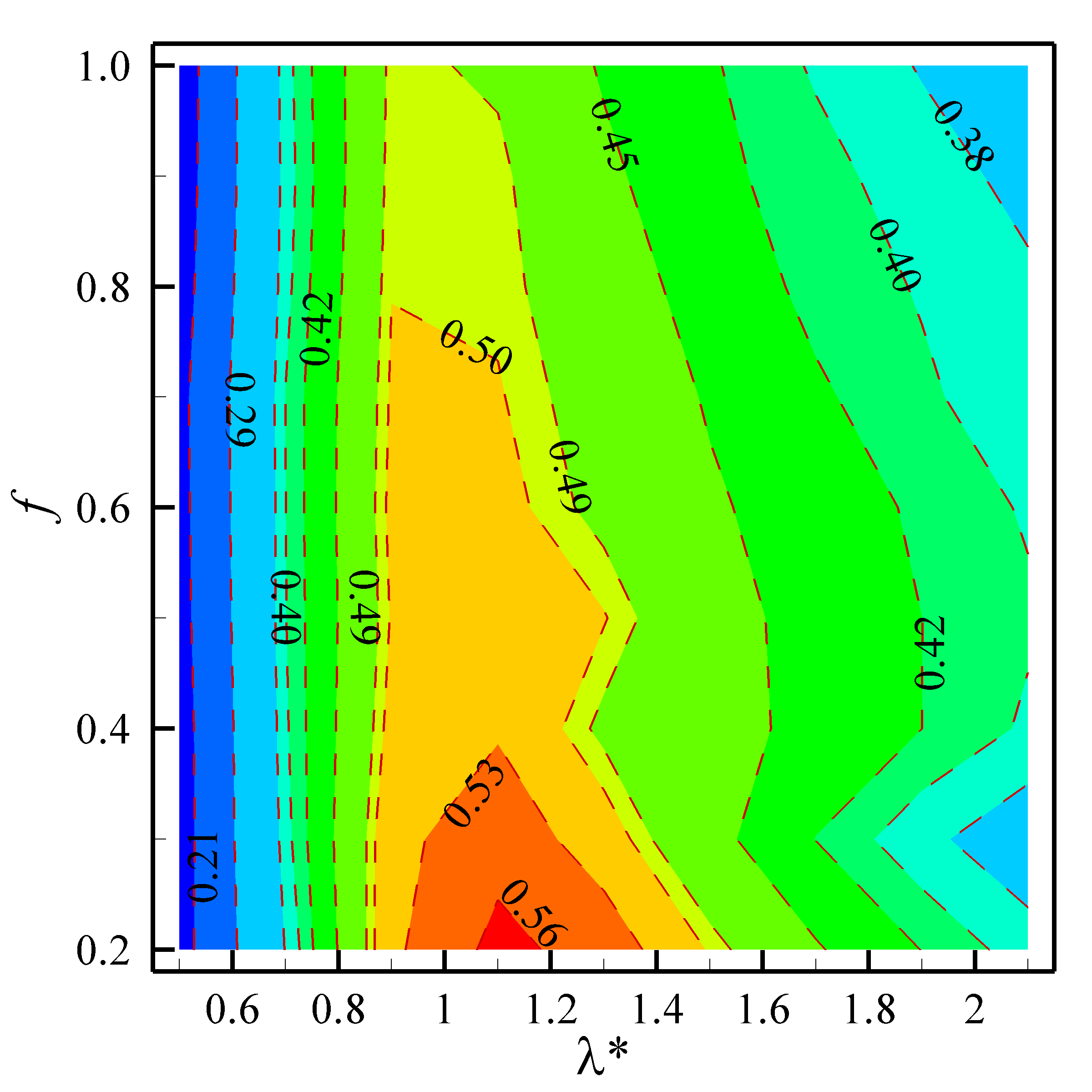}}
\subfigure[]{\includegraphics[width=0.48\linewidth]{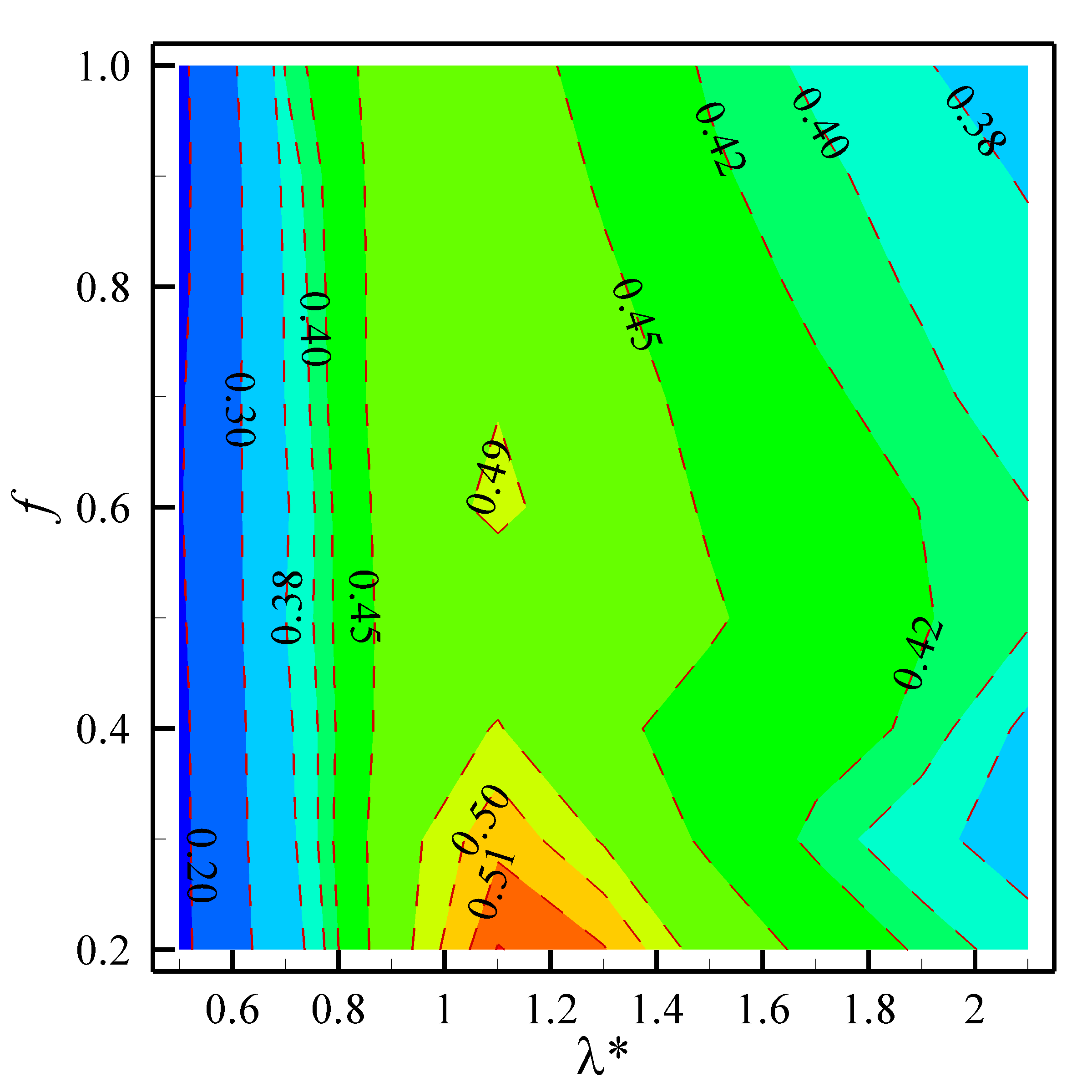}}
\subfigure[]{\includegraphics[width=0.48\linewidth]{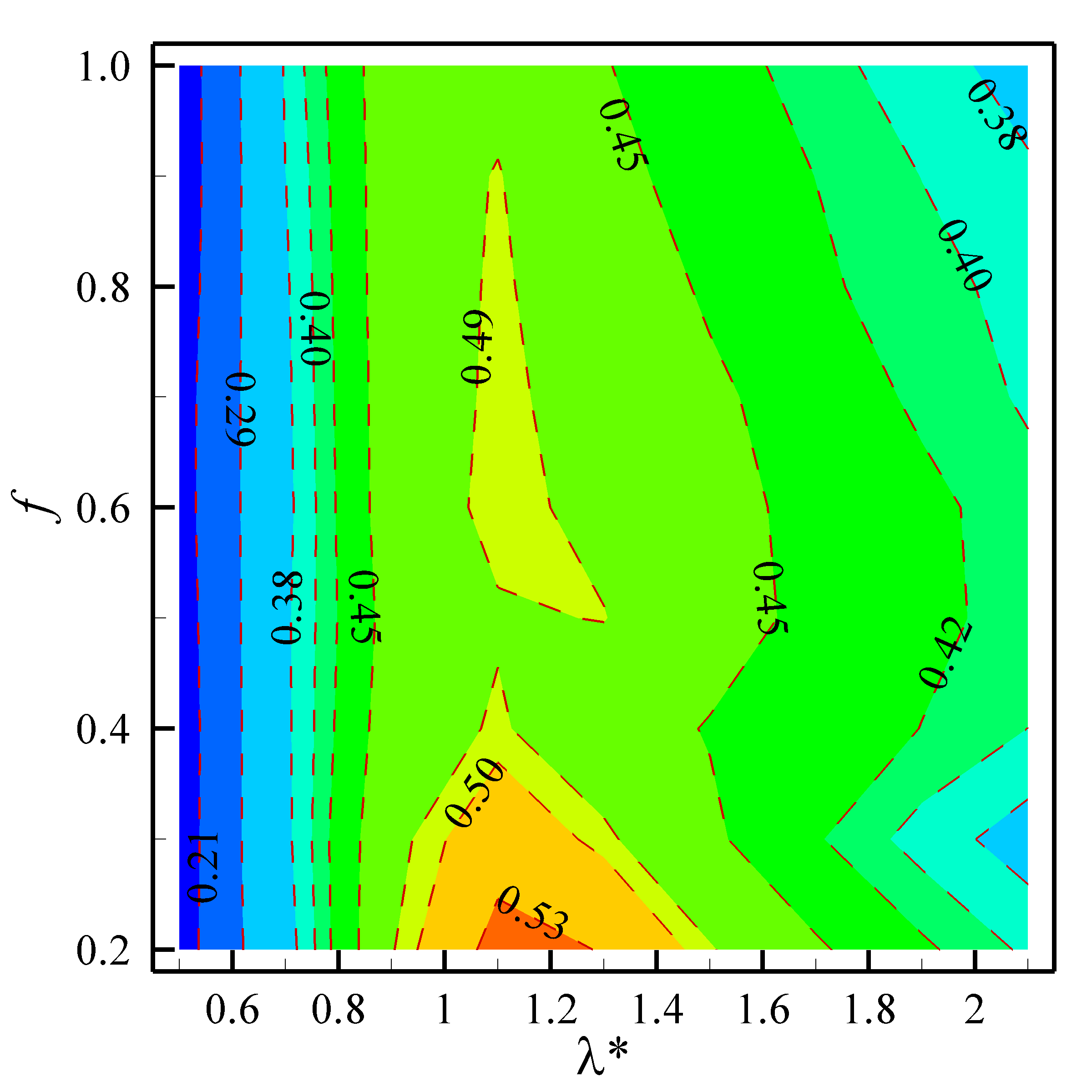}}
\subfigure[]{\includegraphics[width=0.48\linewidth]{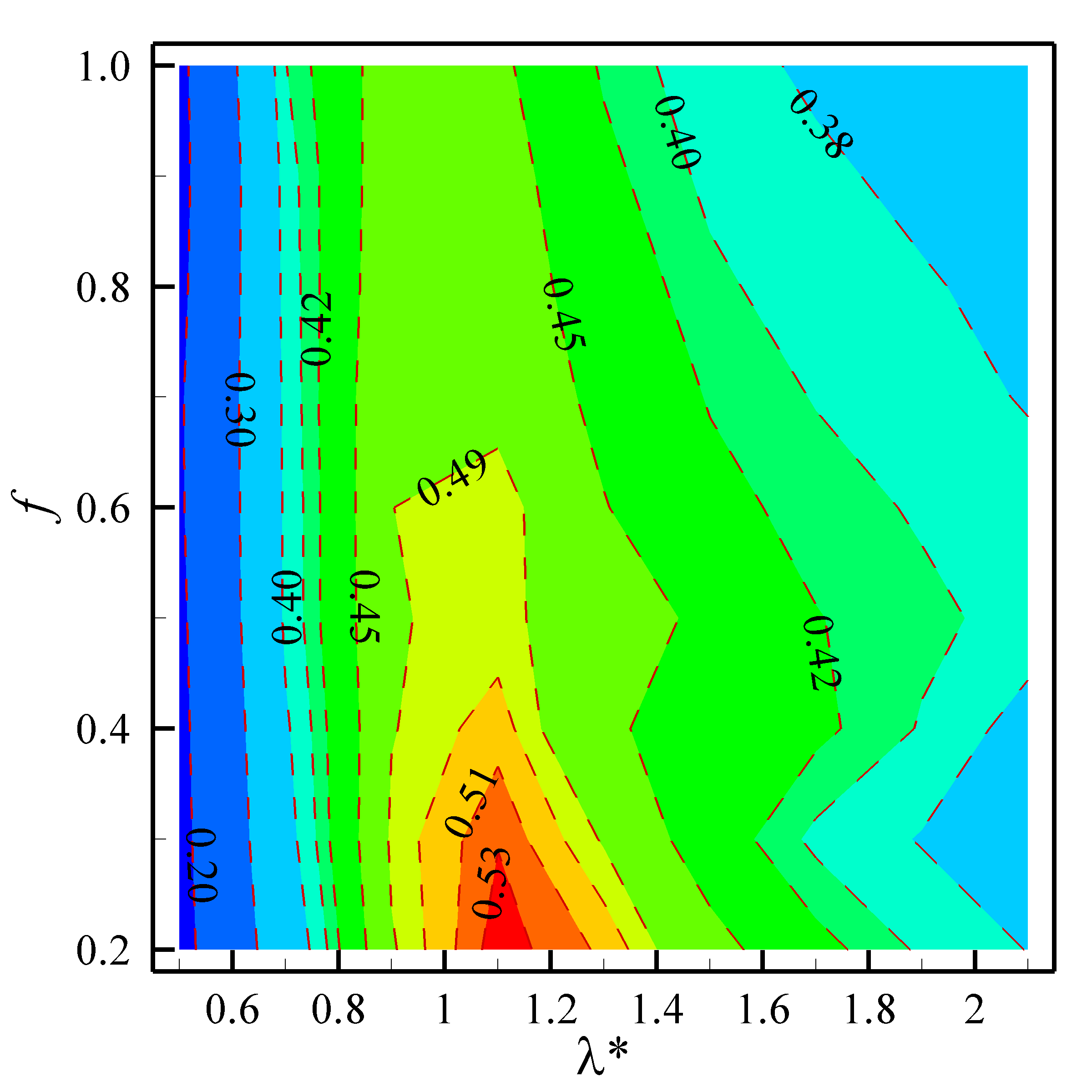}}
\caption{Contour map of the propulsive efficiency in the frequency–wavelength space for (\textbf{a}) $1^{st}$ best, (\textbf{b}) $2^{nd}$ best, (\textbf{c}) $3^{rd}$ best, and (\textbf{d}) $4^{th}$ best profiles at $\mbox{Re}=1000$.} 
\label{fig:effOpt}
\end{center}
\end{figure}

\ed{It is important to emphasize that the ranking of the optimized profiles reported in Table~\ref{tab:Parf} is determined by their peak propulsive efficiency at specific wavelength--frequency combinations. Accordingly, the profiles are locally optimal under their respective kinematic conditions. However, such a ranking does not fully characterize the overall performance of these profiles across the entire frequency--wavelength domain. To obtain deeper insight into their global performance characteristics, additional simulations are performed for the four best optimized profiles across the frequency–wavelength parameter space. The resulting propulsive efficiency contours, presented in Fig.~\ref{fig:effOpt}, provide a comprehensive assessment of the performance of each optimized profile beyond its locally optimal operating point. 

It is observed that the optimized profile (denoted as the $1^{\mathrm{st}}$ best profile in Table~\ref{tab:Parf}), hereafter referred to as the \emph{optimal profile}, attains a maximum propulsive efficiency of approximately $57\%$ at a low undulation frequency of $f = 0.2$. Moreover, the optimal profile sustains a high efficiency level, ranging from $49\%$ to $57\%$ (corresponding to an enhancement of roughly $16\%$--$35\%$ relative to the baseline swimming modes), over a broad region of the frequency--wavelength parameter space, indicating robust performance across a wide range of kinematic conditions. All other optimized profiles exhibit qualitatively similar efficiency trends, with elevated performance concentrated at low frequencies and moderate wavelengths; however, their peak efficiencies are comparatively lower and the regions of sustained high performance are more limited. Based on this global assessment across the entire frequency--wavelength parameter space, the optimized profiles are hereafter ranked accordingly.

It should also be noted that the present study considers an idealized representation of body kinematics. In real fish, physiological and biomechanical constraints may limit the ability to achieve the optimized body profiles identified in this work. Furthermore, the present analysis is based on two-dimensional simulations; therefore, three-dimensional swimming effects, such as spanwise flow structures and body--fin interactions, are not accounted for and may influence the optimal swimming characteristics.}

\begin{figure}[!ht]
\begin{center}
\includegraphics[width=\linewidth]{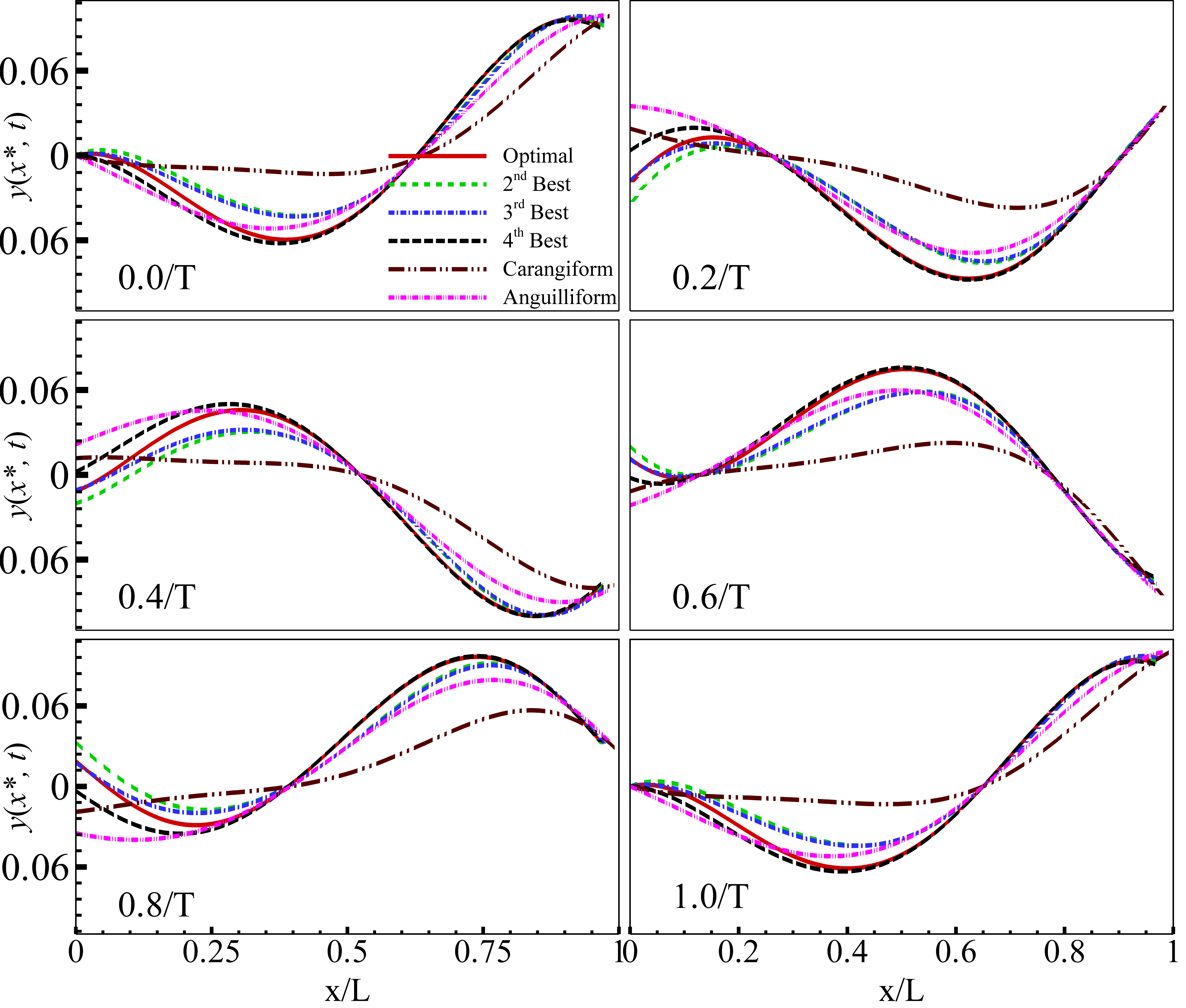}
\caption{Time evolution of the backbone (centerline) kinematics of the top three optimized swimming profiles compared with the conventional anguilliform and carangiform modes at five different phases within one undulation cycle ($t/T = 0.0$, $0.2$, $0.4$, $0.6$, $0.8$, and $1.0$). The optimized profiles exhibit notable differences in head and tail motions compared to the anguilliform profile.} 
\label{fig:backbone}
\end{center}
\end{figure}

To investigate the differences in the kinematic behavior between the optimized and conventional swimming modes, the backbone (centerline) undulation of the NACA 0012 airfoil over six time instances within one oscillation cycle is presented in Fig.~\ref{fig:backbone}. The optimized profiles exhibit similar undulatory patterns, with the main distinction being a more pronounced head motion in the optimal profile. Compared to the anguilliform profile, the optimal profile aligns closely over a significant region that spans approximately 30\% to 87\% of the body length. However, at the leading and trailing edges, notable differences are evident. The optimal profile exhibits head motion in the opposite direction to that of the anguilliform profile, and the tail amplitude is marginally lower, with the anguilliform tail reaching a maximum lateral amplitude near 0.1. These distinctive features, specifically the out-of-phase head motion, may \edt{contribute} to the superior propulsive performance exhibited by the optimized profiles. The observed increase in efficiency is likely associated with these deviations from the anguilliform pattern in both head and tail kinematics. To further examine this hypothesis, a detailed analysis of the wake structures and force distributions is presented in the following discussion.


\begin{table}
\centering
\caption{Detail of governing parameters and their relevant specifications}
\begin{tabular}{cccccccccc}
\hline
 Design &  Epoch & $\omega_1$ & $\omega_2$ & $\omega_3$ & $\omega_4$ & $\omega_5$ & $\lambda^*$ & $f$ & $\eta$\%\\
  \hline
 \hline
1$^{st}$ Best & 160 & -0.866 & 0.2500 & 0.1875 & 0.0406 & 0.1083 & 1.3 & 0.2 & 54.76 \\
2$^{nd}$ Best & 141 & -0.8660 & 0.0 & 0.2165 & 0.0 & 0.2500 & 1.3 & 0.2 & 53.81 \\
3$^{rd}$ Best & 130 & -0.8660 & 0.0 & 0.2165 & -0.0312 & 0.2165 & 1.2 & 0.2 & 53.57 \\
4$^{th}$ Best & 143 & -0.8660 & 0.2500 & 0.1875 &-0.0406 & 0.1083 & 1.3 & 0.2 & 52.87
\end{tabular}

\label{tab:Parf}
\end{table}

\begin{figure}[!ht]
\begin{center}
\includegraphics[width=\linewidth]{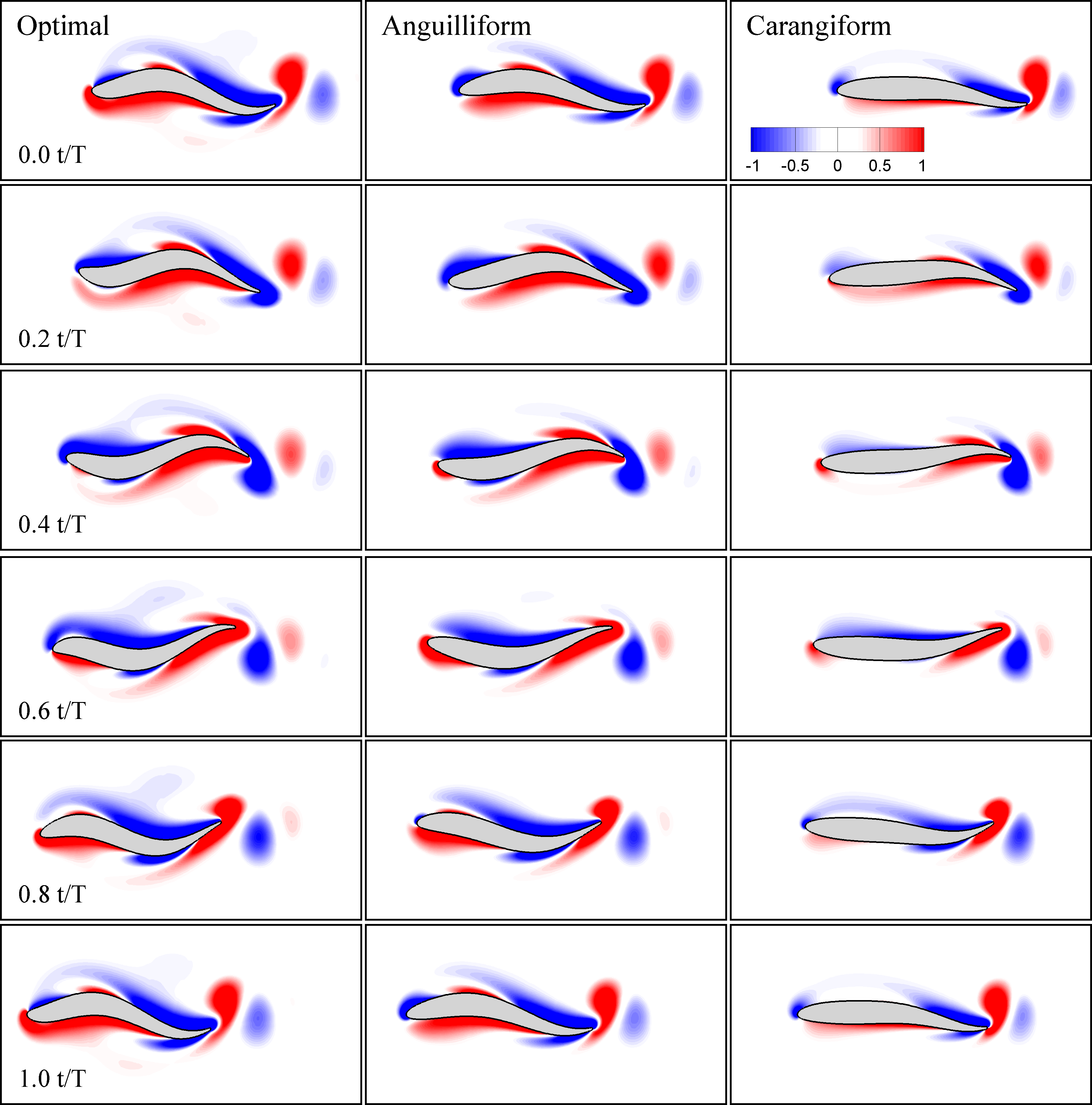}
\caption{Instantaneous vorticity contours, ranging from -1 to 1, illustrating the wake topology for the optimal ($\eta$=57\%), anguilliform ($\eta$=42\%), and carangiform ($\eta$=36\%) swimming profiles.} 
\label{fig:wake}
\end{center}
\end{figure}

The wake topology of the optimal profile, along with those of the anguilliform and carangiform profiles for comparison at $f=0.2$ and $\lambda^*=1.1$, is presented in Fig.~\ref{fig:wake} for over a complete oscillating cycle. Although the global wake structures appear qualitatively similar across all three cases, a distinct difference is evident in the vortex roll-up behavior. The optimal profile exhibits a more pronounced and coherent rolling of vortices shed from both the lower and upper surfaces, particularly in the mid-to-posterior region of the body. This feature is relatively weaker in the anguilliform profile and virtually absent in the carangiform case. This enhanced vortex roll-up in the optimal configuration can be linked to the unique kinematic features identified in the backbone motion (see Fig.~\ref{fig:backbone}). The optimal profile exhibits reduced tail excursion and an out-of-phase head motion relative to the anguilliform profile. These features modify the pressure distribution and vorticity generation along the body, promoting stronger and more organized vortex formation. Consequently, the resulting wake pattern in the optimal case supports more efficient momentum transfer to the fluid, which aligns with the significant reduction in the work required to undulate the body and the corresponding increase in propulsive efficiency observed in the performance metrics.

\begin{figure}[!ht]
\begin{center}
\includegraphics[width=\linewidth]{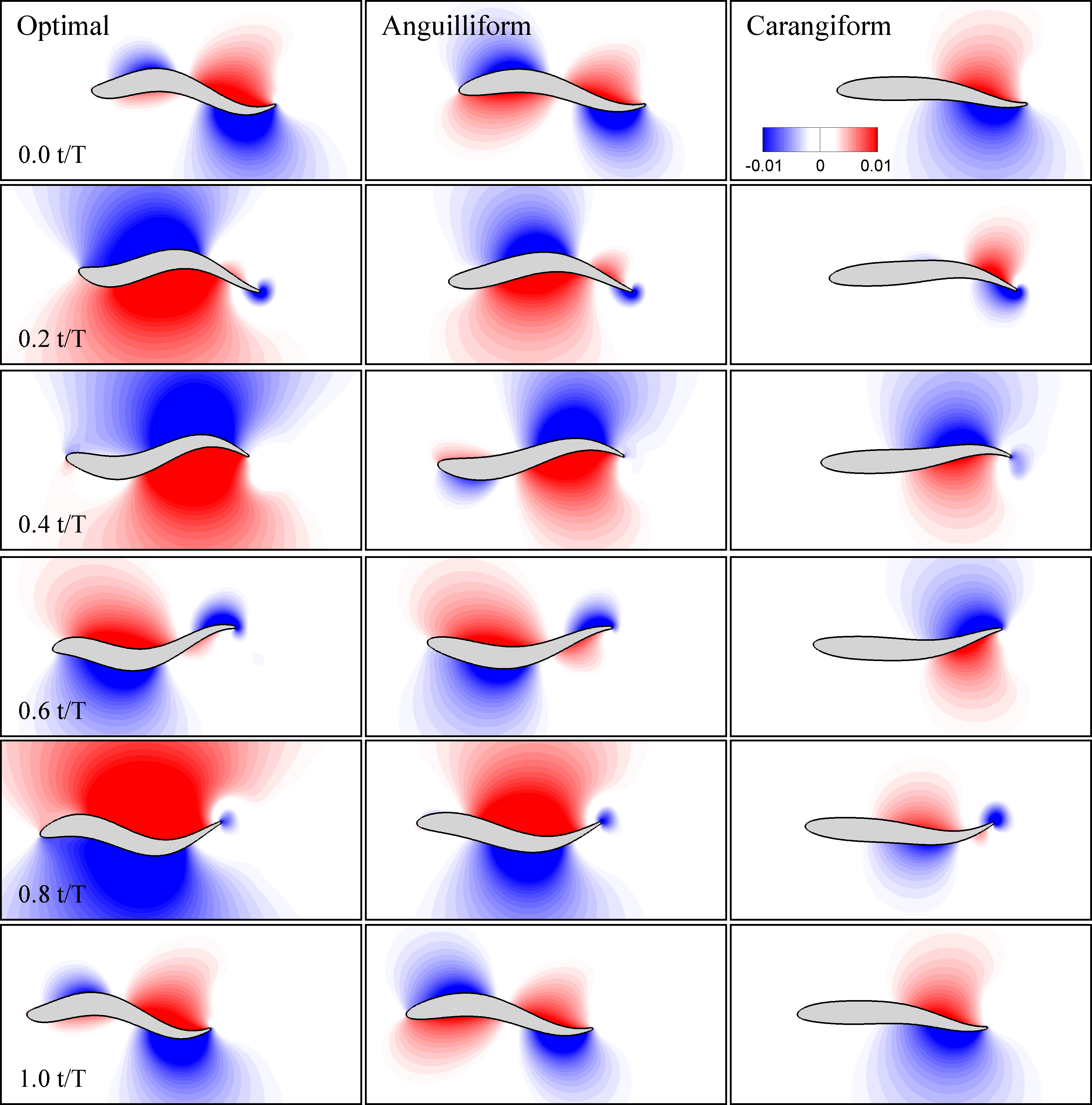}
\caption{Instantaneous pressure contours over one undulatory cycle for the optimal, anguilliform, and carangiform profiles, shown in the left, middle, and right columns, respectively. The optimal profile (left column) exhibits significantly higher pressure magnitudes, particularly near the head and mid-body regions, due to its out-of-phase head motion. In contrast, the anguilliform (middle column) and carangiform (right column) profiles display weaker pressure variations, with the carangiform profile showing minimal pressure near the head.} 
\label{fig:presswake}
\end{center}
\end{figure}

\begin{figure}[!ht]
\begin{center}
\includegraphics[width=0.8\linewidth]{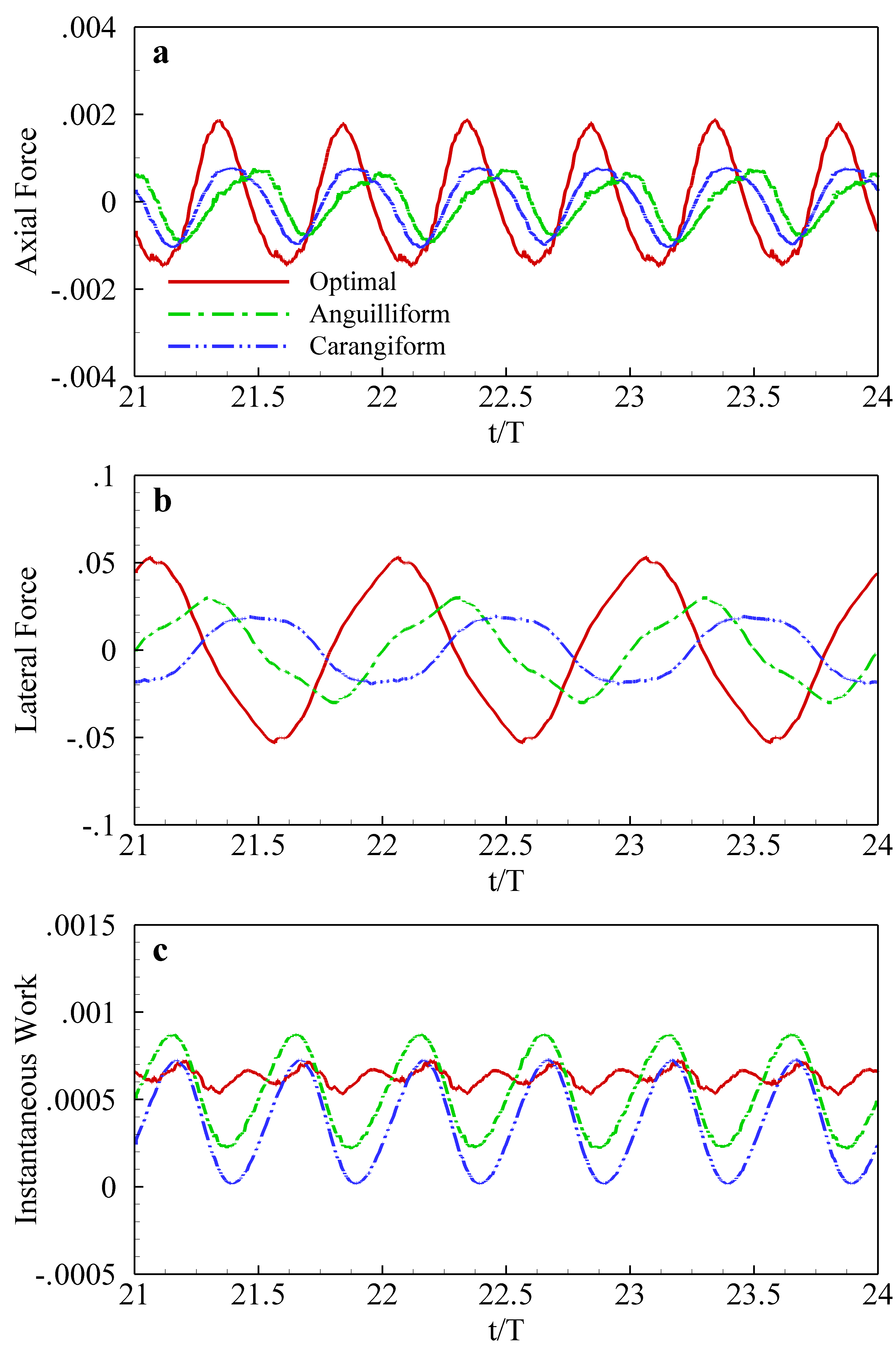}
\caption{Instantaneous variations of hydrodynamic quantities over three undulatory cycles for the optimal, anguilliform, and carangiform profiles: (a) axial force, (b) lateral force, and (c) instantaneous work performed by the foil. The optimal profile (solid red line) exhibits significantly higher axial and lateral force magnitudes with minimal phase lag. 
}
\label{fig:forces}
\end{center}
\end{figure}

\begin{figure}[!ht]
\begin{center}
\subfigure{\includegraphics[width=0.48\linewidth]{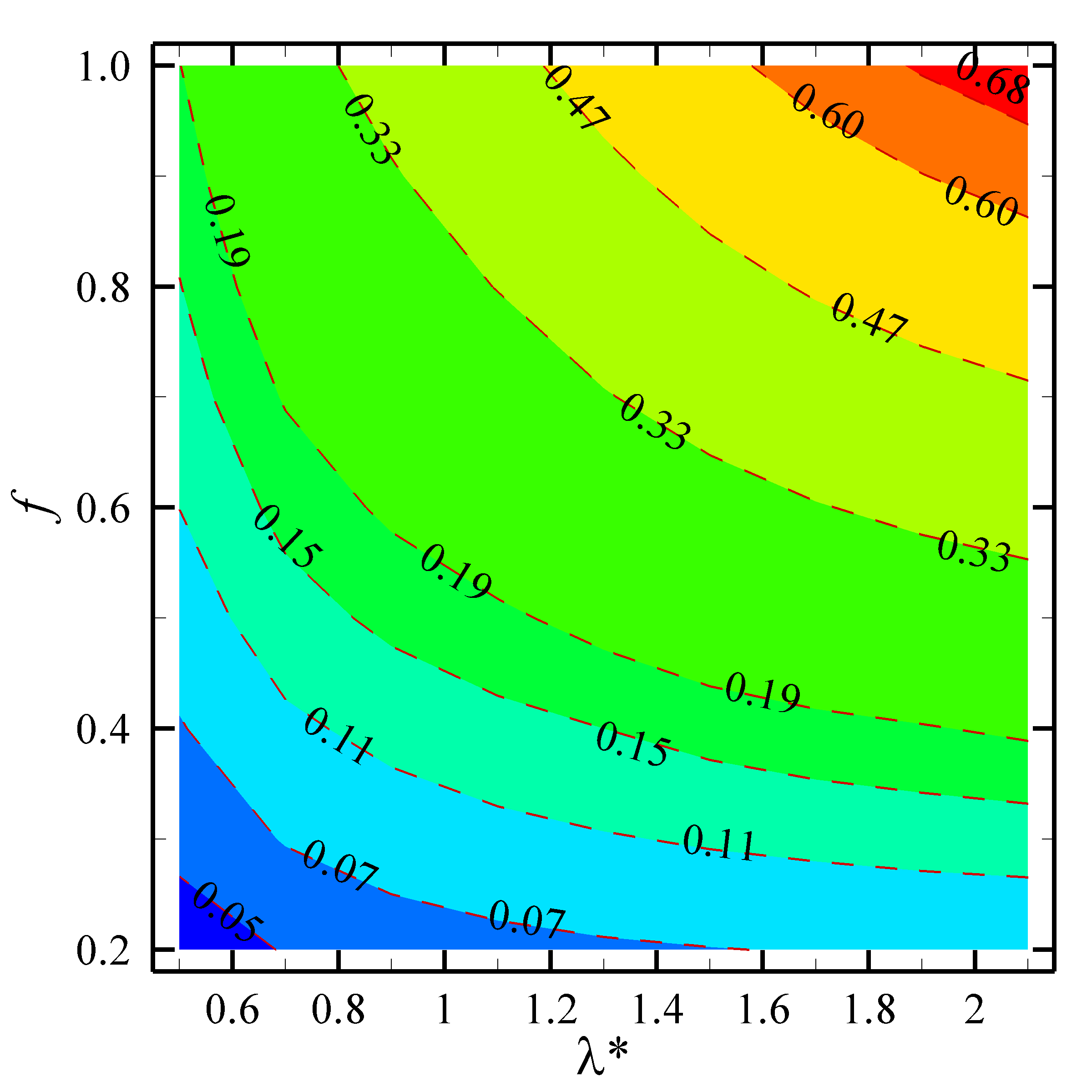}}
\subfigure{\includegraphics[width=0.48\linewidth]{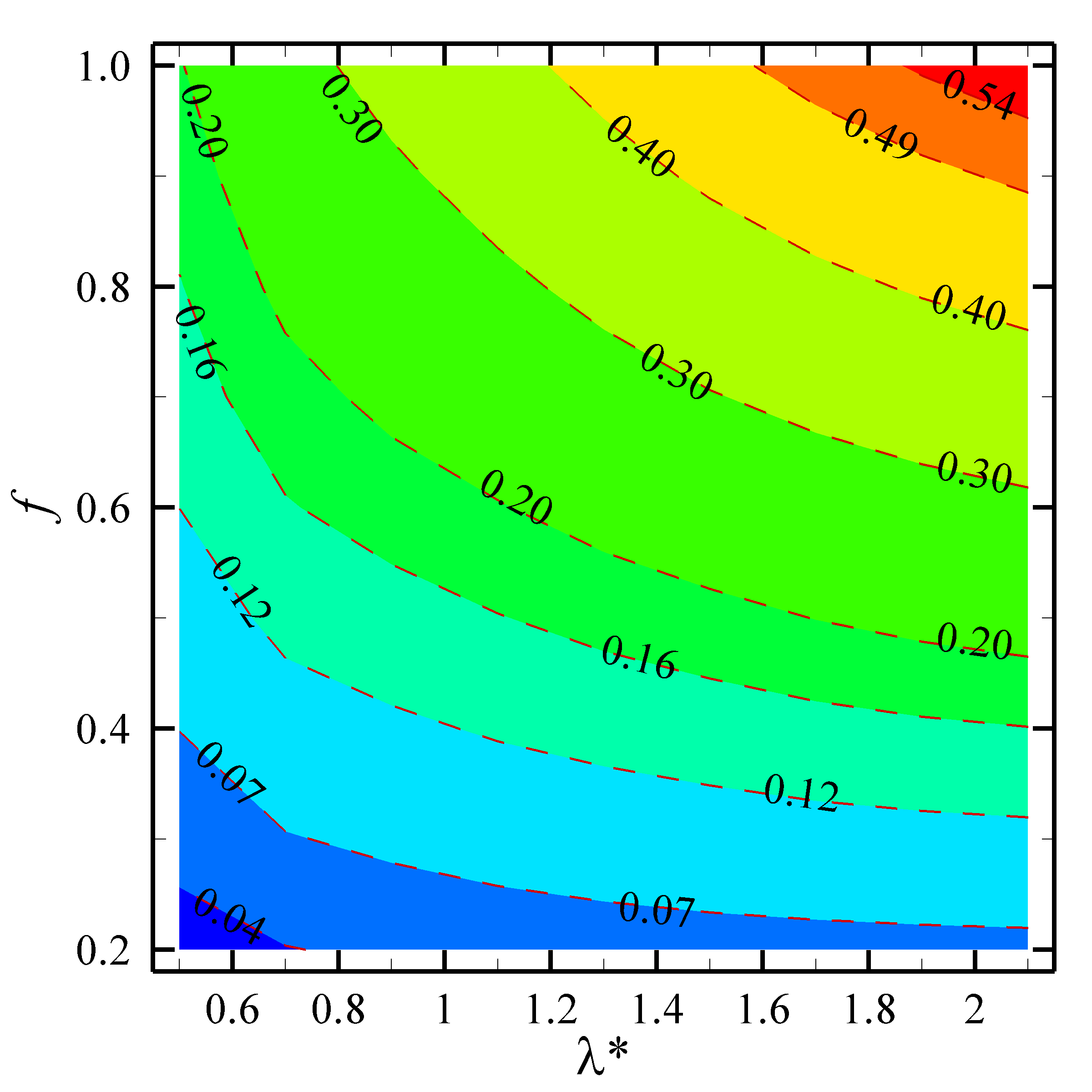}}
\subfigure{\includegraphics[width=0.48\linewidth]{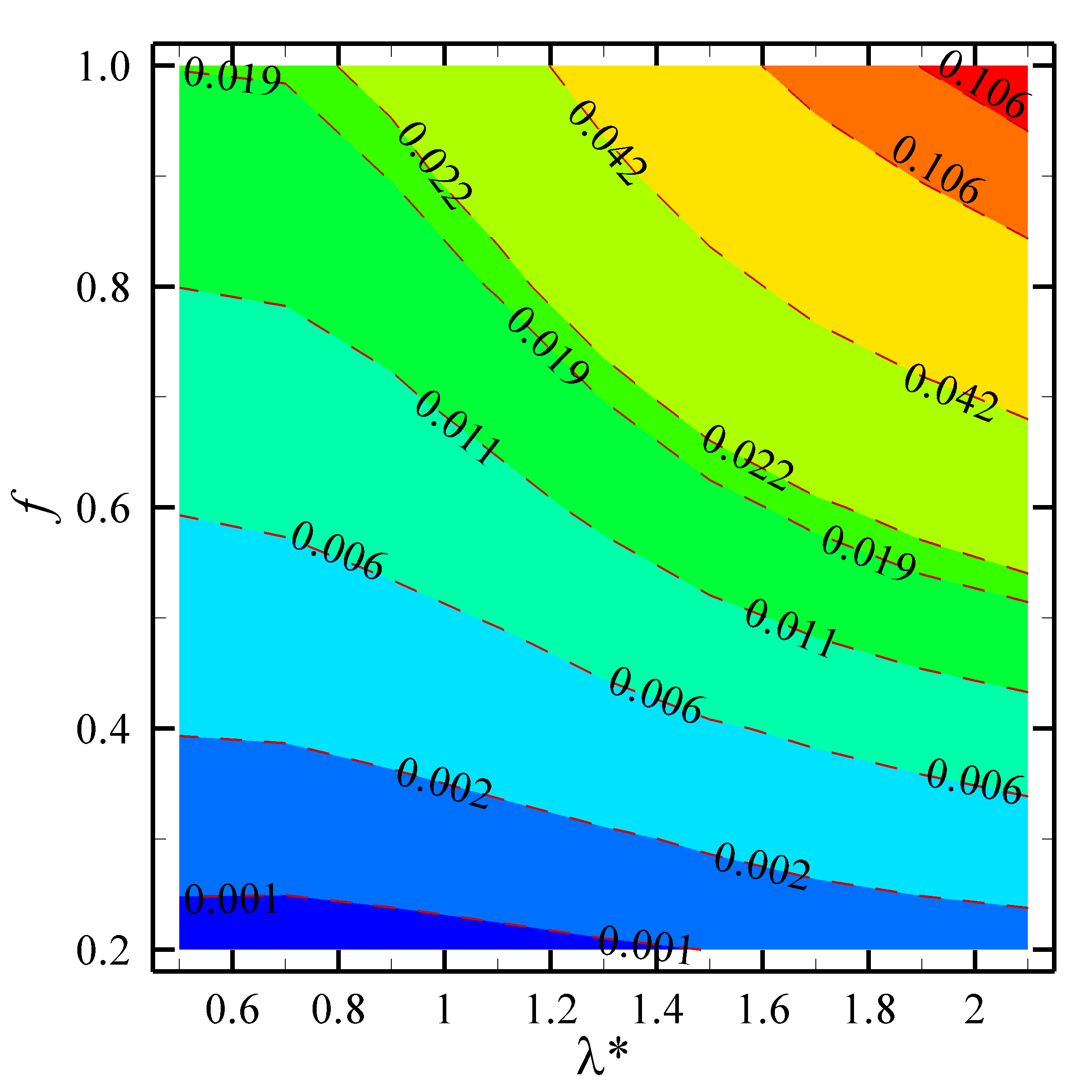}}
\subfigure{\includegraphics[width=0.48\linewidth]{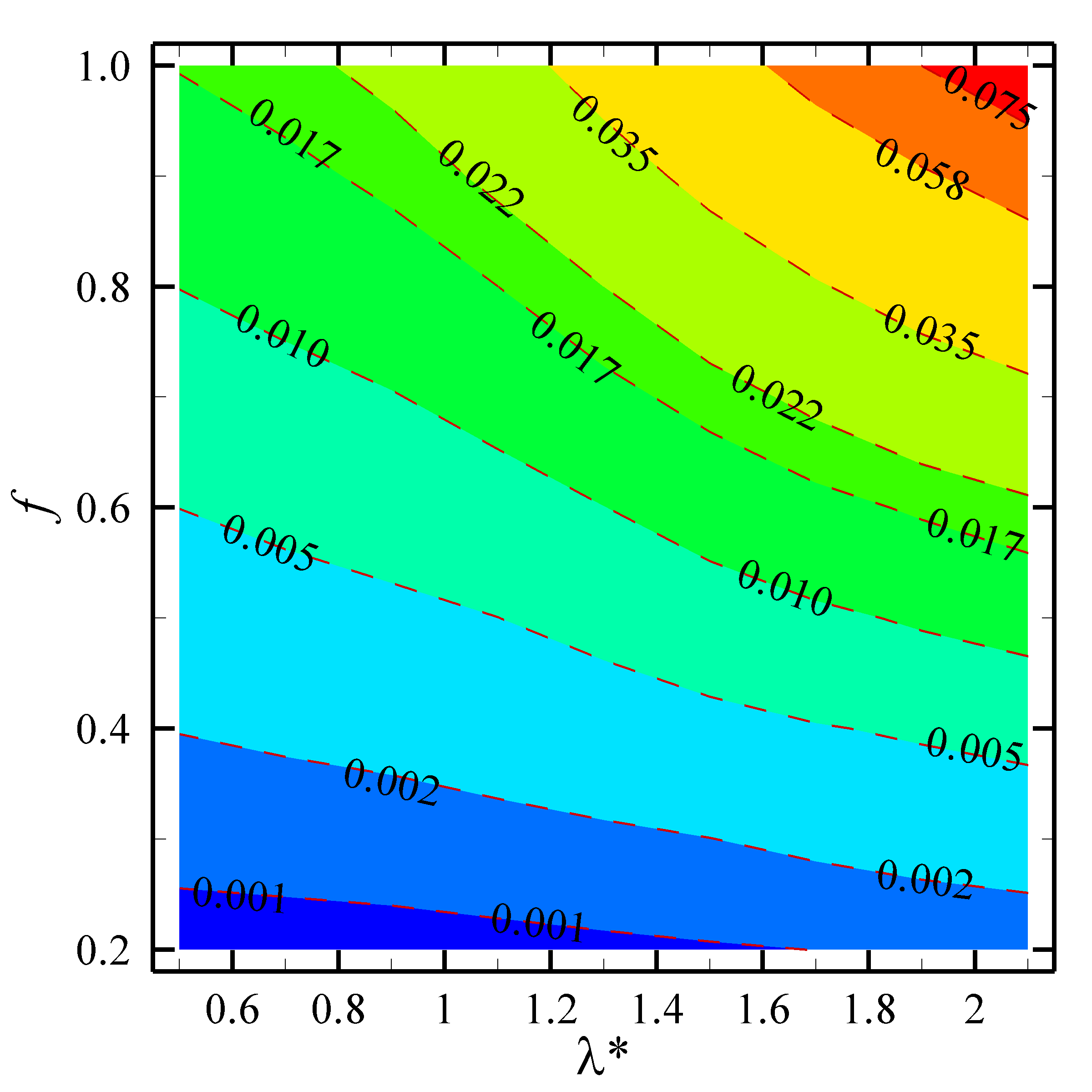}}
\caption{Comparison of contour maps for the time-averaged propulsive velocity (top row) and mean work consumption (bottom row) across the frequency–wavelength domain. Results for the optimal swimming profile are shown in the left column, and those for the conventional anguilliform profile are shown in the right column.
} 
\label{fig:velwork}
\end{center}
\end{figure}

Figure~\ref{fig:presswake} presents the instantaneous pressure contours for one undulatory cycle for the optimal profile, shown alongside the anguilliform and carangiform profiles for a comparative analysis. Distinct differences in the pressure distribution are evident among the three profiles, particularly near the anterior and posterior regions. These variations arise due to the out-of-phase head motion exhibited by the optimal profile compared to the anguilliform case. The overall pressure magnitude is higher in the optimal profile, which indicates stronger flow-body interaction and better hydrodynamics performance. In contrast, the anguilliform profile displays relatively lower pressure levels, while the carangiform profile exhibits minimal pressure variation near the head. The stronger flow–body interaction in the optimal profile is further supported by the hydrodynamic performance metrics shown in Fig.~\ref{fig:forces}, which presents the instantaneous variations in axial force, lateral force, and the work performed by the foil over one undulatory cycle. \ed{The optimal profile exhibits significantly higher force magnitudes with minimal phase lag. Although the mean work consumption of the optimal profile is slightly higher than that of the conventional profiles, it achieves faster propulsion, resulting in improved overall efficiency. In other words, the optimizer identifies a profile that prioritizes higher propulsive speed. The contour maps of mean work and propulsive speed for both the optimal and anguilliform profiles are presented in Fig.~\ref{fig:velwork}. It is observed that the optimal profile achieves a higher propulsive speed than the anguilliform case, albeit at the expense of increased work consumption, as illustrated in the region $0.9 \le \lambda^* \le 1.5$ in Fig.~\ref{fig:velwork}, where enhanced efficiency is observed. Importantly, the relative increase in propulsive speed slightly exceeds the corresponding increase in work, indicating a more favorable performance trade-off. Consequently, this imbalance leads to an overall improvement in propulsive efficiency, rendering the optimal profile superior to the anguilliform profile.
}

\begin{figure}[!ht]
\begin{center}
\includegraphics[width=\linewidth]{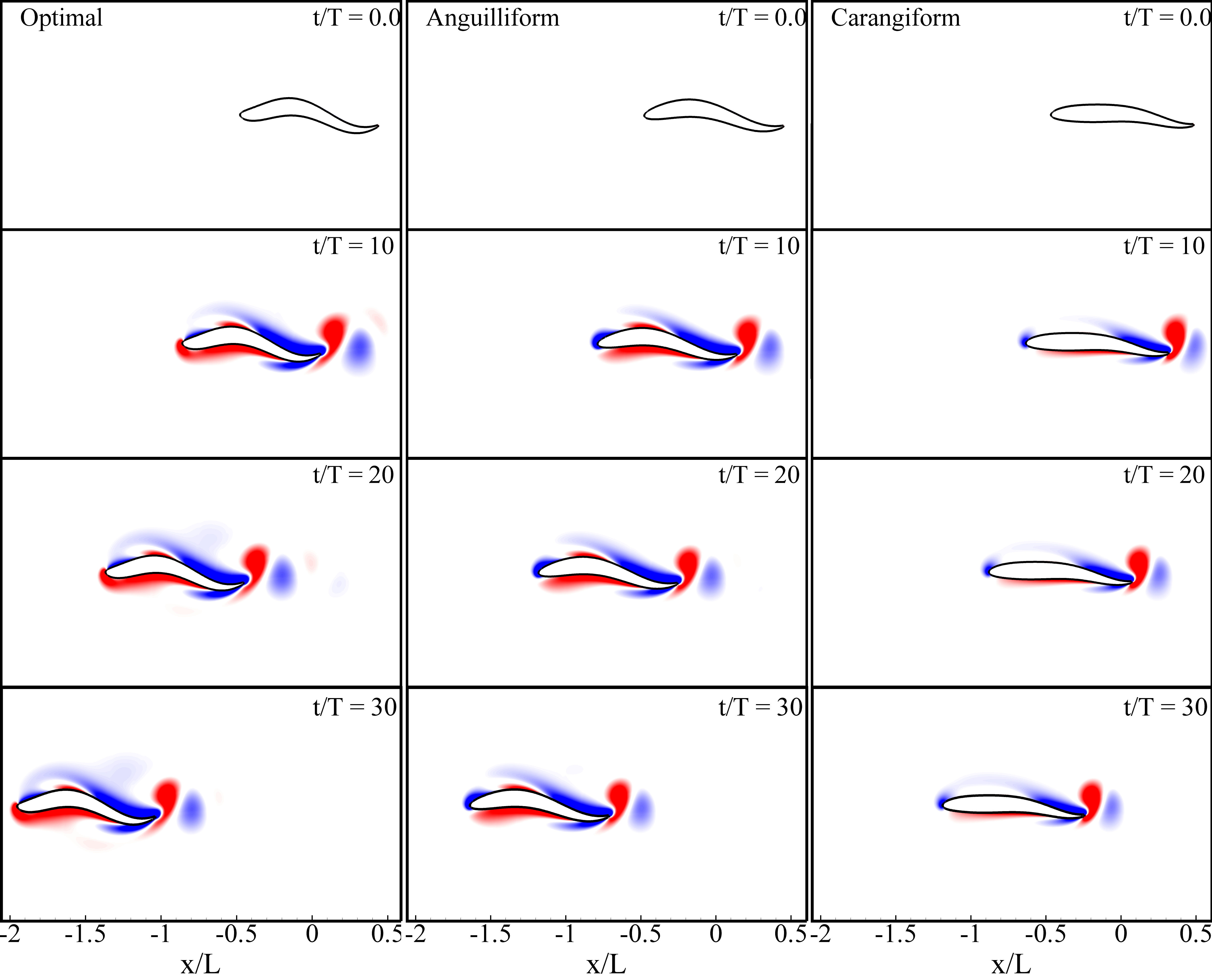}
\caption{Instantaneous vorticity contours for the optimal, anguilliform, and carangiform swimming modes at four time instants ($t/T = 0, 10, 20,$ and $30$).} 
\label{fig:fishspeed}
\end{center}
\end{figure}

\ed{To further elucidate the flow–body interaction mechanisms underlying the enhanced performance of the optimal profile, instantaneous vorticity contours for the optimal, anguilliform, and carangiform swimming modes are presented at four representative time instants ($t/T = 0, 10, 20,$ and $30$) in Fig.~\ref{fig:fishspeed}. The snapshots reveal clear differences in wake topology and vortex evolution among the three cases. In particular, the optimal profile generates a more coherent and spatially organized reverse von Kármán vortex street, characterized by stronger and more consistently shed vortical structures. This organized wake indicates more effective momentum transfer to the fluid, resulting in higher thrust production. 

These flow characteristics are consistent with the trends observed in Fig.~\ref{fig:velwork}, where the optimal profile achieves higher propulsive speed at the cost of moderately increased work consumption. Importantly, the enhanced coherence and strength of the wake structures indicate that the additional input work is more effectively converted into useful propulsion. As a result, the relative increase in propulsive speed slightly exceeds the corresponding increase in work, leading to a more favorable performance trade-off. This fluid-dynamic evidence further substantiates that the optimal profile attains superior propulsive efficiency compared to the anguilliform profile.}

\subsection{Force Characteristics}
\begin{figure}[!ht]
\includegraphics[width=0.48\linewidth]{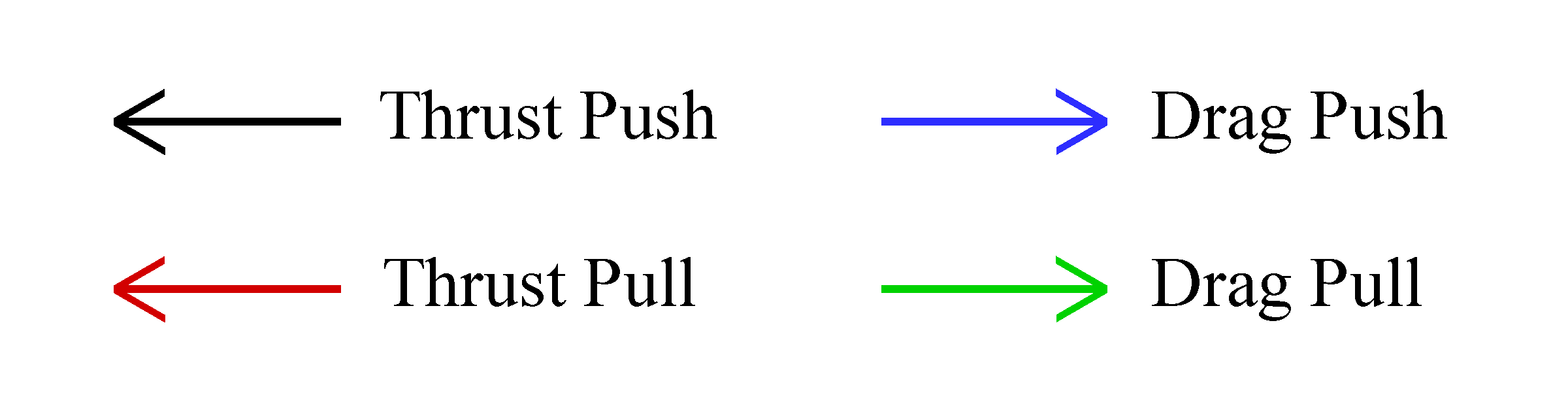}
\begin{center}
\subfigure{\includegraphics[width=0.48\linewidth]{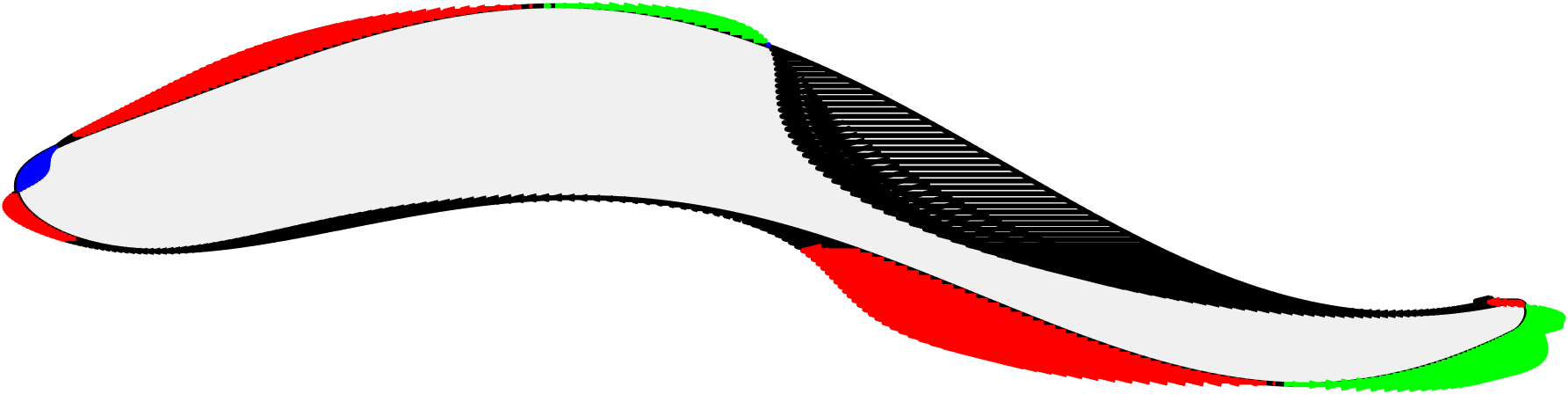}}
\subfigure{\includegraphics[width=0.48\linewidth]{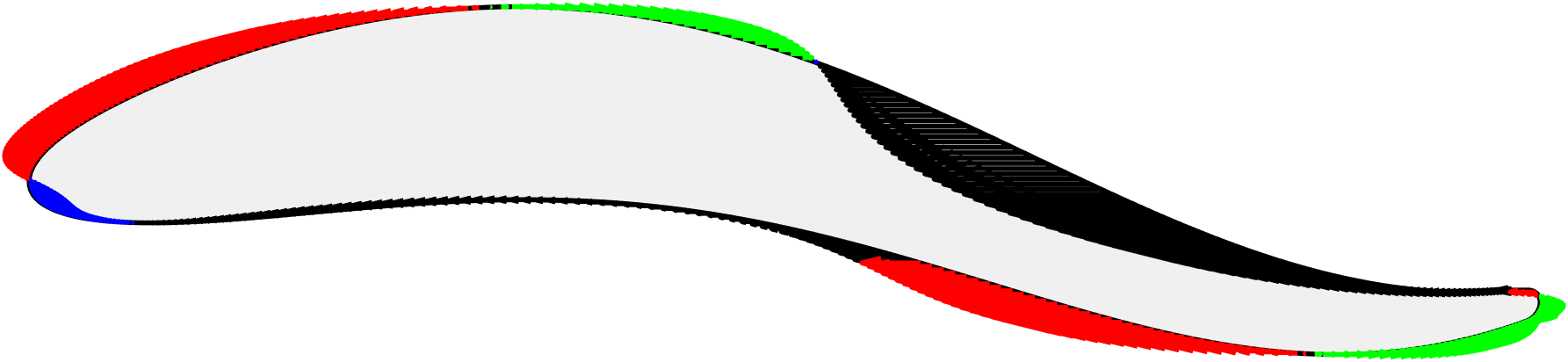}}
\subfigure{\includegraphics[width=0.48\linewidth]{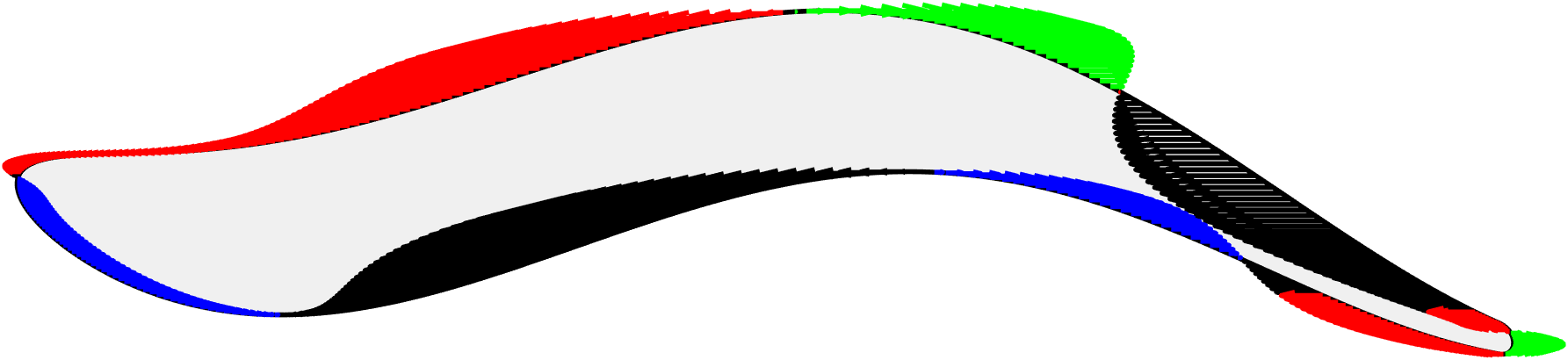}}
\subfigure{\includegraphics[width=0.48\linewidth]{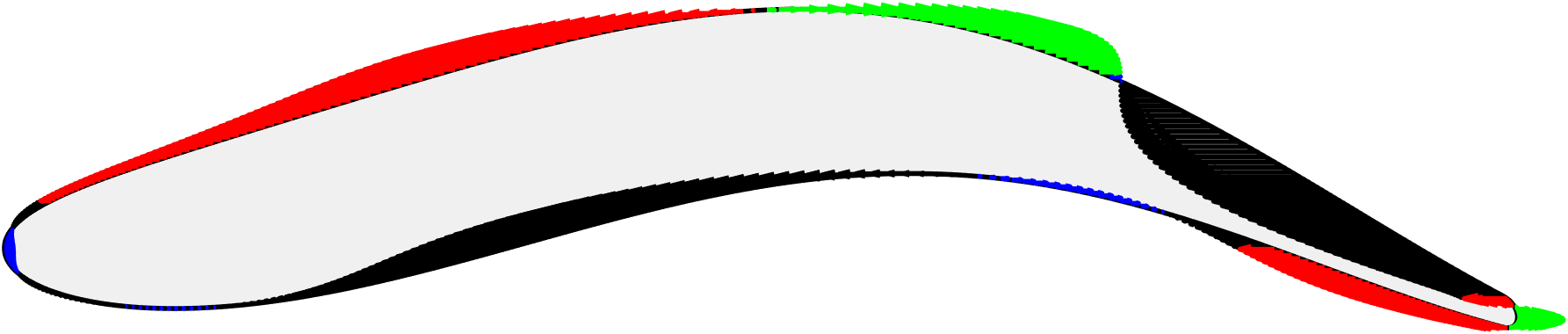}}
\subfigure{\includegraphics[width=0.48\linewidth]{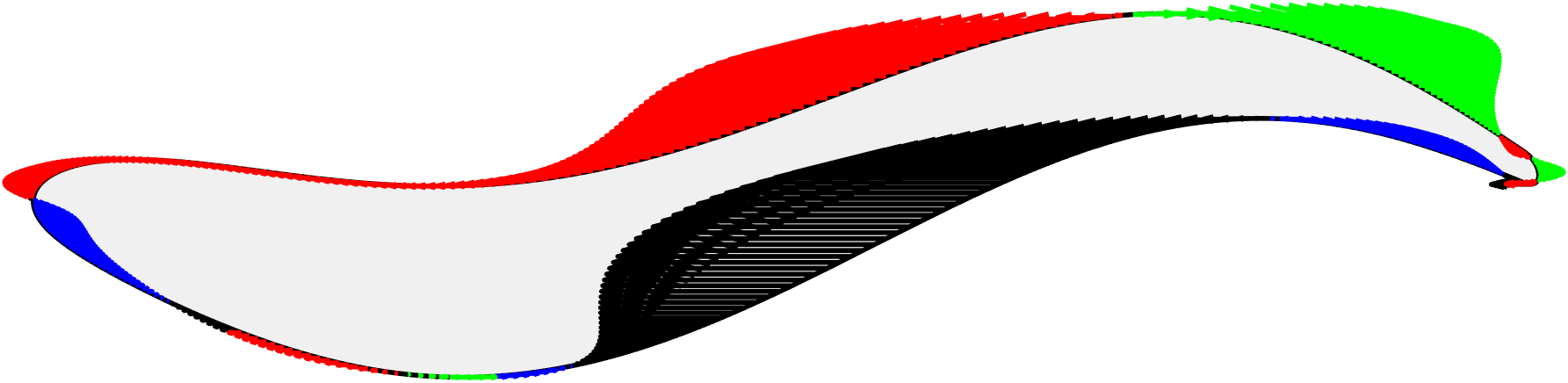}}
\subfigure{\includegraphics[width=0.48\linewidth]{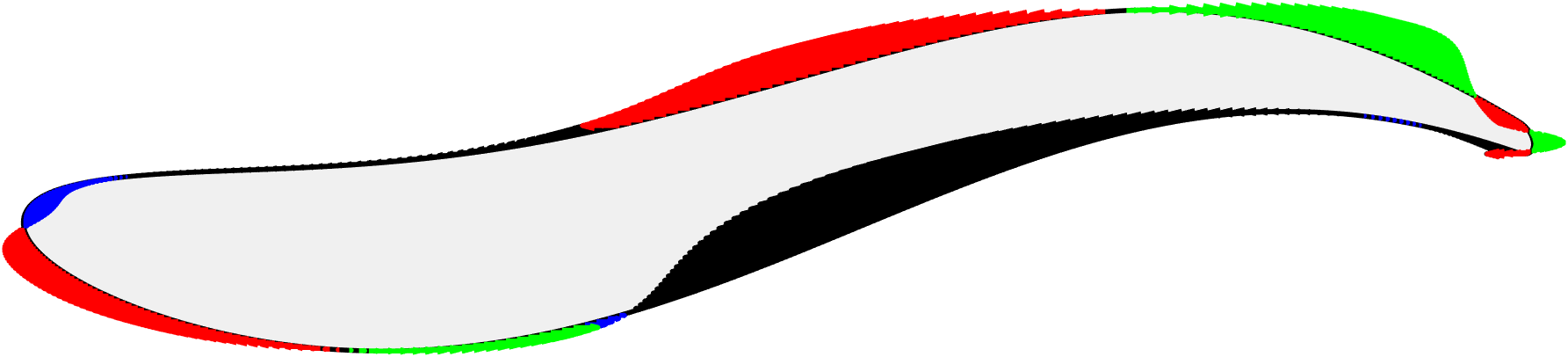}}
\subfigure{\includegraphics[width=0.48\linewidth]{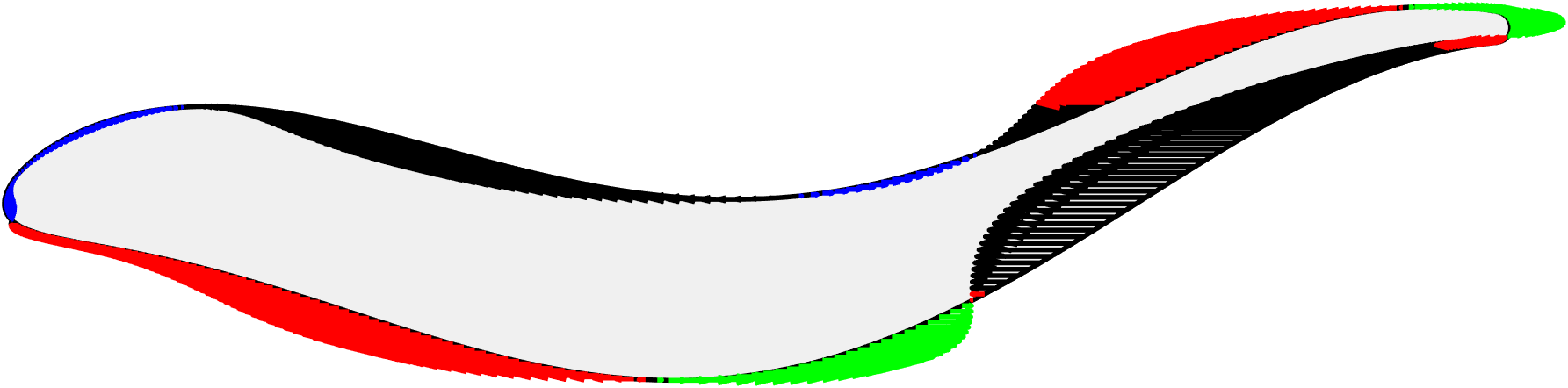}}
\subfigure{\includegraphics[width=0.48\linewidth]{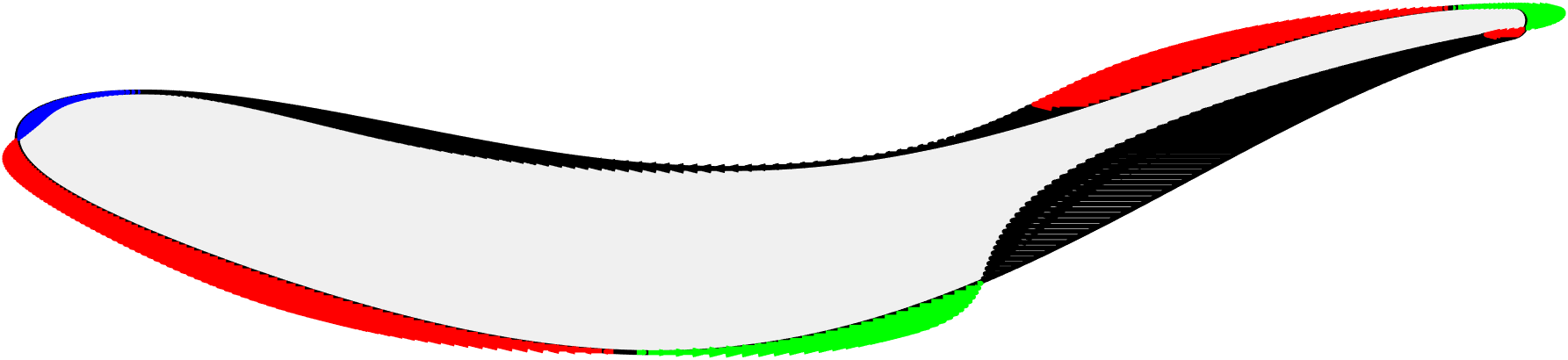}}
\subfigure{\includegraphics[width=0.48\linewidth]{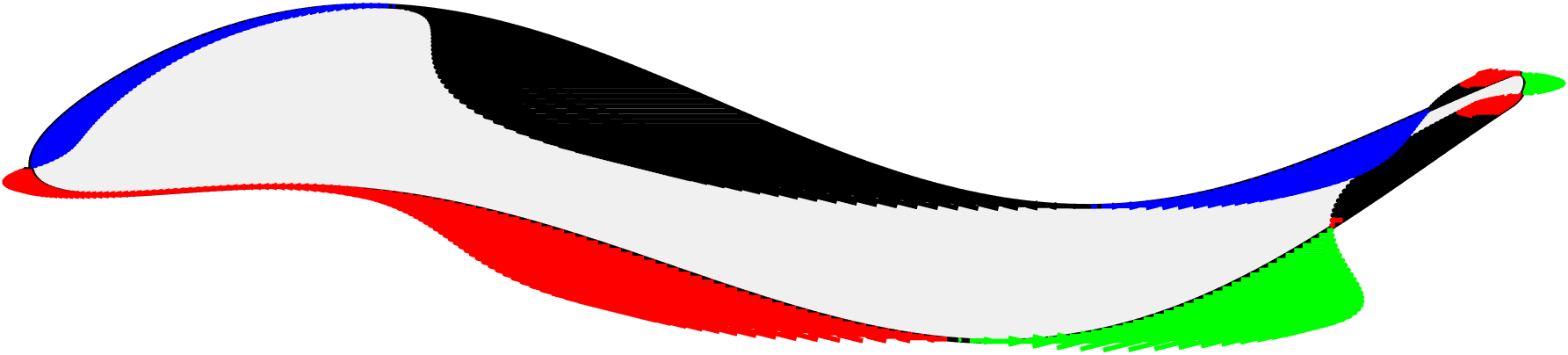}}
\subfigure{\includegraphics[width=0.48\linewidth]{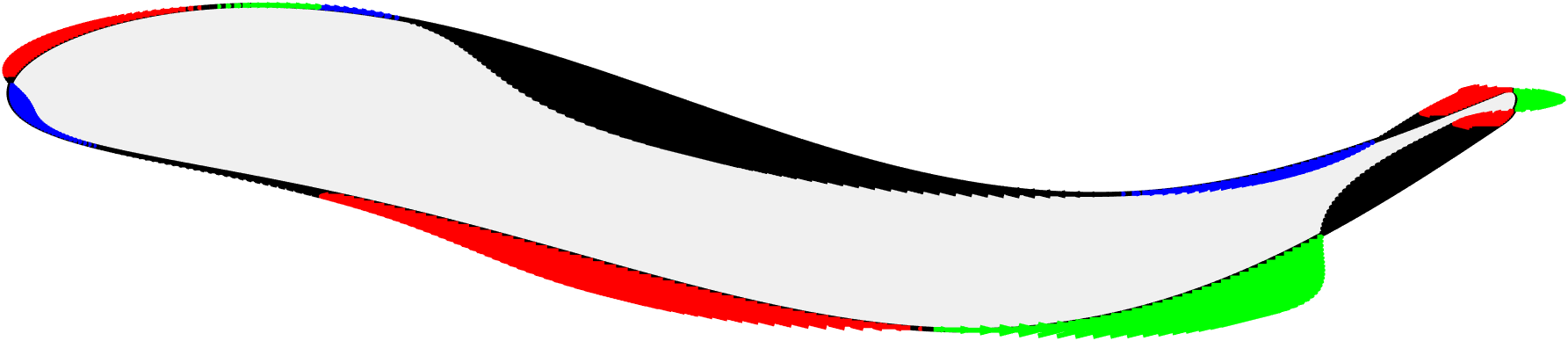}}
\caption{Temporal snapshots of surface force decomposition along the optimal (left column) and anguilliform (right column) profiles over one undulatory cycle. Arrows indicate the local surface force direction and magnitude, decomposed into thrust-push, thrust-pull, drag-push, and drag-pull components.} 
\label{fig:axial}
\end{center}
\end{figure}

It becomes essential to investigate the distribution of flow-induced forces in greater detail, particularly how these forces contribute to propelling the body. The interaction between these surface forces and the local flow velocity governs the instantaneous energy exchange and determines the overall energy expenditure. Consequently, a detailed analysis of the spatial and temporal distribution of work along the undulating body is critical for elucidating the mechanisms that lead to high propulsive efficiency. To this end, the axial force comprising both pressure and shear stress contributions is first decomposed into local thrust and drag components, corresponding to the positive and negative values of the local axial force, respectively. These components are further resolved into push and pull forces \citep{du2019thrust}, based on positive and negative pressure relative to the ambient, and are visualized using directional vectors, as illustrated in Fig.~\ref{fig:axial}, at five representative time instants over a complete undulatory cycle. From this point onward, the comparison is restricted to the optimal and anguilliform profiles, for $f=0.2$ and $\lambda^* = 1.1$, as the optimal profile shares greater kinematic similarity with the anguilliform mode.

\begin{figure}[!ht]
\begin{center}
\subfigure[]{\includegraphics[width=0.48\linewidth]{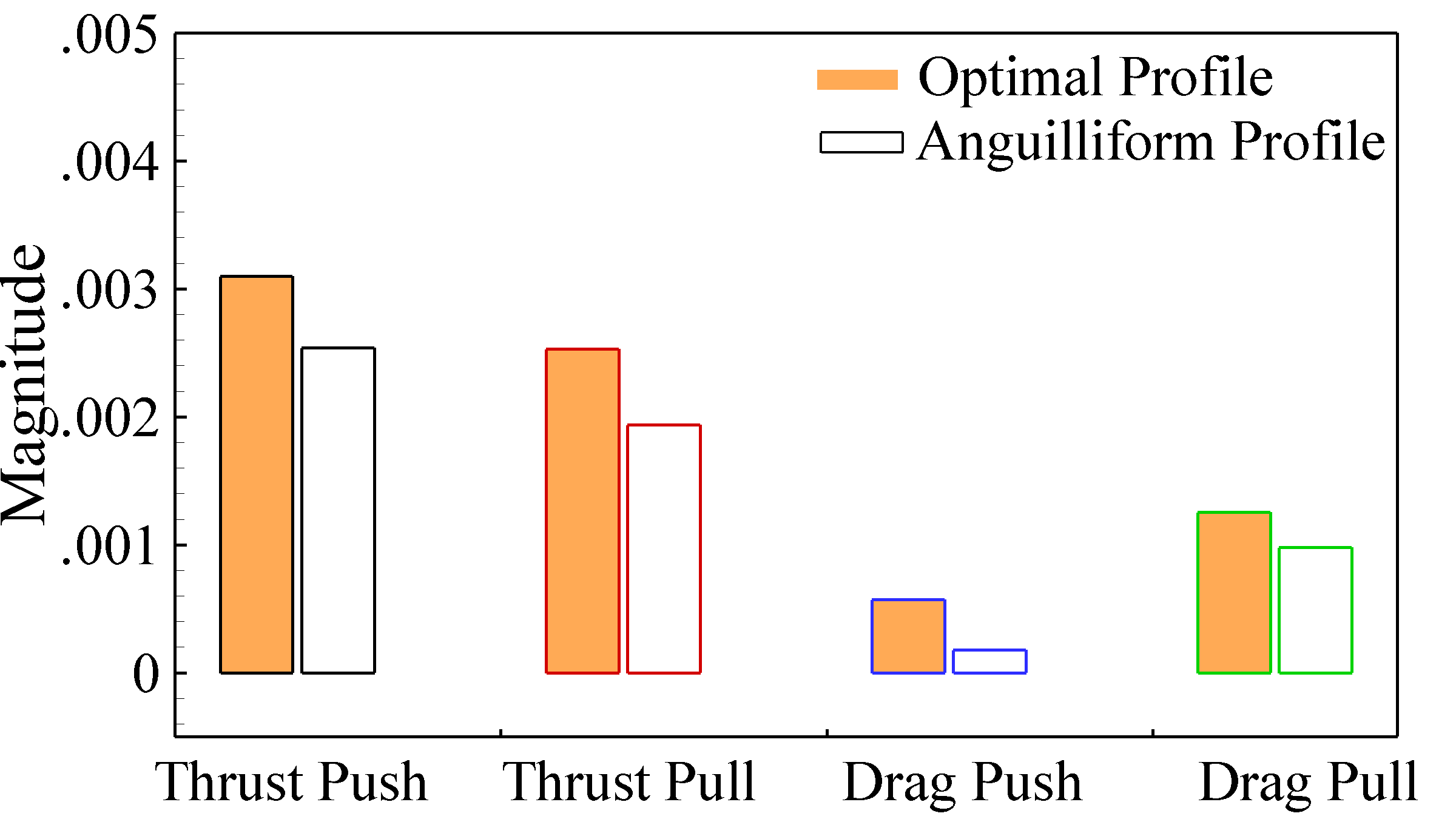}}
\subfigure[]{\includegraphics[width=0.48\linewidth]{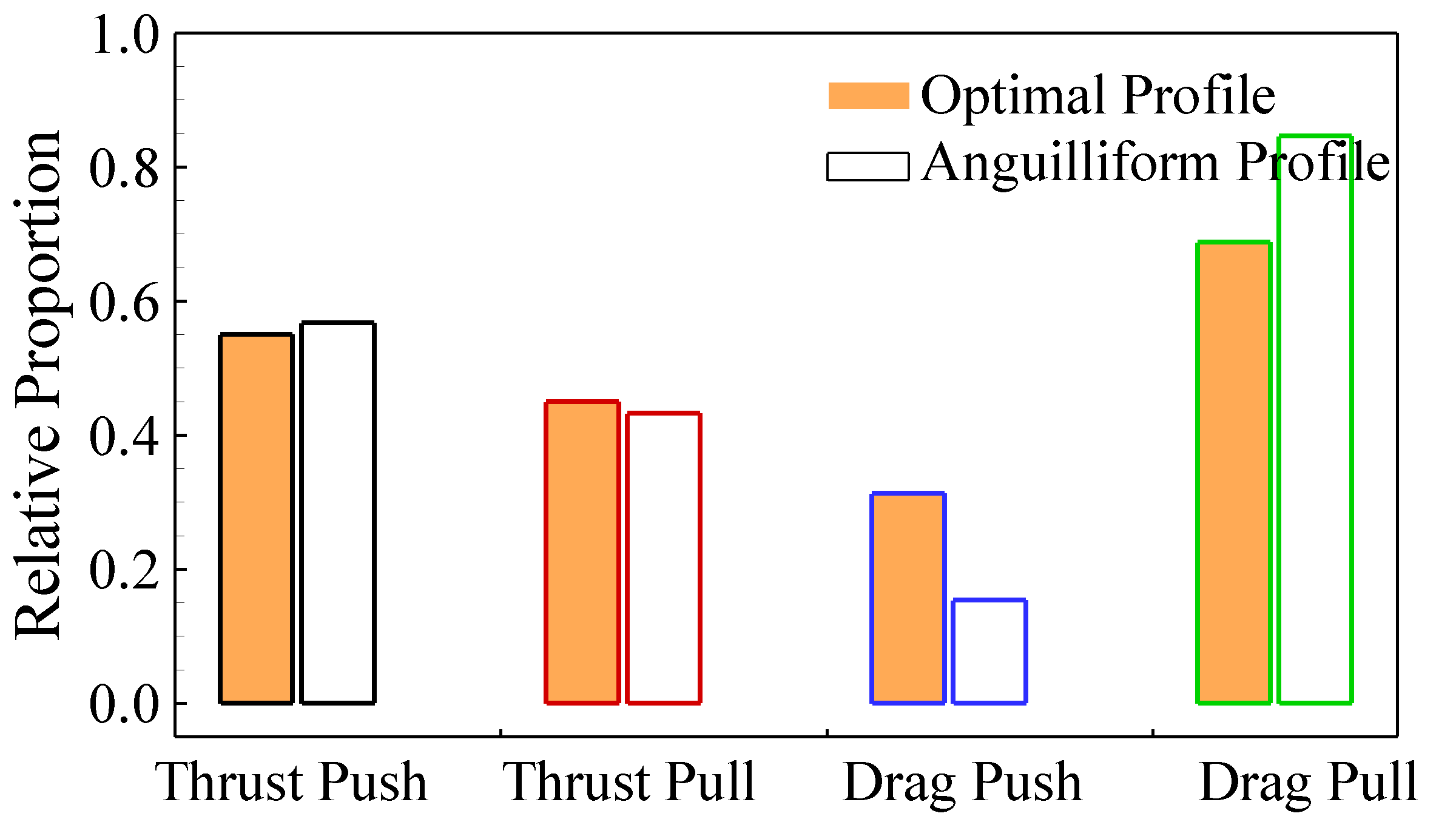}}
\caption{(\textbf{a}) Comparison of the magnitude of surface force components, thrust-push, thrust-pull, drag-push, and drag-pull, between the optimal and anguilliform profiles, aggregated over one undulatory cycle. The optimal profile exhibits noticeably higher amplitudes in all four components, indicating stronger flow-body interactions.
(\textbf{b}) Relative contribution of push- and pull-type forces in both thrust and drag for the optimal and anguilliform profiles. The optimal profile shows a dominant push-type contribution, whereas the anguilliform motion is primarily driven by pull-type forces, emphasizing the more efficient propulsion strategy in the optimized case. \textbf{Note:} All force values are computed using 100 snapshots sampled uniformly over one complete undulatory cycle.} 
\label{fig:ThrustDrag}
\end{center}
\end{figure}

In the optimal profile, the anterior region along with the subsequent 15–30\% of the body length experiences a relatively stronger thrust pull compared to the anguilliform mode at several time instants during the undulatory cycle. In addition to the observed thrust pull, the anterior region of the optimal profile also encounters additional drag forces during certain phases of the undulatory cycle. Since the motion from the midbody to the posterior region is largely similar between the optimal and anguilliform profiles, no significant qualitative differences are observed in the spatial distributions of thrust and drag components or in the corresponding push and pull forces. However, the magnitudes of all four force components are noticeably higher in the optimal profile, specifically in the 66-95\% region, and thus, lead to faster propulsion. This comparison is also illustrated in Fig.~\ref{fig:ThrustDrag}a, where the bar graphs demonstrate the enhanced force amplitudes associated with the optimized motion. Furthermore, the thrust and drag forces in the optimal and anguilliform profiles are predominantly governed by push-and pull-type contributions, respectively. This is illustrated in Fig.~\ref{fig:ThrustDrag}b, where the relative proportions of the push and pull forces are presented as bar plots, highlighting the marginally dominant push-type thrust contribution in the optimized configuration. As a consequence, it is revealed that the anterior and posterior regions of the optimal profile play a pivotal role in its superior performance, exhibiting motion characteristics and force distributions. 

\begin{figure}[!ht]
\includegraphics[width=0.48\linewidth]{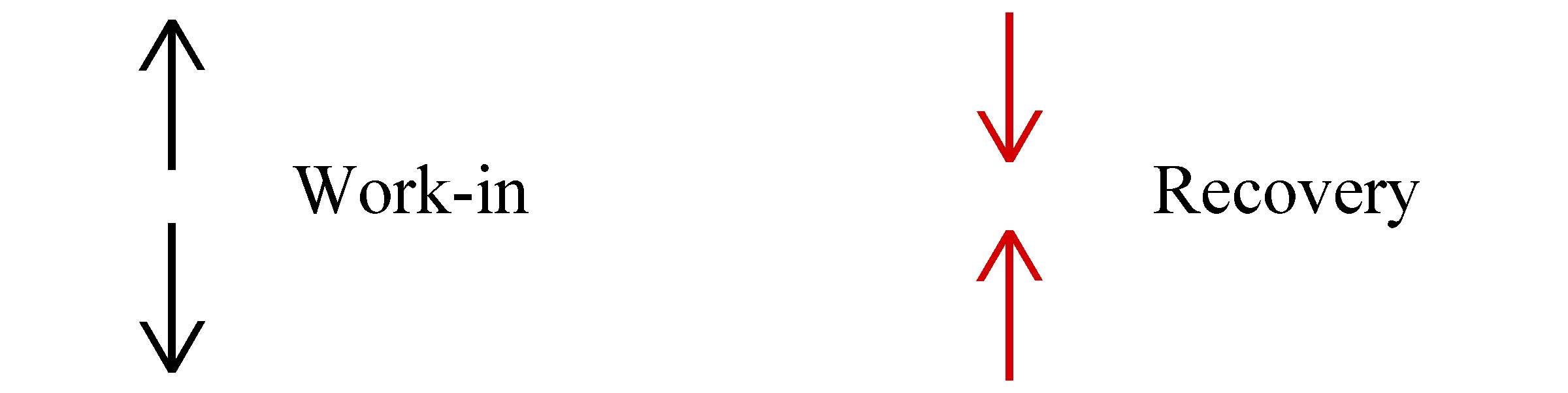}
\begin{center}
\subfigure{\includegraphics[width=0.48\linewidth]{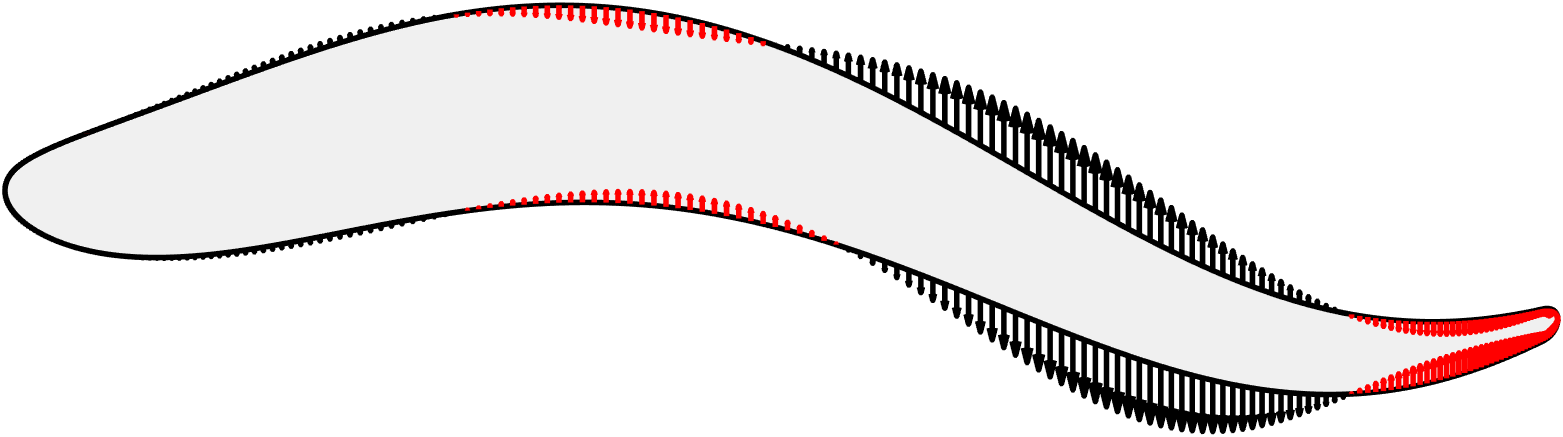}}
\subfigure{\includegraphics[width=0.48\linewidth]{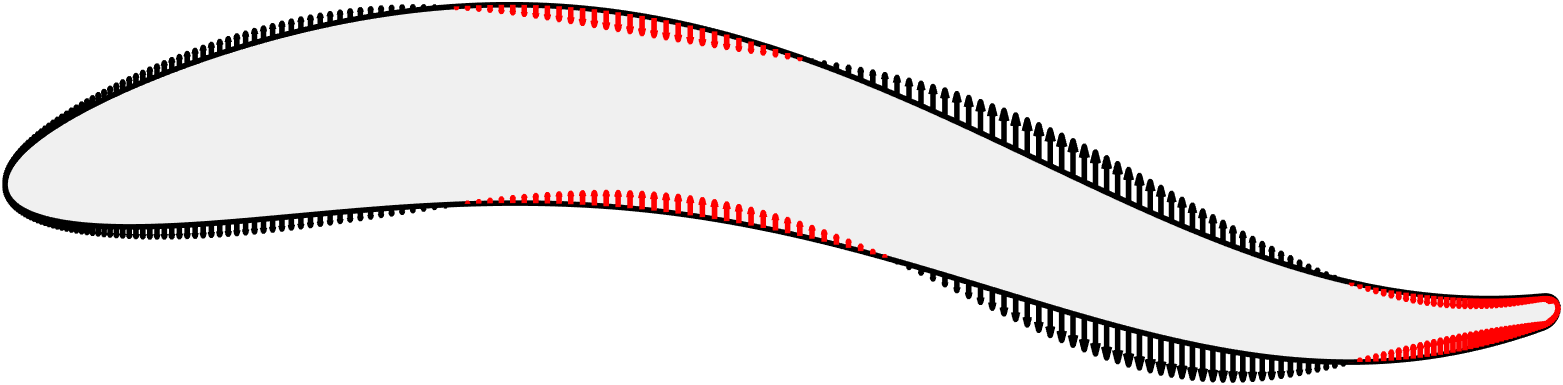}}
\subfigure{\includegraphics[width=0.48\linewidth]{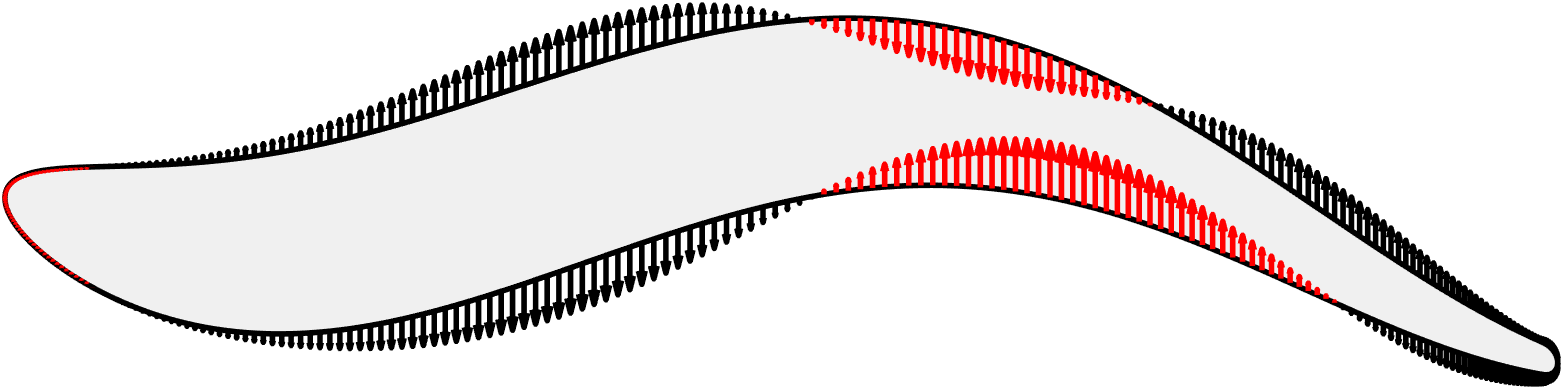}}
\subfigure{\includegraphics[width=0.48\linewidth]{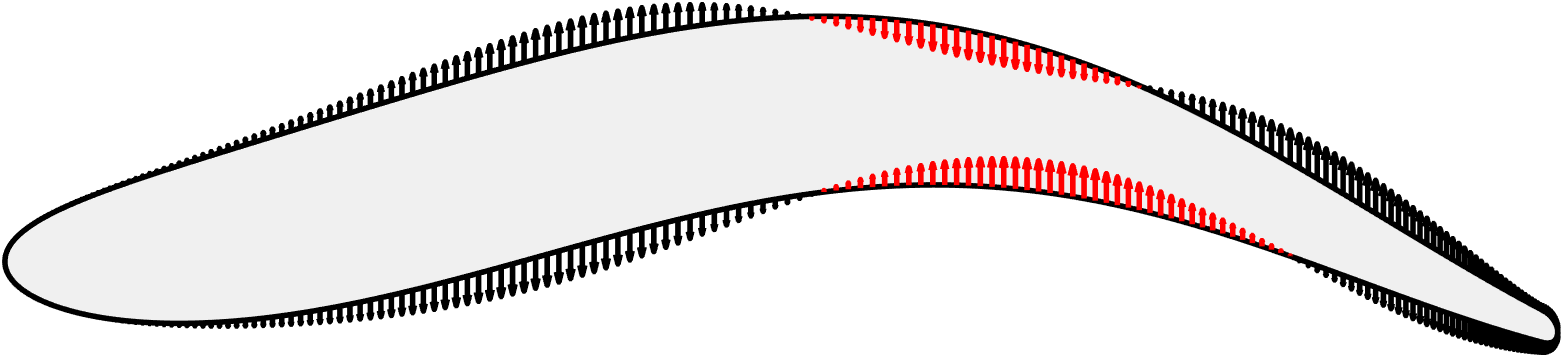}}
\subfigure{\includegraphics[width=0.48\linewidth]{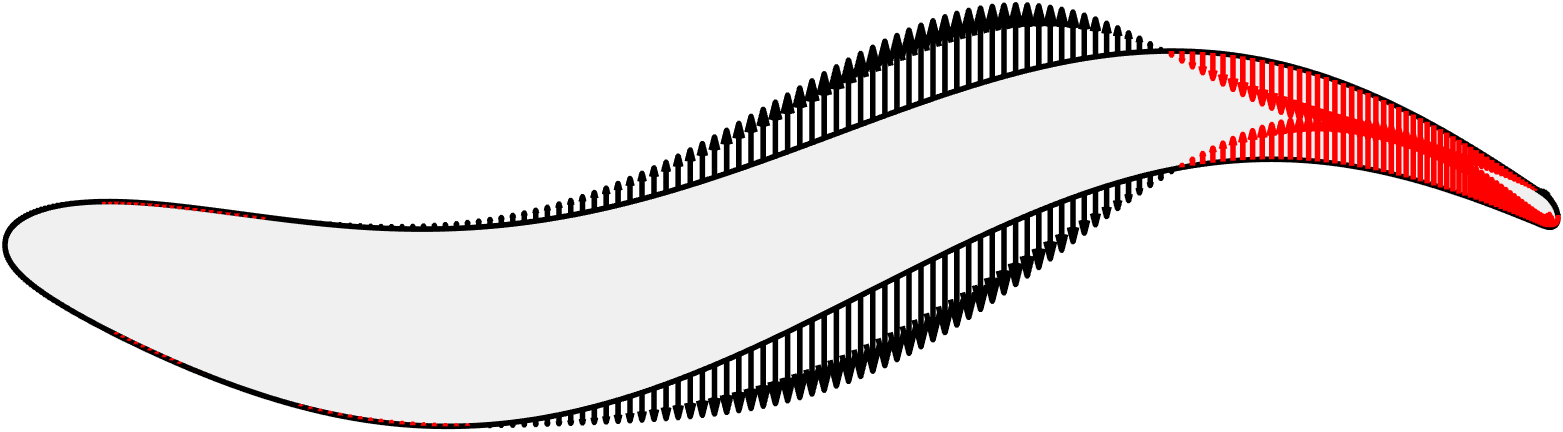}}
\subfigure{\includegraphics[width=0.48\linewidth]{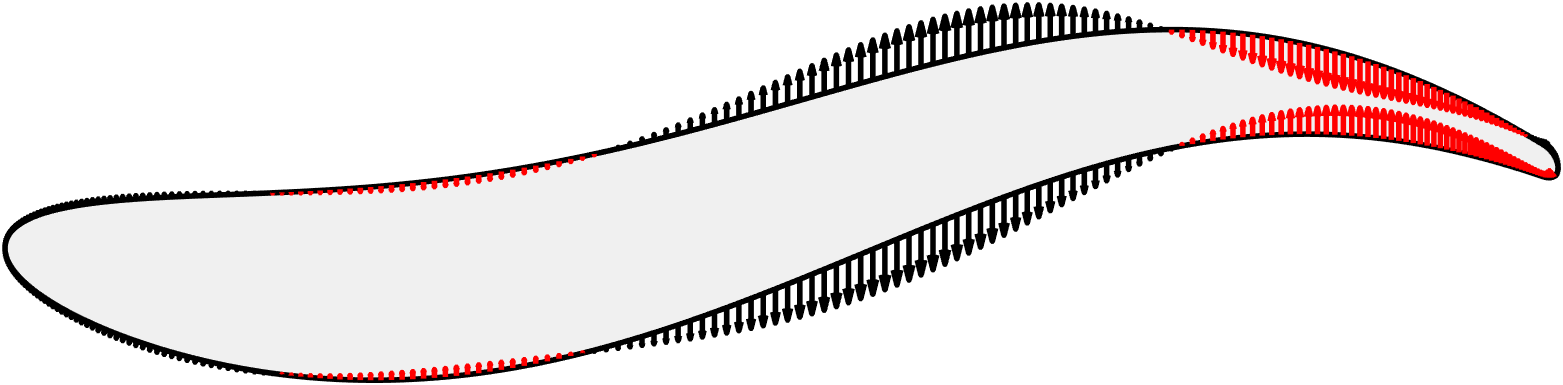}}
\subfigure{\includegraphics[width=0.48\linewidth]{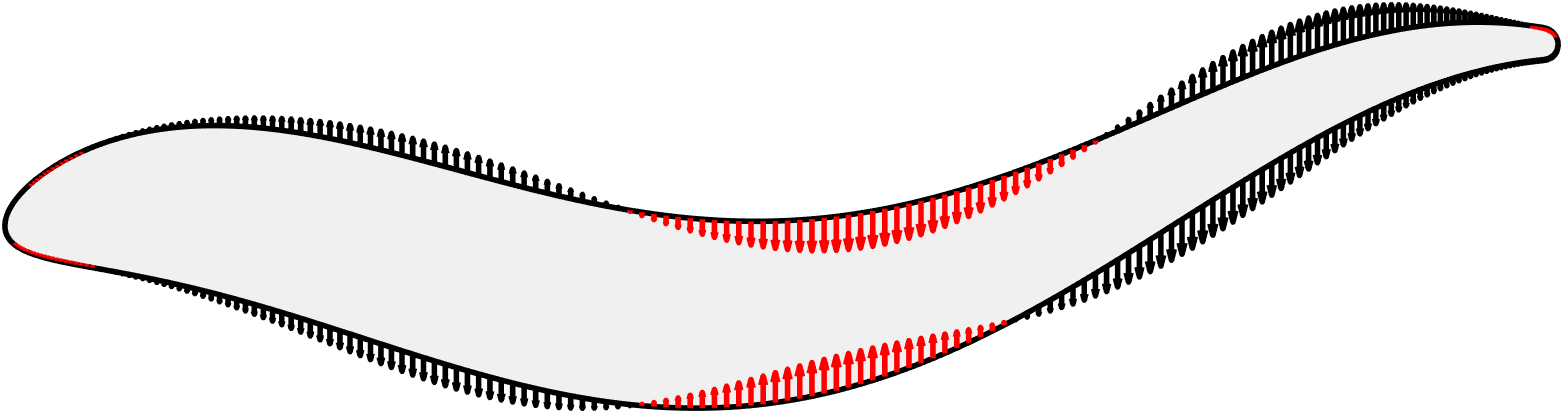}}
\subfigure{\includegraphics[width=0.48\linewidth]{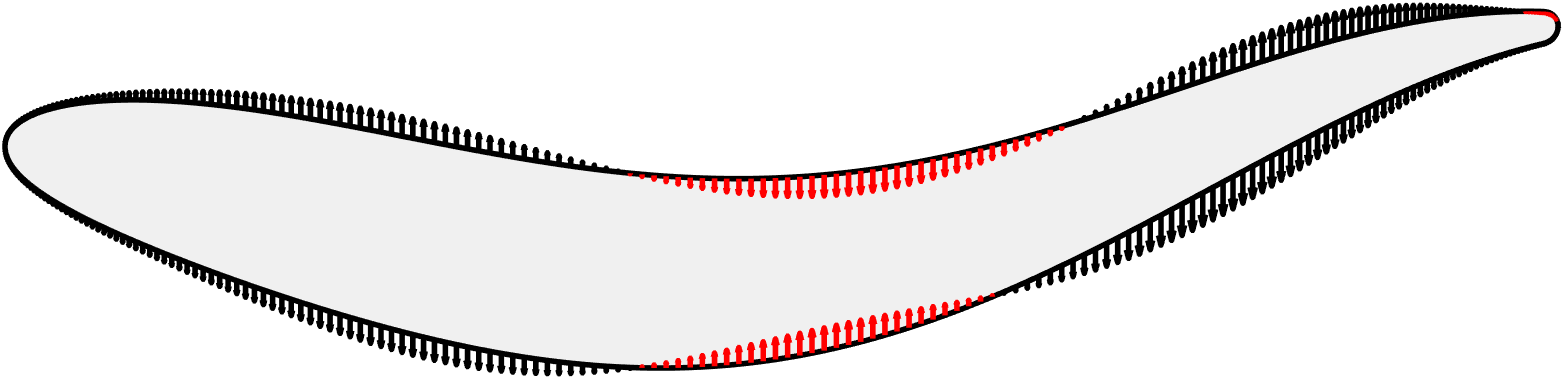}}
\subfigure{\includegraphics[width=0.48\linewidth]{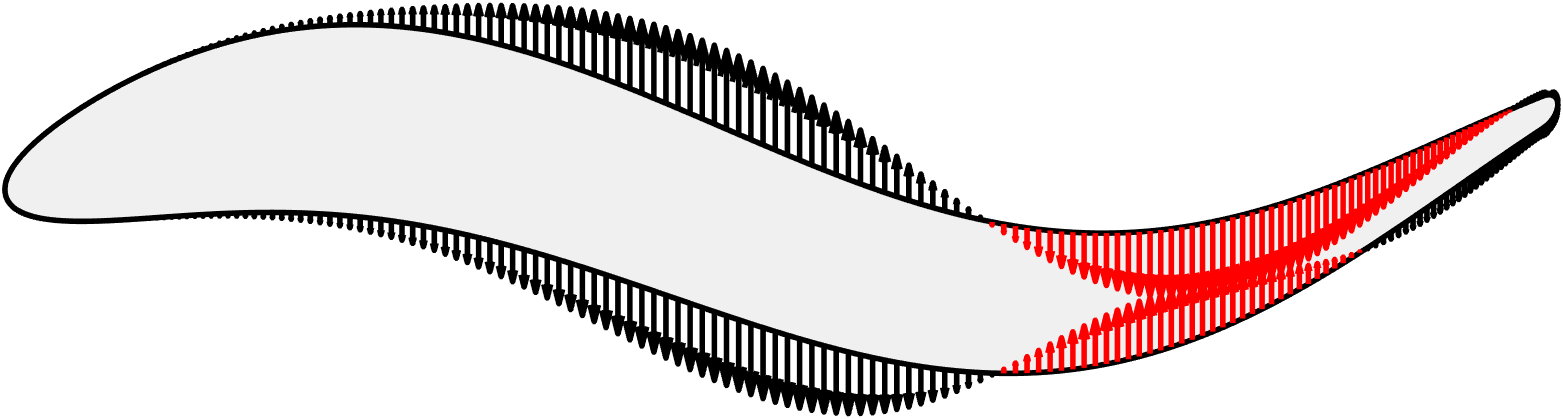}}
\subfigure{\includegraphics[width=0.48\linewidth]{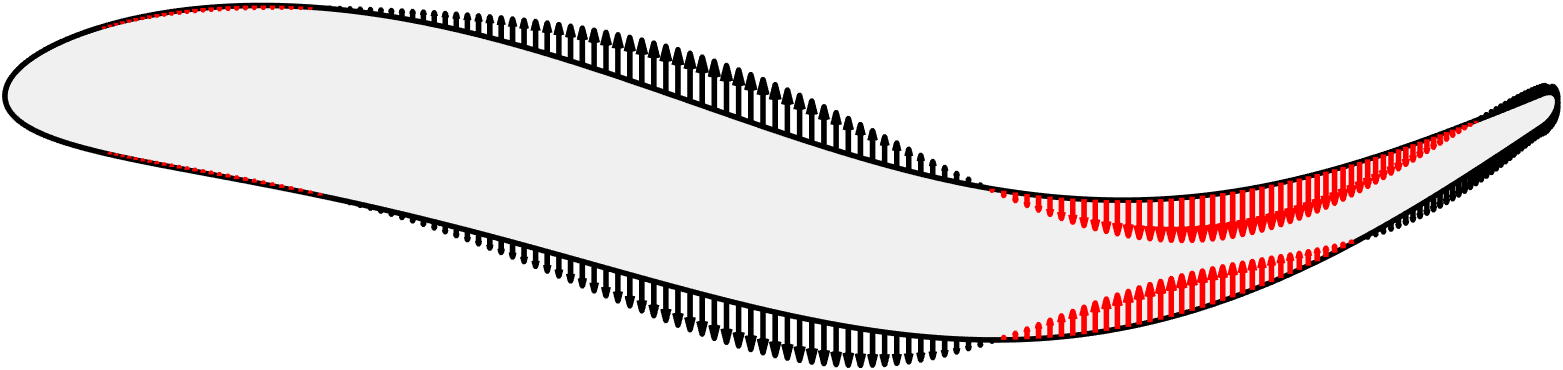}}
\caption{Instantaneous surface work distribution on the optimal (left column) and anguilliform (right column) swimming profiles at five representative time instants over one undulatory cycle. Vertical arrows indicate local work (or energy) transfer: outward arrows represent positive work (work done by the foil on the surrounding fluid), and inward arrows represent negative work (work done by the fluid on the foil). This decomposition highlights the spatiotemporal interplay between energy input and recovery along the body surface.} 
\label{fig:work}
\end{center}
\end{figure}

\begin{figure}[!ht]
\begin{center}
\subfigure[]{\includegraphics[width=0.48\linewidth]{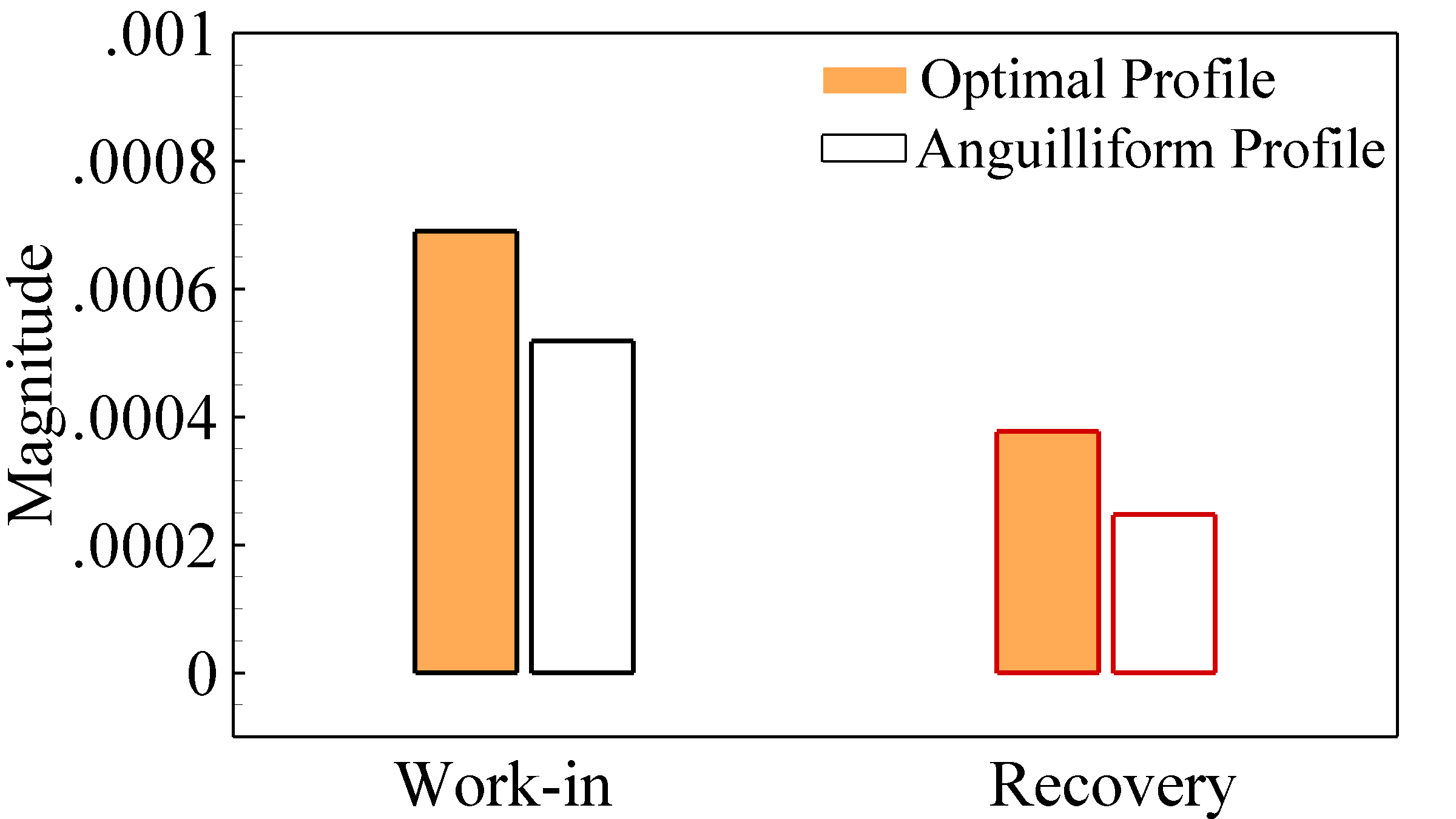}}
\subfigure[]{\includegraphics[width=0.48\linewidth]{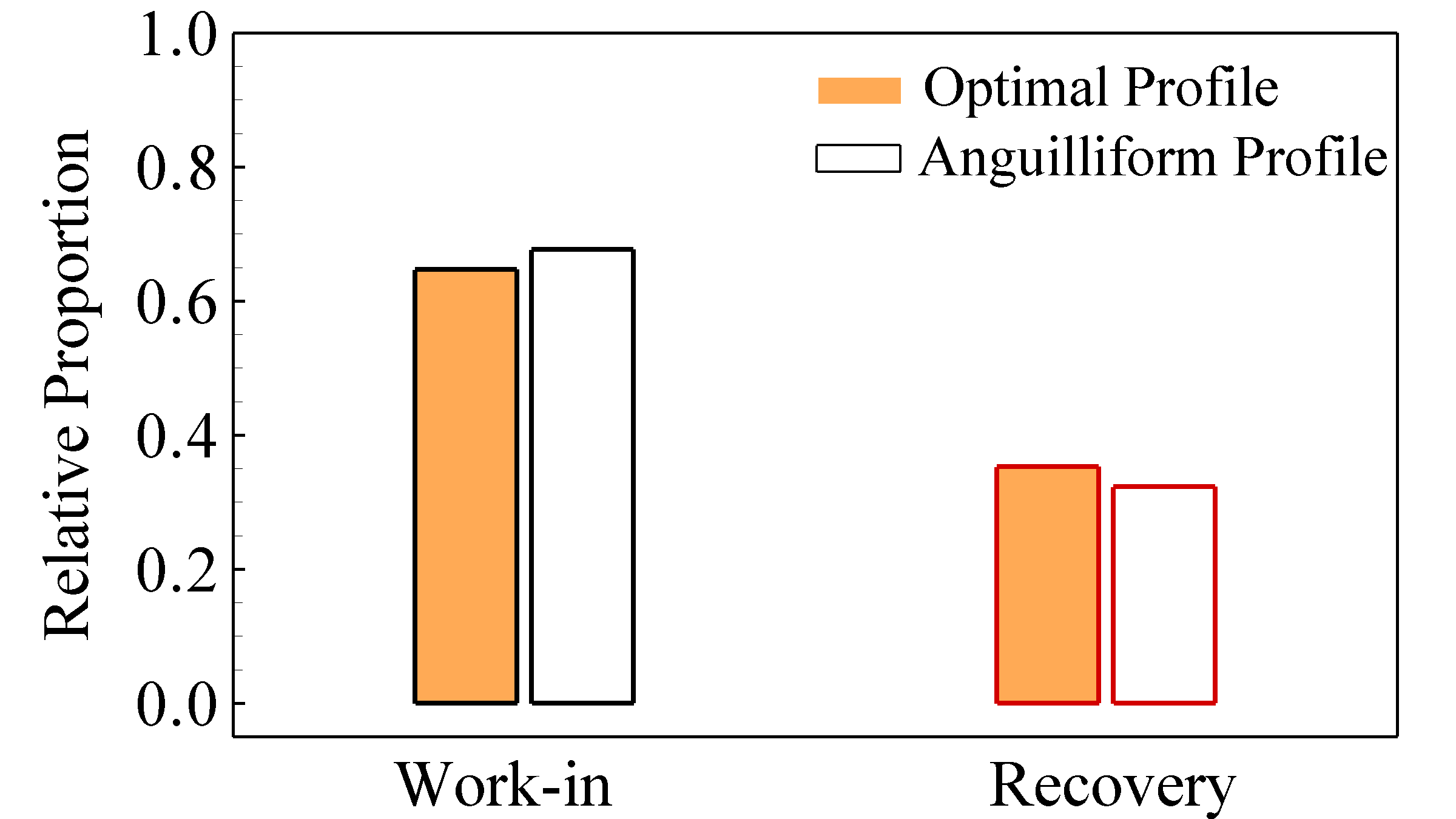}}
\subfigure[]{\includegraphics[width=0.48\linewidth]{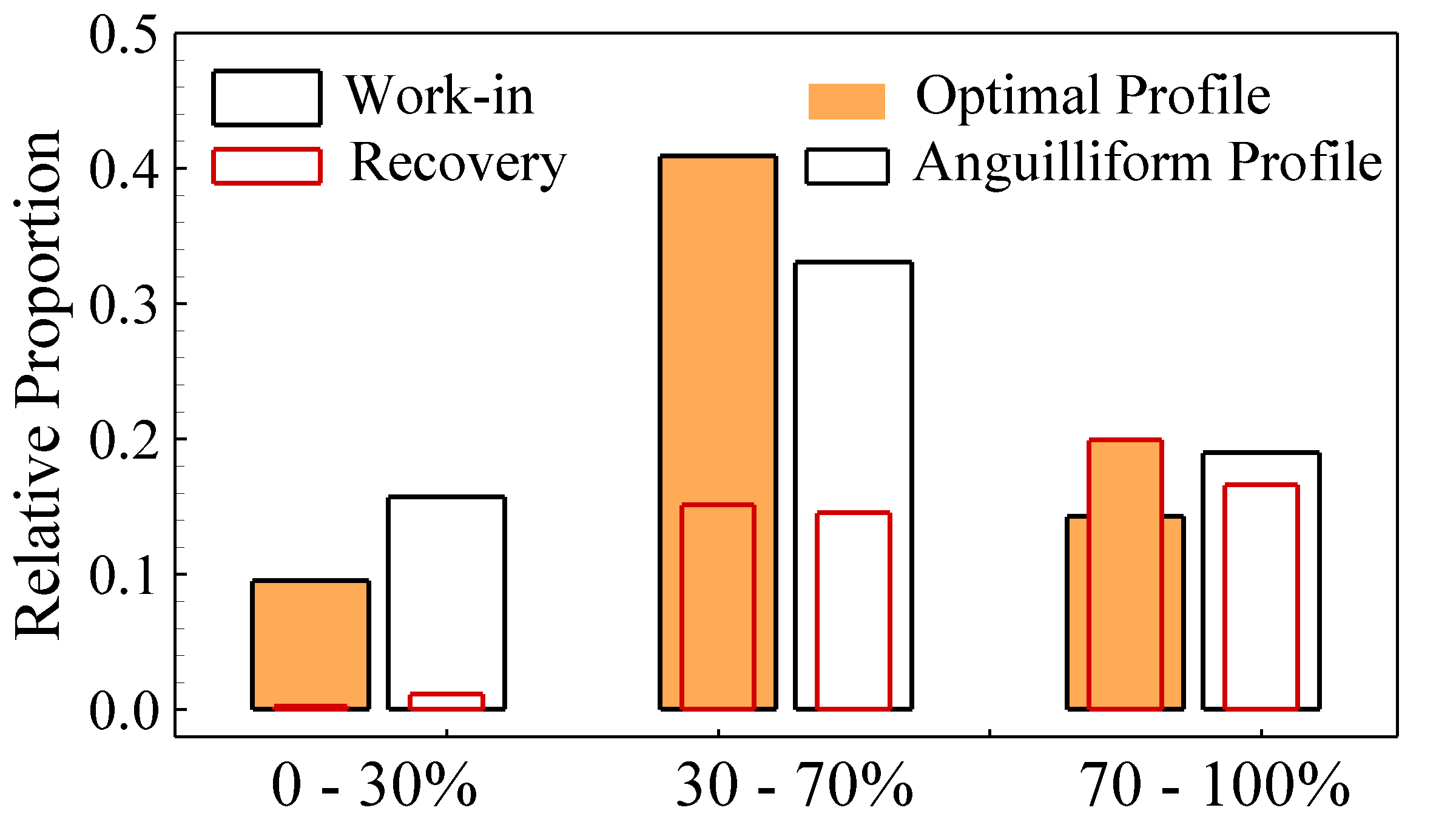}}
\caption{(\textbf{a}) Integrated magnitudes of positive (work-in) and negative (recovery) work components within a complete undulatory cycle for the optimal and anguilliform profiles. (\textbf{b}) Corresponding relative percentages of work-in and recovery. (\textbf{c}) Relative proportion of work-in and recovery at three distinct segments of the body: anterior (0–30\% of body length), midbody (30–70\%), and posterior (70–100\%). \ed{It is important to note that thicker bars are used to represent the work-in, while thinner bars indicate the recovered work.}} 
\label{fig:WorkDecomp}
\end{center}
\end{figure}

Despite achieving faster propulsion, the optimal profile exhibits reduced relative work (or energy) consumption associated with body undulation. To investigate this characteristic, a similar decomposition of surface work is performed, distinguishing between positive work (or work-in) and negative work (or recovery). This decomposition, visualized at five representative time instants during an undulating cycle using vertical arrows in Fig.~\ref{fig:work}, illustrates the distribution of surface work along the body. Outward-pointing arrows indicate work-in, where the foil performs work on the surrounding fluid, while inward-pointing arrows represent recovery, signifying regions where the fluid performs work on the foil, thereby assisting the undulatory motion in terms of energy recovery. Again, it is evident that the anterior and posterior regions are primarily responsible for the differences in work-in and recovery between the profiles. Both the work-in and recovery components are found to be significantly higher in the optimal profile compared to the anguilliform profile. Figure~\ref{fig:WorkDecomp}a illustrates the integrated magnitudes of these components, computed over 100 time instants within one undulatory cycle. These bar plots clearly show that the optimal profile not only performs greater positive work but also benefits from enhanced energy recovery. Moreover, the relative proportion between work-in and recovery, depicted in Fig.~\ref{fig:WorkDecomp}b, reveals that the anguilliform profile exhibits a higher ratio of work-in to recovery compared to the optimal profile. It suggests that the optimal profile requires relatively less mechanical effort based on its superior performance to sustain its undulatory motion, which in turn contributes to its enhanced propulsive efficiency.

To identify the contribution of different body segments to the work-in and energy recovery during an undulatory cycle, the body is divided into three distinct regions: anterior (0–30\% of body length), midbody (30–70\%), and posterior (70–100\%). The work-in and recovery contributions from each region are then analyzed over a complete undulation cycle. The relative proportions contributing to work-in and energy recovery across the three defined body regions for both the optimal and anguilliform profiles are presented in Fig.~\ref{fig:WorkDecomp}c using bar graphs. A detailed comparison of the regional energy dynamics reveals distinct performance characteristics across the three body segments. In the anterior region, the optimal profile performs less mechanical work on the surrounding fluid compared to the anguilliform profile, and although the corresponding energy recovery is also lower, the net energy balance is more favorable. It indicates that the optimal profile requires reduced mechanical effort to undulate its anterior portion. In the midbody region, both profiles exhibit similar levels of energy recovery. However, the anguilliform profile exhibits slightly lower work-in, highlighting more efficient energy exchange in this segment by anguilliform profile. In contrast, the posterior region of the optimal profile demonstrates a substantial increase in energy recovery and a decrease in work-in relative to the anguilliform profile. This highlights the posterior segment as a key contributor to energy reutilization and overall propulsive efficiency in the optimal profile. Thus, the optimal profile emerges as an efficient swimmer by leveraging reduced mechanical effort in the anterior region and enhanced energy recovery in the posterior region, collectively contributing to its superior propulsive performance.

\clearpage
\section{Conclusions}\label{sec13}

This work introduced a morphing-based parametric design framework, integrated with Bayesian optimization, to identify energetically efficient undulatory swimming profiles. By expressing the body deformation as a linear combination of diverse baseline shapes, the proposed method was able to explore a broad design space that includes both conventional and unconventional swimming kinematics. \ed{The optimization process reveals a best-performing profile that attains a peak propulsive efficiency of approximately $57\%$ at a low undulation frequency of $f = 0.2$ and maintains high efficiency levels between $49\%$ and $57\%$ over a broad region of the kinematic space, corresponding to an overall enhancement of roughly $16\%$--$35\%$ compared to conventional anguilliform and carangiform swimming kinematics. Our results also explain that the optimal profile is closely aligned with the anguilliform swimming mode, with the distinction that the anterior motion exhibits a phase lag. It can be concluded that the nature-inspired anguilliform swimming mode, with modifications introduced through the proposed optimization strategy, leads to superior performance.}

The hydrodynamic analysis \edt{demonstrate} that the optimal profile reduces resistive drag and strategically redistributes work input and energy recovery along the body. The decomposition of hydrodynamic work highlights a more favorable phase relationship between fluid forces and body motion, which enables effective energy recapture and minimizes unnecessary power expenditure. Wake topology examination further demonstrated that the optimized swimmer generates coherent and stable vortex structures, which indicates improved momentum transfer. The findings emphasize the capability of morphing-based design, coupled with surrogate-assisted Bayesian optimization, to uncover novel swimming gaits with superior propulsive efficiency. This approach offers valuable insight into the hydrodynamic principles underlying efficient undulatory propulsion and holds promise for the development of next-generation bio-inspired underwater vehicles.






\section*{Acknowledgements}

\edt{MSU Khalid acknowledges the funding support from the Natural Sciences and Engineering Research Council of Canada (NSERC) through the Discovery grant program. The computational work and simulations reported here were performed on the supercomputing clusters administered and managed by the Digital Research Alliance of Canada.} The authors would also like to acknowledge the invaluable comments regarding arc-length preservation from a peer-reviewer on an earlier draft of this paper, which led to significant improvement of our work.





\bigskip







\bibliographystyle{plainnat}  
\bibliography{references}

\end{document}